%% file: 202411XX_arxiv_rmueller_tensor_model_for_us.tex
\documentclass[journal,11pt,left = 1.25in, right = 1.25in, top = 1in, bottom = 1in, onecolumn,final,]{IEEEtran}

\usepackage{amssymb}
\usepackage{amsthm}
\usepackage{mymacros}
\usepackage{hyperref}
\usepackage{cleveref}
\usepackage{glossaries}
\usepackage{enumitem}
\usepackage{pgfgantt}
\usepackage[subrefformat=parens,labelformat=simple]{subcaption}
\usepackage{floatrow}
\usepackage{stfloats}
\newfloatcommand{capbtabbox}{table}[][\FBwidth]
\usepackage{pgfplots}
\usepackage{tikz}
\usetikzlibrary{shapes.geometric, arrows, shadows}
\usetikzlibrary{fit,backgrounds}
\usetikzlibrary{shadows.blur}
\usetikzlibrary{spy}
\usepackage{physics}
\usepackage{tikz-3dplot}
\usepackage[outline]{contour}
\tikzstyle{vector}=[-stealth,thick,line cap=round]
\tikzset{>=latex}
\usepackage{xcolor}
\usepackage{soul}
\usepackage{tabularx,multirow}
\usepackage{algorithm}
\usepackage{algpseudocode}
\usepackage{bm}
\usepackage{mathtools}
\usepackage{siunitx}
\usepackage[short]{optidef}
\usepackage[normalem]{ulem}

\Crefname{equation}{}{}
\Crefname{figure}{Figure}{Figures}
\Crefname{section}{Section}{Sections}
\Crefname{appendix}{}{}
\Crefname{algorithm}{Procedure}{}

\DeclareRobustCommand{\titlebm}[1]{\bm{#1}}


\let\OLDthebibliography\thebibliography
\renewcommand\thebibliography[1]{
	\OLDthebibliography{#1}
	\setlength{\parskip}{0pt}
}

\newtheorem{prop}{Proposition}
\renewenvironment{proof}[1][Proof]{%
	\par\pushQED{\qed}\normalfont\topsep6pt\trivlist
	\item[\hskip\labelsep\itshape #1:]\ignorespaces
}{%
	\popQED\endtrivlist
}
\newcommand{\sketchofproof}[1]{\textit{Sketch of Proof for}#1}

\setacronymstyle{long-short}
\newacronym{ALS}{ALS}{alternating least squares}
\newacronym{BCD}{BCD}{block coordinate descent}
\newacronym{CANDECOMP}{CANDECOMP}{canonical decomposition}
\newacronym{longplural=capacitive micromachined ultrasonic transducers}{CMUT}{capacitive micromachined ultrasonic transducer}
\newacronym{CPD}{CPD}{canonical polyadic decomposition}
\newacronym{DFT}{DFT}{discrete Fourier transform}
\newacronym{doa}{DOA}{direction-of-arrival}
\newacronym[longplural=digital signal processors]{DSP}{DSP}{digital signal processors}
\newacronym{EM}{EM}{electromagnetic}
\newacronym{ESPRIT}{ESPRIT}{estimation of signal parameters via rotational invariance technique}
\newacronym{FFT}{FFT}{fast Fourier transform}
\newacronym{fov}{FOV}{field-of-view}
\newacronym[longplural=field-programmable gate arrays]{FPGA}{FPGA}{field-programmable gate array}
\newacronym[longplural=graphics processing units]{GPU}{GPU}{graphics processing unit}
\newacronym{KKT}{KKT}{Karush-Kuhn-Tucker}
\newacronym{lambda-half}{$\lambda/2$}{half-wavelength}
\newacronym{LTI}{LTI}{linear time-invariant}
\newacronym{LS}{LS}{least-squares}
\newacronym{MEMS}{MEMS}{microelectromechanical systems}
\newacronym[longplural = multiple measurement vectors]{MMV}{MMV}{multiple measurement vector}
\newacronym[longplural = mean complementary normalized cross-correlations]{MCNCC}{MCNCC}{mean complementary normalized cross-correlation}
\newacronym{MIMO}{MIMO}{multiple-input multiple-output}
\newacronym{MP}{MP}{matching pursuit}
\newacronym{m-t-av}{$m$-$t$-av}{$m$-$t$-average}
\newacronym{MUSIC}{MUSIC}{multiple signal classification}
\newacronym{OMP}{OMP}{orthogonal matching pursuit}
\newacronym{OLS}{OLS}{orthogonal least squares}
\newacronym{PARAFAC}{PARAFAC}{parallel factor}
\newacronym[longplural = piezoelectric ultrasonic transducers]{PUT}{PUT}{piezoelectric ultrasonic transducer}
\newacronym{RMSE}{RMSE}{root mean squared error}
\newacronym{t-av}{$t$-av}{$t$-average}
\newacronym[longplural = uniform linear arrays]{ULA}{ULA}{uniform linear array}
\newacronym{URA}{URA}{uniform rectangular array}
\newacronym{US}{US}{ultrasound}
\newacronym{voi}{VOI}{volume-of-interest}
\newacronym{1D}{1D}{one-dimensional}
\newacronym{2D}{2D}{two-dimensional}
\newacronym{3D}{3D}{three-dimensional}

\tikzset{rect/.style={rectangle, rounded corners, minimum width=4.5cm, minimum
		height=1cm,text centered, draw=black, fill=blue!10,blur shadow},
	arrow/.style={thick,->,>=stealth}}

\usepackage{graphicx}
\usepackage{orcidlink}
\usepackage{textcomp}
\usepackage{cite}
\usepackage{todonotes}
\usepackage{CJK}
\newcommand{\orcid}[1]{\href{https://orcid.org/#1}{\includegraphics[width=8pt]{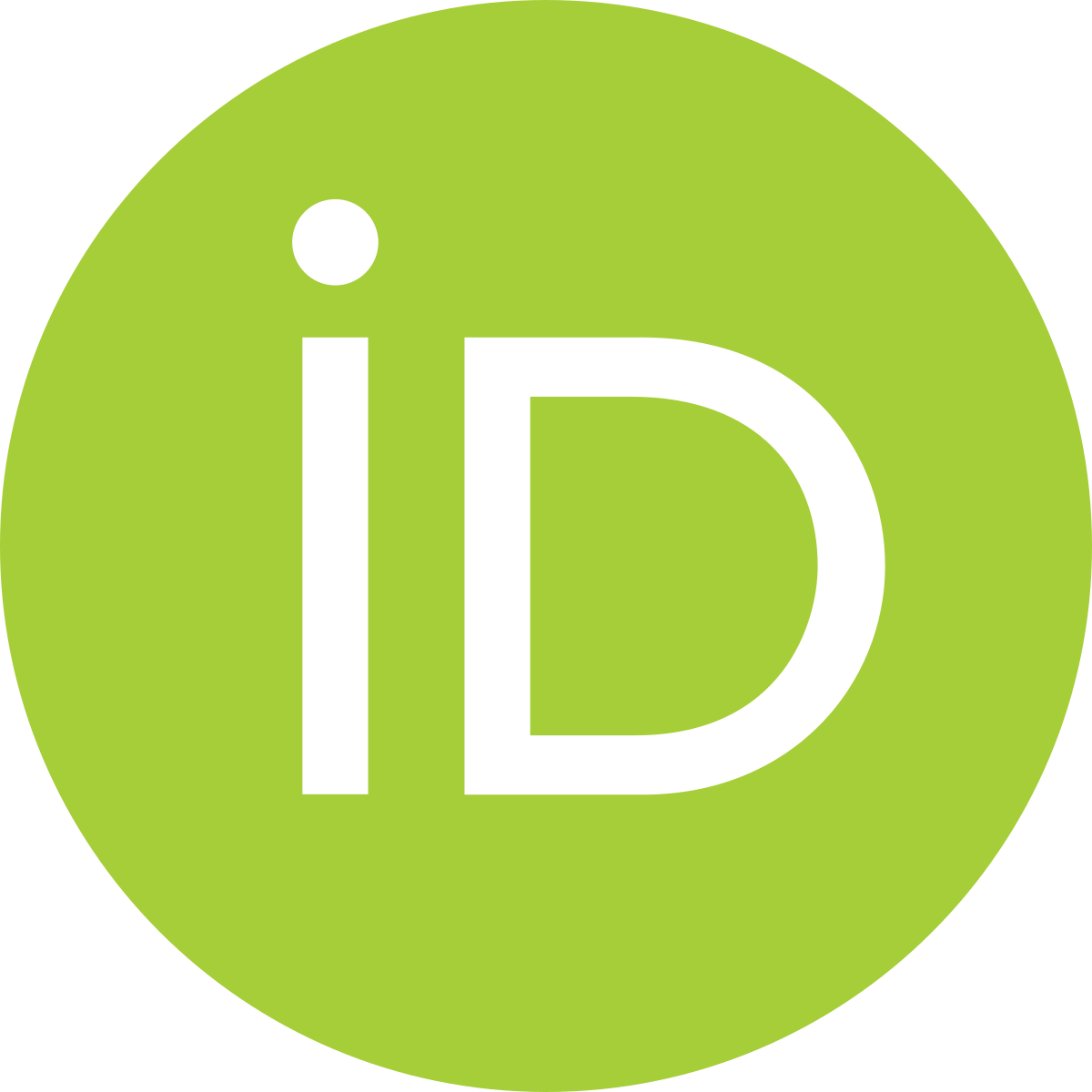}}}

\newcommand\copyrighttext{%
	\footnotesize Accepted manuscript licensed under CC BY-NC-ND 4.0. To view a copy of this license, visit \href{https://creativecommons.org/licenses/by-nc-nd/4.0/}{https://creativecommons.org/licenses/by-nc-nd/4.0/}. Formal publication by Elsevier B.V. available via the following DOI: \href{https://doi.org/10.1016/j.sigpro.2024.109812}{10.1016/j.sigpro.2024.109812}.
}
\makeatletter
\def\ps@IEEEtitlepagestyle{
	\def\@oddfoot{\mycopyrightnotice}
	\def\@evenfoot{}
}
\def\mycopyrightnotice{
	{\footnotesize
		\begin{minipage}{\textwidth-2\fboxsep}%
			\centering%
			\noindent\fbox{\parbox{\linewidth}{\copyrighttext}}
		\end{minipage}
	}
}

\begin{document}
\begin{CJK*}{UTF8}{gbsn}
	
\title{A tensor model for the calibration of air-coupled ultrasonic sensor arrays in 3D imaging}

\author{Raphael~M\"uller,
	Gianni~Allevato,
	Matthias~Rutsch,
	Christoph~Haugwitz,
	Tianyi~Liu~(刘添翼),
	Mario~Kupnik,
	Marius~Pesavento%
	\thanks{This work has received funding from the German Federal Ministry of Education and Research (BMBF) within the project ``Open6GHub'' under grant number 16KISK014.}%
	\thanks{Raphael M\"uller \orcid{0000-0003-4614-9663}, Tianyi Liu (刘添翼) \orcid{0000-0001-8338-1651}, and Marius Pesavento \orcid{0000-0003-3395-2588} are with the Communication
		Systems Group, TU Darmstadt, Darmstadt, Germany
		(e-mail: \{{raphael.mueller}, {tianyi.liu}, {pesavento}\}@nt.tu-darmstadt.de).}%
	\thanks{Gianni Allevato \orcid{0000-0002-8852-6773} is with Hottinger Br\"uel \& Kj{\ae}r GmbH, Darmstadt, Germany.}%
	\thanks{Matthias Rutsch \orcid{0000-0001-6705-9793} is with Robert Bosch GmbH, Leonberg, Germany.}%
	\thanks{Christoph Haugwitz \orcid{0000-0001-6756-9988} and Mario Kupnik \orcid{0000-0003-2287-4481} are with the Measurement and Sensor Technology Group, TU Darmstadt, Darmstadt, Germany
		(e-mail: \{{christoph.haugwitz}, {mario.kupnik}\}@tu-darmstadt.de).}
}

\maketitle

\begin{abstract}
	Arrays of ultrasonic sensors are capable of 3D imaging in air and an affordable supplement to other sensing modalities, such as radar, lidar, and camera, i.e.\@ in heterogeneous sensing systems. However, manufacturing tolerances of air-coupled ultrasonic sensors may lead to amplitude and phase deviations. Together with artifacts from imperfect knowledge of the array geometry, there are numerous factors that can impair the imaging performance of an array. We propose a reference-based calibration method to overcome possible limitations. First, we introduce a novel tensor signal model to capture the characteristics of \glspl{PUT} and the underlying multidimensional nature of a \gls{MIMO} sensor array. Second, we formulate and solve an optimization problem based on this model to obtain the calibrated parameters of the array. Third, we assess both our model and the commonly used analytic model using real data from a 3D imaging experiment.	The experiment reveals that our array response model we learned with calibration data yields an imaging performance similar to that of the analytic array model, which requires perfect array geometry information.
\end{abstract}

\glsresetall

\section{Introduction}
\label{sec:introduction}
			
\IEEEPARstart{W}{hen} considering imaging with \gls{US}, the first thing that comes to mind is medical imaging \cite{huangFastReductionSpeckle2013,yuEnvelopeSignalBased2012,kiymikUltrasoundImagingBased1997}. With the advent of cost-efficient high-power sensors, however, \gls{3D} ultrasonic imaging, which combines \gls{2D} direction finding and range estimation, has been extended to operate in air \cite{harputUltrasonicPhasedArray2008}. In general, air-coupled ultrasonic imaging builds on phased array technology and uses similar image formation principles as medical sonography and \gls{MIMO} radar, e.g.\@ beamforming techniques \cite{synnevagAdaptiveBeamformingApplied2007} and synthetic apertures \cite{jensenSyntheticApertureUltrasound2006}.

The main advantages of \gls{US} over other air-coupled sensing modalities are its low deployment cost, its robustness against particles in the air, e.g.\@ smoke, dust, or rainfall, as well as its ability to operate independently of the lighting conditions \cite{leightonWhatUltrasound2007,mohammedPerceptionSystemIntelligent2020}. In addition, acoustic waves offer distinct imaging characteristics compared to \gls{EM} waves or light due to the different system parameters such as the wave propagation speed or material reflectivity. This makes air-coupled \gls{US} attractive for heterogeneous sensing systems where different sensing modalities can complement each other as frequently encountered in autonomous driving \cite{patoleAutomotiveRadarsReview2017} or, more recently, in healthcare \cite{fangReviewEmergingElectromagneticacoustic2022}.

The major obstacle in air-coupled \gls{US} is the poor coupling of acoustic waves into air, as well as the high attenuation and the low propagation speed of \gls{US} in air as compared to fluids or tissue \cite{chimentiReviewAircoupledUltrasonic2014,haller13CompositesUltrasonic1992,lionettoAircoupledUltrasoundNovel2007,bassAtmosphericAbsorptionSound1990}. In contrast to medical sonography, the requirement for large coverage areas of airborne \gls{US}, along with the reduced speed of sound in air lead to large propagation delays, thereby slowing down the image frame rate. Additionally, the larger wavelengths of \gls{US} systems in air lead to a reduced angular resolution compared to medical \gls{US} if arrays of the same aperture size are considered. Thus, air-coupled \gls{US} faces a variety of challenges that distinguish it from medical \gls{US}. The latter is also comparatively mature whereas research on \gls{3D} imaging with an \gls{US} array in air has only recently become a study subject as well \cite{strakowskiUltrasonicObstacleDetector2006}.

Air-coupled ultrasonic sensors, also referred to as transducers, are realized with different materials and principles, resulting in varying degrees of sensitivity, mechanical robustness, and costs \cite{dahlApplicationsAirborneUltrasound2014}. The inherent properties of the materials used and the production tolerances of low cost transducers lead to variations in the directivity pattern and frequency responses. Furthermore, differing resonance frequencies in the transmit and receive mode, respectively, result in nonreciprocal transceiver characteristics of \glspl{PUT} such that no common array manifold for transmission and reception exists. These and other factors, e.g.\@ aging or exposure of an imaging system to environmental conditions such as temperature fluctuations, urge the need to learn the transducer array responses from training data, particularly within complex heterogeneous sensing systems where describing the array response analytically is often impractical.

In this work, we consider a bistatic \gls{MIMO} imaging system with a \gls{URA} of \gls{PUT} elements as a prototype \gls{US} system for \gls{3D} imaging in air. \glspl{PUT} are a popular choice for this as they deliver a high output pressure at their resonance frequency while being available for a low unit price, making them attractive for large array configurations in various applications \cite{suzukiAUTD3ScalableAirborne2021,leggUltrasonicArraysRemote2020,rekhiWirelessPowerTransfer2017,marzoHolographicAcousticElements2015}. Concomitant disadvantages include a comparatively small relative bandwidth as well as fluctuations in the element directivity and phase response \cite{gellyComparisonPiezoelectricThickness2003}. As \glspl{PUT} are often too large to be placed at \gls{lambda-half} spacing, the array is embedded into a waveguide for grating-lobe-free beamforming \cite{jagerAircoupled40KHZUltrasonic2017a}. While the waveguide offers protection in harsh environments and enables flexible array shape designs, it introduces further uncertainties in the array response and increases the overall complexity of the imaging system. Therefore, it is necessary to investigate whether these uncertainties adversely affect the imaging performance when using the commonly employed analytic point source model, and to explore how anomalous array responses can be accurately modeled.

We propose a novel multidimensional array model to capture artifacts and uncertainties from imperfections in both the individual transducer elements as well as the array configuration. The model parameters are learned from reference-based calibration data. High-resolution \gls{3D} imaging is accomplished with the virtual array concept where the aperture is spanned by the Kronecker product of the receive array and multiple, spatially displaced transmit sensors. Each transmitter emits \gls{US} pulses into a sparse scenery such that at the receive array, target echoes are recorded in the image data tensor $\bm{\mathcal{Y}}$. During image formation, indexed positions $\hat{p}$ of each potential target are estimated from $\bm{\mathcal{Y}}$ based on the spatial signatures corresponding to $P$ candidate positions that have been learned previously at the calibration stage. For each source position $\hat{p}$, a corresponding gain $\bm{\hat{h}}_{\hat{p}}$ is estimated, whose magnitude is displayed in the final image (\Cref{fig:data-flowchart}).
			
\begin{figure}
	\centering
	\begin{tikzpicture}[node distance=2.75cm,text width=10.5em]
		\node (calibdata) [rect, label={[font=\small\sffamily,name=label1,xshift=6em,yshift=-.5em]below left:{\scriptsize \mbox{Calibration (array response learning)}}}] {\footnotesize Calibration data $\left\{ \bm{\mathcal{Y}}_p \right\}_{p=1}^P$};
		\node (optimization) [rect, right = 5em of calibdata] {\footnotesize \mbox{Parameter estimates} $\left\{ \bm{\hat{g}}_{n, \text{Tx}} \right\}_{n=1}^N$, $\left\{ \bm{\hat{g}}_{m, \text{Rx}} \right\}_{m=1}^M$, $\left\{ \bm{\hat{a}}_{p, \text{Tx}}, \, \bm{\hat{a}}_{p, \text{Rx}}, \, \bm{\hat{c}}_p \right\}_{p=1}^P$};
		\node (imaging) [rect, above right = .75em and 5em of optimization] {\footnotesize Image data $\bm{\mathcal{Y}}$};
		\node (estimation) [rect, right = 5em of optimization, label={[font=\small\sffamily,name=label2,xshift=-4.5em,yshift=-0.5em]below right:{\scriptsize Imaging (target localization)}}] {\footnotesize Localized signal gains $\bm{\hat{h}}_{\hat{p}}$};
		
		\draw [arrow] (calibdata) -- node[midway,above,xshift=3.65em,yshift=-1.05em] {\tiny Decomposition \\ \vspace{-.5em} modified BCD} (optimization);
		\draw [arrow] (optimization) -- node[midway,above,xshift=4.25em,yshift=-.25em] {\tiny Array \\ response \\ \vspace{-1.0em} dictionary} (estimation);
		\draw [arrow] (imaging) -- node[midway,right,yshift=-.25em] {\tiny Data matching} (estimation);
		
		\begin{scope}[on background layer]
			\tikzset{myfit/.style={draw,dashed,gray,rounded corners,fill=yellow!30,
					inner sep=7.5pt}}
			\node[myfit,fit=(calibdata) (optimization) (label1.west)]{};
			\node[myfit,fit=(imaging) (estimation) (label2.west)]{};
		\end{scope}
	\end{tikzpicture}
	\caption{Flowchart for the proposed calibration and imaging steps: calibration data $\{ \bm{\mathcal{Y}}_p \}_{p=1}^P$ from $P$ distinct reference positions, characterized by 3D coordinates, is measured to estimate the parameters $\{ \bm{\hat{g}}_{n, \text{Tx}} \}_{n=1}^N$, $\{ \bm{\hat{g}}_{m, \text{Rx}} \}_{m=1}^M$, and $\{ \bm{\hat{a}}_{p, \text{Tx}}, \, \bm{\hat{a}}_{p, \text{Rx}}, \, \bm{\hat{c}}_p \}_{p=1}^P$. With those, targets in an unknown scene are localized from image data $\bm{\mathcal{Y}}$ by selecting target position estimates ${\hat{p}}$ from the $P$ candidate positions of the array response dictionary learned during calibration. Finally, the target signal gain $\bm{\hat{h}}_{\hat{p}}$ of each target detected is estimated.}
	\label{fig:data-flowchart}
\end{figure}

\subsection{State of the art and own contributions}
			
Array calibration addresses the problem of modeling errors in the array response, most commonly due to unknown sensor amplitudes and phases, mutual coupling (electrical, mechanical, and acoustic cross-talk) between the array elements, or inaccurate knowledge of their relative positions \cite{vibergCalibrationArrayProcessing2009}. Remedies for these imperfections differ based on the availability of additional calibration measurements. Non-reference-based, or \emph{autocalibration}, methods do not require calibration targets at known locations and perform calibration and target localization either alternately or jointly on the same data using parametric models, e.g., the sensor locations, the complex gains, or the coupling matrix are jointly estimated during the image formation \cite{wuSelfCalibrationDirectPosition2020,ngSensorarrayCalibrationUsing1996,vibergBayesianApproachAutocalibration1994,weissArrayShapeCalibration1989}. Calibration is achieved by exploiting the specific geometry of the array \cite{hanCalibratingNestedSensor2015,liuEigenstructureMethodEstimating2011,liuDOAEstimationUniform2009}, especially in the case of partly calibrated arrays \cite{liuClutterbasedGainPhase2019,liaoDirectionFindingPartly2012,parvaziDirectionofarrivalEstimationArray2011} and higher-order statistics of the associated measurements \cite{wanFourthorderDirectionFinding2020}, or other concepts such as low-rank modeling or sparsity \cite{huangLowrankRowsparseDecomposition2023,geissAntennaArrayCalibration2021,taghizadehAdHocMicrophone2015}. However, the parametric approach of joint calibration and imaging requires a structured model with only few degrees of freedom. In the context of air-coupled \gls{US}, Ramamohan \emph{et al.}\@ recently investigated joint calibration and \gls{1D} \gls{doa} estimation with a linear array assuming a highly structured array manifold \cite{ramamohanSelfcalibrationAcousticScalar2023}. 

Contrary to autocalibration, reference-based calibration methods are more flexible in terms of error modeling and suggest superior imaging performance of a stationary system after calibration \cite{ollierRobustCalibrationRadio2017,pierreExperimentalPerformanceCalibration1991,paulrajDirectionArrivalEstimation1985}. Other recent studies such as \cite{nanzerDistributedPhasedArrays2021,heidenreichJoint2DDOA2012,wijnholdsMultisourceSelfcalibrationSensor2009} reveal that the majority of calibration methods in the literature, both reference-based and non-reference-based, consider linear measurement models, even though the underlying imaging principles are of a multilinear nature, comprising a multidimensional data tensor with a transmitter, receiver, frequency, and time dimension, respectively. In such a case, using multilinear algebra for dictionary learning of the multidimensional data is more appropriate \cite{tosicDictionaryLearning2011}. Recently, Guo \emph{et al.} demonstrated such a multilinear approach to the calibration of a bistatic \gls{MIMO} radar where a rank-one \gls{CPD} model was considered for each source \cite{guoTensorbasedAngleArray2019}. Similar approaches were later pursued by Sun \emph{et al.} \cite{sunSpaceTimerangeClutter2024} and Chen \emph{et al.} \cite{chenAngleEstimationBased2024}. In our previous work \cite{mullerDictionarybasedLearning3Dimaging2020}, we applied low-rank tensor factorization to calibration data and exploited sparsity in the image formation \cite{kusheParallelSparseRegularization2019}. That approach with a generic tensor model is designed for arrays composed of identical transceiver elements with identical transmit and receive magnitude responses.
			
In this work, we take the particular manufacturing tolerances and transceiver characteristics of \gls{PUT}s into account and propose a novel tensor signal model that captures the multidimensional structure of an air-coupled ultrasonic sensor array for \gls{MIMO} imaging. We learn the parameters of our proposed tensor signal model jointly from measurements of the entire \gls{voi}, i.e.\@ the angular \gls{fov} in a given range interval. In a testbed deployed in our anechoic chamber at TU Darmstadt, we additionally assess the performance of our proposed method with a phased array for \gls{3D} imaging where the \glspl{PUT} are combined with a 3D-printed waveguide. In summary, the key contributions of this article are:
\begin{enumerate}[label={(\arabic*)}]
	\item We introduce a novel tensor signal model that accounts for varying element transceiver characteristics in air-coupled ultrasonic \gls{MIMO} sensor arrays with nonuniform transmitter and receiver magnitude responses and nonuniform element patterns. This makes our model also of interest for other array processing applications, including sonar and radar.
	\item We learn the array response from reference-based calibration measurements using targets at known locations. An optimization problem is formulated to jointly estimate all tensor model parameters.
	\item We apply a unique parameterization to the proposed tensor model and derive a modified \gls{BCD} method that exhibits highly parallel, closed-form solutions of the corresponding subproblems in each set of block variables. We prove that the modified \gls{BCD} method converges to a stationary point of the calibration problem.
	\item We test the proposed calibration method using synthetic data and real measurements recorded with prototype hardware. A \gls{3D} imaging example serves as a proof of concept for the usability of the dictionary of array responses learned from calibration measurements.
	\item We compare the imaging quality of our proposed tensor model after calibration with the conventional analytic point source model, which in the case of our testbed data reveals to be a valid model.
\end{enumerate}
			
\subsection{Notation and outline}
			
We use the italic letter $x$, the boldface lowercase italic letter $\bm{x}$, the boldface uppercase italic letter $\bm{X}$, and the boldface uppercase calligraphic italic letter $\bm{\mathcal{X}}$ to denote a scalar, a vector, a matrix, and a general multi-way array, respectively. The $(i_1, i_2, \ldots)$-th scalar element of $\bm{\mathcal{X}}$ is denoted by $\elem{\bm{\mathcal{X}}}_{i_1,i_2,\ldots}$ and a colon instead of a fixed index indicates the selection of all entries of the corresponding mode. Subtensors with all but one index fixed (fibers) are assumed to be column vectors and subtensors with all but two indices fixed (slices) are treated as matrices. Unfolding the tensor $\bm{\mathcal{X}}$ to its mode $w$ is denoted by $\unfold{w}{\bm{\mathcal{X}}}$ and yields matrix $\bm{X}_{(w)}$. The all-one vector of length $I$ is denoted by $\bm{1}_I$ and $\vectorize{\cdot}$ returns its multidimensional argument reshaped into a vector, while $\diag{\cdot}$ returns a square matrix with the elements of the vector argument on the main diagonal and zero everywhere else, respectively. Symbols $\hadamul$, $\kron$, $\khatri$ and $\outerprod$ denote Hadamard multiplication, Kronecker multiplication, column-wise Kronecker (Khatri-Rao) multiplication, and outer tensor multiplication, respectively. The complex conjugate, transpose, and Hermitian transpose are represented by $(\cdot)^\ast$, $(\cdot)^\mathsf{T}$, and $(\cdot)^\mathsf{H}$, respectively, while $(\cdot)^{(i)}$ denotes the $i$-th iterate of a variable. Taking the Euclidean norm or the Frobenius norm is indicated by $\norm{\cdot}_2$ and $\norm{\cdot}_\mathsf{F}$, respectively. Finally, $\arg(\cdot)$ returns the principal angle of its complex argument while operators $\reval{\cdot}$ and $\imval{\cdot}$ return the real and imaginary part of their argument, respectively.

The rest of this article is organized as follows: \Cref{sec:methodology} introduces the tensor signal model that is used to formulate the estimation problem for calibration in \Cref{sec:methodology-calibration}, where it is solved with a modified \gls{BCD} method. The resulting parameter estimates are then used for imaging in \Cref{sec:methodology-imaging}. Results from simulations as well as experiments with real data are presented in \Cref{sec:results-discussion} and concluding remarks with future research directions are given in \Cref{sec:conclusions}.
			
\section{Signal Model for Calibration}
\label{sec:methodology}
			
Consider a single and static point target at a position indexed by integer $p$ in the far field of a co-located \gls{MIMO} transceiver array. In polar coordinates, the position of this target can be, e.g., characterized by $(r_p, \vartheta_p, \varphi_p)$, where $r_p$, $\vartheta_p$, and $\varphi_p$ represent the target range, azimuth angle and elevation angle, respectively. The target is repeatedly exposed to \gls{US} pulses emitted from one of the transceiver elements. In total, $N$ different transmitters are used for excitation and, w.l.o.g., we index the reference sensor with $n=1$. Under the narrowband assumption, the spatial displacement of a transmitter element w.r.t.\@ the reference sensor yields a corresponding phase shift in the waveform at the point target. At position $p$, the phase shifts of all $N$ transmitter elements relative to the reference sensor are characterized by the transmit steering vector $\bm{a}_{p, \text{Tx}} \in \mathbb{C}^{N}$ \cite{liuRobustDetectionMIMO2019}. Furthermore, the complex gain $\elem{\bm{a}_{p, \text{Tx}}}_n$ also captures the propagation delay and the path attenuation experienced by the $n$-th transmitter as a result from additional transceiver structures, such as an attached waveguide \cite{rutschWaveguideAircoupledUltrasonic2021}.

According to the Huygens-Fresnel principle, the wave impinging on a target initiates a reflecting wave that travels back to the transceiver array. Hence in the receiver path, the far field point target is viewed as a point source transmitting an echo signal to the array. For an arbitrary array geometry, the $M$-element array response to the echo signal originating from a target at position $p$ is described by a general receive steering vector $\bm{a}_{p, \text{Rx}} \in \mathbb{C}^{M}$ \cite{vibergIntroductionArrayProcessing2014}. Similar to $\bm{a}_{p, \text{Tx}}$, each receive steering vector consists of receiver-dependent complex gains based on individual characteristics, e.g.\@ point responses, propagation delays, and path attenuations.

\begin{figure}[t]
	\centering
	\begin{tikzpicture}
		\node (image) at (0,0) {
			\includegraphics[height=0.2\textheight]{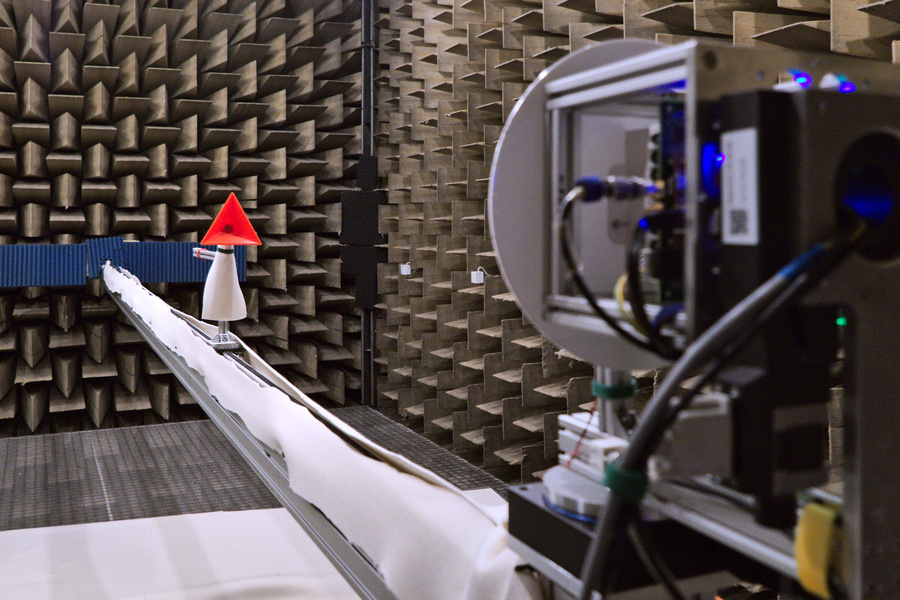}
		};
		\draw[latex-, very thick,white] (-1.75,.75) -- (-2.35,1.75);
		\draw (-3,1.75) node[right,black,fill=white]{\tiny Calibration target};
		\draw[latex-, very thick,white] (1.4,0.6) -- (0.6,-1.35);
		\draw (-0.5,-1.45) node[right,black,fill=white]{\tiny Array incl. waveguide};
		\draw (-0.5,-1.8) node[right,black,fill=white,align=center]{\tiny and rigid baffle};
	\end{tikzpicture}
	\caption{Measurement setup in an anechoic chamber to obtain data for the calibration technique proposed in \Cref{sec:methodology-calibration}.}
	\label{fig:calibration-setup}
\end{figure}

For the calibration of the measurement system, a reflector is set up at distinct but known positions $p = 1, \ldots, P$ in an anechoic chamber (\Cref{fig:calibration-setup}). Compared to its surrounding, the reflector provides a strong echo that approximates a point source in free field. Sequentially, each transmitter $n= 1,\ldots, N$ fires a series of $T$ acoustic pulses, also referred to as bursts \cite{allevatoRealtime3DImaging2021}. Each burst emerges from the physical operating principle of a \gls{PUT}, i.e.\@ after applying multiple periods of a sinusoidal electrical signal with constant excitation frequency to the active transducer.
			
After reflection at the target object, the incoming echo pulse is measured at the receive array and converted back from the acoustic to the electrical domain based on the receiver characteristics of the respective \gls{PUT} elements. Thus, the data acquisition principle is similar as in \gls{MIMO} radar or medical \gls{US}. However, passband signals in air-coupled \gls{US} can be directly sampled without downconversion due to the comparably low carrier frequency of $f_0 = \SI{40}{kHz}$. After analog-to-digital conversion of the sampled echo signal per transmitter and burst, a rectangular window of length $N_\text{rect}=2^{12}$ is applied to each of the $M$ receive sequences to filter out undesired reflections from near field clutter or ghost targets from multipath propagation. In the following, we consider measurements in the frequency domain by applying the \gls{DFT} to the received time signals.

The \glspl{PUT} regarded in this work have a narrow relative bandwidth of about $\SI{3}{\%}$, i.e., most of the signal energy is spread across the frequency range from \SIrange{39.4}{40.6}{kHz}. Based on the sampling frequency $f_\text{s} = \SI{195}{kHz}$ of the measurement system, we select $L = 24$ frequency bins equally spaced by $N_\text{rect}^{-1} f_\text{s}\approx \SI{48}{Hz}$ that contain the significant signal power around the average receive resonance frequency. The measurements of each transmitter $n=1,\ldots,N$ are then arranged in a three-way array of dimension $M \times L \times T$, similar to the data tensor used in \gls{MIMO} radar \cite{melvinSpacetimeAdaptiveProcessing2014}. As in radar, air-coupled \gls{US} imaging also uses the short time measurements for range estimation. In the following, we introduce a novel tensor signal model that accounts for nonidentical magnitude responses of distinct transceiver elements in a sensor array.

In the frequency band of interest, we model the transmit and receive characteristics (transfer functions) of the $n$-th transmitter and the $m$-th receiver on the $L$ dominant \gls{DFT} bins by the real-valued frequency response vectors $\bm{g}_{n, \text{Tx}}$ and $\bm{g}_{m, \text{Rx}}$. The \gls{US} transceivers can be operated in transmit or receive mode with a fast switch between both modes. We remark, however, that due to manufacturing tolerances, the frequency responses of different sensors are generally different and also the transmit and receive responses within the same sensor generally differ due to the distinct resonance frequencies for both operating modes \cite{allevatoUltrasonicPhasedArrays2023}.
			
The propagation time of a burst affects the phase of the \gls{DFT} of the receive pulse proportionally to the range of the target at position $p$. For a point target, the phase increment over frequency is generally linear, however, for generality we allow arbitrary phase increments in the phase response $\bm{c}_p \in \mathbb{C}^L$, which is modeled with a unit modulus property, i.e.\@ $\abs{ \elem{ \bm{c}_p }_\ell } = 1$, $\ell=1,\ldots,L$. We combine $\bm{g}_{n, \text{Tx}}$, $\bm{g}_{m, \text{Rx}}$ and $\bm{c}_p$ into the frequency vector
\begin{equation}
	\label{eq:frequency-vector}
	\bm{b}_{n,m,p} = \bm{g}_{n, \text{Tx}} \hadamul \bm{g}_{m, \text{Rx}} \hadamul \bm{c}_p
\end{equation}
of the pulse from the $n$-th transmitter to the $m$-th receiver after reflection from the target at position $p$, which models the input-output relationship of the transceivers as a \gls{LTI} system.

Lastly, the modulation of each pulse sequence, the path attenuation between the reference sensors of the transmit and receive array, and the reflectivity of the point target at position $p$ are all modeled by the signal gain vector $\bm{h}_p \in \mathbb{C}^T$. Thus $[\bm{h}_p]_t$, i.e.\@ the $t$-th entry of the gain vector $\bm{h}_p$, does not only describe particular system attributes such as the amplitude and the phase of the $t$-th transmit pulse but also the acoustic cross-section of the scatterer. 

We assume, w.l.o.g., that the signal gain vector does not change between different transmit pulses for the same target, i.e.\@ $\bm{h}_p = [\bm{h}_p]_1 \bm{1}_T$. In our approach, however, we admit arbitrary, possibly nonidentical entries in $\bm{h}_p$ to model artifacts of the data acquisition system such as sample offsets that occur during reception as well as variations in the pulse amplitudes.
For simplicity, we assume that all transmitters use the same pulse waveform (envelope) so that $\bm{h}_p$ is independent of the transmitter index $n$.

Using all of the parameters introduced above, we model each pulse that travels from transmitter $n$ to receiver $m$ as the trilinear term in \cref{eq:frequency-vector} weighted by scalars $\elem{\bm{a}_{p, \text{Tx}}}_n \in \mathbb{C}$ and $\elem{\bm{a}_{p, \text{Rx}}}_m\in \mathbb{C}$. Evidently, this expression is also dependent on the target position indexed by $p$. A pulse sequence over $T$ distinct time instants, referred to as snapshots, over the $L$ frequency bins of interest then yields the $L \times T$ rank-$1$ matrix
\begin{equation}
	\label{eq:m-n-slice}
	\elem*{\bm{a}_{p, \text{Tx}}}_n \, \elem*{\bm{a}_{p, \text{Rx}}}_m \left( \bm{g}_{n, \text{Tx}} \hadamul \bm{g}_{m, \text{Rx}} \hadamul \bm{c}_p \right) \bm{h}_p^\mathsf{T}.
\end{equation}
Furthermore, the collection of all receive measurements initiated by all transmitters yields a four-way array $\bm{\mathcal{Y}}_p \in \mathbb{C}^{N \times M \times L \times T}$ for every calibration target position $p = 1,\ldots,P$. We model the $p$-th tensor as
\begin{equation}
	\label{eq:array-output-model}
	\bm{\mathcal{Y}}_p = \sum_{n=1}^{N} \sum_{m=1}^{M} \left( \bm{e}_n \hadamul \bm{a}_{p, \text{Tx}} \right) \outerprod \left( \bm{e}_m \hadamul \bm{a}_{p, \text{Rx}} \right) \outerprod \left( \bm{g}_{n, \text{Tx}} \hadamul \bm{g}_{m, \text{Rx}} \hadamul \bm{c}_p \right) \outerprod \bm{h}_p + \bm{\mathcal{Z}}_p, 
\end{equation}
where $\bm{e}_i$ is the standard basis vector of compatible dimension with entries $\elem{ \bm{e}_i }_j = 1$ for $i = j$ and $\elem{ \bm{e}_i }_j = 0$ for $i \neq j$, and $\bm{\mathcal{Z}}_p \in \mathbb{C}^{N \times M \times L \times T}$ represents additive measurement noise. 
The basis vectors impose a masking on all but one entry in $\bm{a}_{p, \text{Tx}}$ and $\bm{a}_{p, \text{Rx}}$, respectively, such that \cref{eq:array-output-model} is composed of the slices in \cref{eq:m-n-slice}.
			
\subsection{Relation to other tensor models}
\label{sec:related-models}

According to \cref{eq:array-output-model}, the receive measurements $\{\bm{\mathcal{Y}}_p\}_{p=1}^P$ from $P$ target positions are coupled by the magnitude responses $\{\bm{g}_{n, \text{Tx}}\}_{n=1}^N$ and $\{\bm{g}_{m, \text{Rx}}\}_{m=1}^M$.
A special case arises if all transceivers have identical transmit and receive magnitude responses, i.e.\@ $\bm{g}_{n, \text{Tx}} = \bm{g}_{n', \text{Tx}}$ and $\bm{g}_{m, \text{Rx}} = \bm{g}_{m', \text{Rx}}$ for $n, n' = 1, \ldots, N$ and $m, m' = 1, \ldots, M$, respectively. In this instance, we may introduce a common transmit magnitude response $\bm{g}_{\text{Tx}}$ for all $n$ and a common receive magnitude response $\bm{g}_{\text{Rx}}$ for all $m$, respectively, such that the model in \cref{eq:array-output-model} reduces to
\begin{equation}
	\label{eq:rank-one-model}
	\bm{\widetilde{\mathcal{Y}}}_p = \bm{a}_{p, \text{Tx}} \outerprod \bm{a}_{p, \text{Rx}} \outerprod \left( \bm{g}_{\text{Tx}} \hadamul \bm{g}_{\text{Rx}} \hadamul \bm{c}_p \right) \outerprod \bm{h}_p + \bm{\mathcal{Z}}_p.
\end{equation}
For a single position $p$, the noise-free part of \cref{eq:rank-one-model} admits a rank-$1$ \gls{CPD}, which roots in the \gls{CANDECOMP} \cite{carrollAnalysisIndividualDifferences1970} and the \gls{PARAFAC} model \cite{harshmanFoundationsPARAFACProcedure1970}. Measurement tensors of different target positions, however, still share a common transmit and magnitude response under \cref{eq:rank-one-model}, which is not addressed by the general positionwise rank-$1$ \gls{CPD} approximation
\begin{equation}
	\label{eq:rank-one-approximation}
	\bm{\mathcal{Y}}_p \approx \bm{a}_{p, \text{Tx}} \outerprod \bm{a}_{p, \text{Rx}} \outerprod \bm{b}_p \outerprod \bm{h}_p.
\end{equation}
Unlike our proposed tensor model, the rank-$1$ \gls{CPD} approximation \cref{eq:rank-one-approximation} is in this case associated with a model mismatch.
For studies on the multi-rank \gls{CPD} of a measurement tensor, where each rank-$1$ component corresponds to the signal from one of multiple targets, we refer the reader to \cite{kusheParallelSparseRegularization2019}. 

Another interesting link to multilinear algebra is given by the block-term decomposition of a third-order tensor \cite{delathauwerDecompositionsHigherorderTensor2008}. If we rearrange our $P$ fourth-order tensors $\{\bm{\mathcal{Y}}_p\}_{p=1}^P$ from \cref{eq:array-output-model} into the coupled third-order tensors
\begin{equation}
	\label{eq:block-term-tensor}
	\bm{\mathcal{Y}}_{\text{virt},p} = \begin{bmatrix}
		\elem*{\bm{\mathcal{Y}}_p}_{1,:,:,:}, & \ldots, & \elem*{\bm{\mathcal{Y}}_p}_{N,:,:,:}
	\end{bmatrix}^\mathsf{T} \in \mathbb{C}^{NM \times L \times T}
\end{equation}
for $p = 1,\ldots, P$ with a virtual array spanned by $NM$ transceiver elements, we can factorize the measurements into
\begin{equation}
	\label{eq:block-term-decomp}
	\bm{\mathcal{Y}}_{\text{virt},p} = \left( \bm{A}_p \bm{B}_p^\mathsf{T} \right) \outerprod \bm{h}_p
\end{equation}
for $p = 1,\ldots, P$ using the diagonal matrix 
\begin{equation}
	\label{eq:Ap}
	\bm{A}_p = \diag{ \bm{a}_{p, \text{Tx}} \kron \bm{a}_{p, \text{Rx}} } \in \mathbb{C}^{NM \times NM}
\end{equation}
for $p = 1,\ldots, P$, which is formed from the virtual array steering vector corresponding to the target at position $p$, and the frequency matrix 
\begin{equation}
	\label{eq:factor-matrix-b}
	\bm{B}_p = \left(\bm{c}_p^\mathsf{T} \khatri \bm{G}_\text{Tx}^\mathsf{T}  \khatri \bm{G}_\text{Rx}^\mathsf{T} \right)^\mathsf{T} \in \mathbb{C}^{L \times NM}
\end{equation}
for $p = 1,\ldots, P$ with
\begin{equation}
	\label{eq:Gtx}
	\bm{G}_\text{Tx} = \begin{bmatrix}
		\bm{g}_{1,\text{Tx}}, &\ldots, &\bm{g}_{N,\text{Tx}}
	\end{bmatrix} \in \mathbb{C}^{L \times N}
\end{equation}
and
\begin{equation}
	\label{eq:Grx}
	\bm{G}_\text{Rx} = \begin{bmatrix}
		\bm{g}_{1,\text{Rx}}, &\ldots, &\bm{g}_{M,\text{Rx}}
	\end{bmatrix} \in \mathbb{C}^{L \times M}.
\end{equation}

The factorization in \cref{eq:block-term-decomp} is similar to the rank-$(I,I,1)$ block-term decomposition of \cref{eq:block-term-tensor} given by $\bm{\mathcal{Y}}_{\text{virt},p} = \sum_{r=1}^{R} ( \bm{\tilde{A}}_{p,r} \bm{\tilde{B}}_{p,r}^\mathsf{T} ) \outerprod \bm{\tilde{h}}_{p,r}$ with $\bm{\tilde{A}}_{p,r} \in \mathbb{C}^{NM \times I}$ and $\bm{\tilde{B}}_{p,r} \in \mathbb{C}^{L \times I}$ for $I \leq \min(L,NM)$, and tensor rank $R$ \cite{cichockiTensorDecompositionsSignal2015}. In contrast to matrices $\bm{\tilde{A}}_{p,r}$ and $\bm{\tilde{B}}_{p,r}$, which do not exhibit any specific structure, our matrices $\bm{A}_p$ and $\bm{B}_p$ are structured according to \cref{eq:Ap,eq:factor-matrix-b}, respectively. Furthermore, our frequency matrix $\bm{B}_p$ is not guaranteed to have full column rank, e.g.\@ if $L < NM$, but allows a more descriptive insight on the physical behavior of the imaging system. Thus, we prefer \cref{eq:block-term-decomp} over the ``regular'' rank-$(I,I,1)$ decomposition of the tensor in \cref{eq:block-term-tensor} for the considered application. As a final remark, we note that all third-order tensors $\{\bm{\mathcal{Y}}_{\text{virt},p}\}_{p=1}^P$ are coupled by the magnitude response matrices $\bm{G}_\text{Tx}$ and $\bm{G}_\text{Rx}$ in \cref{eq:Gtx} and \cref{eq:Grx}, respectively, which is not captured by the positionwise rank-$(I,I,1)$ decomposition and highlights the novelty of our proposed signal model in \cref{eq:array-output-model}.
			
\subsection{Unfolding of the tensor signal model}
			
When dealing with the tensor model in \cref{eq:array-output-model}, it is often convenient to operate with the corresponding matrix unfoldings instead. Unfolding, also known as matricization, rearranges any $W$-way array $\bm{\mathcal{X}} \in \mathbb{C}^{I_1 \times \dots \times I_W}$  into matrices ${\bm{X}_{(w)} \in \mathbb{C}^{I_w \times I_1 \ldots I_{w-1}I_{w+1} \ldots I_W}}$ for modes $w = 1, \ldots, W$. An intuitive way to matricize a tensor is to cyclically permute its modes.The unfoldings of a four-way array $\bm{\mathcal{X}} \in \mathbb{C}^{I_1 \times I_2 \times I_3 \times I_4}$ are given by
\begin{subequations}
	\label{eq:unfolding}
	\begin{align}
		\unfold{1}*{\bm{\mathcal{X}}} &= \bm{X}_{(1)} \coloneq \begin{bmatrix}
			\elem*{\bm{\mathcal{X}}}_{:,:,1,1}, &\elem*{\bm{\mathcal{X}}}_{:,:,2,1}, &\ldots, &\elem*{\bm{\mathcal{X}}}_{:,:,I_3,1}, &\ldots, &\elem*{\bm{\mathcal{X}}}_{:,:,I_3,I_4}
		\end{bmatrix}, \label{eq:unfolding-mode-one} \\
		\unfold{2}*{\bm{\mathcal{X}}} &= \bm{X}_{(2)} \coloneq \begin{bmatrix}
			\elem*{\bm{\mathcal{X}}}_{1,:,:,1}, &\elem*{\bm{\mathcal{X}}}_{1,:,:,2}, &\ldots, &\elem*{\bm{\mathcal{X}}}_{1,:,:,I_4}, &\ldots, &\elem*{\bm{\mathcal{X}}}_{I_1,:,:,I_4}
		\end{bmatrix}, \label{eq:unfolding-mode-two} \\
		\unfold{3}*{\bm{\mathcal{X}}} &= \bm{X}_{(3)} \coloneq \begin{bmatrix}
			\elem*{\bm{\mathcal{X}}}_{1,1,:,:}, &\elem*{\bm{\mathcal{X}}}_{2,1,:,:}, &\ldots, &\elem*{\bm{\mathcal{X}}}_{I_1,1,:,:}, &\ldots, &\elem*{\bm{\mathcal{X}}}_{I_1,I_2,:,:}
		\end{bmatrix}, \label{eq:unfolding-mode-three} \\
		\unfold{4}*{\bm{\mathcal{X}}} &= \bm{X}_{(4)} \coloneq \begin{bmatrix}
			\left(\elem*{\bm{\mathcal{X}}}_{:,1,1,:}\right)^\mathsf{T}, &\left(\elem*{\bm{\mathcal{X}}}_{:,2,1,:}\right)^\mathsf{T}, &\ldots, &\left(\elem*{\bm{\mathcal{X}}}_{:,I_2,1,:}\right)^\mathsf{T}, &\ldots, &\left(\elem*{\bm{\mathcal{X}}}_{:,I_2,I_3,:}\right)^\mathsf{T}
		\end{bmatrix}. \label{eq:unfolding-mode-four}
	\end{align}
\end{subequations}

Let us further define the factor matrices
\begin{subequations}
	\label{eq:factor-matrices}
	\begin{align}
		\bm{A}_{p, \text{Tx}} &= \diag*{\bm{a}_{p, \text{Tx}}} \kron \bm{1}_M^\mathsf{T} \in \mathbb{C}^{N \times NM}, \label{eq:factor-matrix-a-tx} \\
		\bm{A}_{p, \text{Rx}} &= \bm{1}_N^\mathsf{T} \kron \diag*{\bm{a}_{p, \text{Rx}}} \in \mathbb{C}^{M \times NM}, \label{eq:factor-matrix-a-rx} \\
		\bm{H}_p &= \bm{h}_p \bm{1}_{NM}^\mathsf{T} \in \mathbb{C}^{T \times NM}. \label{eq:factor-matrix-h}
	\end{align}
\end{subequations}
Using the matricization in \cite{koldaTensorDecompositionsApplications2009} and the permutation scheme in \cite{kiersStandardizedNotationTerminology2000}, the following properties w.r.t.\@ our tensor model in \cref{eq:array-output-model} hold for $\bm{\mathcal{Z}}_p = \bm{0}$, i.e.\@ the noiseless case:
\begin{subequations}
	\begin{align}
		\unfold{1}*{\bm{\mathcal{Y}}_p} &= {\bm{Y}_p}_{(1)} = \bm{A}_{p, \text{Tx}} \left( \bm{H}_p \khatri \bm{B}_p \khatri \bm{A}_{p, \text{Rx}} \right)^\mathsf{T} \in \mathbb{C}^{N \times MLT}, \label{eq:mode-one-unfolding} \\
		\unfold{2}*{\bm{\mathcal{Y}}_p} &= {\bm{Y}_p}_{(2)} = \bm{A}_{p, \text{Rx}} \left( \bm{A}_{p, \text{Tx}} \khatri \bm{H}_p \khatri \bm{B}_p \right)^\mathsf{T} \in \mathbb{C}^{M \times NLT}, \label{eq:mode-two-unfolding} \\
		\unfold{3}*{\bm{\mathcal{Y}}_p} &= {\bm{Y}_p}_{(3)} = \bm{B}_p \left( \bm{A}_{p, \text{Rx}} \khatri \bm{A}_{p, \text{Tx}} \khatri \bm{H}_p \right)^\mathsf{T} \in \mathbb{C}^{L \times NMT}, \label{eq:mode-three-unfolding} \\
		\unfold{4}*{\bm{\mathcal{Y}}_p} &= {\bm{Y}_p}_{(4)} = \bm{H}_p \left( \bm{B}_p \khatri \bm{A}_{p, \text{Rx}} \khatri \bm{A}_{p, \text{Tx}} \right)^\mathsf{T} \in \mathbb{C}^{T \times NML}. \label{eq:mode-four-unfolding}
	\end{align}
\end{subequations}
In the following, we estimate the model parameters by applying all four unfoldings on the measurement tensors in a modified \gls{BCD} procedure that leads to parallel alternating univariate optimization problems \cite{sidiropoulosTensorDecompositionSignal2017}.
			
\section{Modified Block Coordinate Descent Procedure}
\label{sec:methodology-calibration}
			
The signal model in \cref{eq:array-output-model} provides us with parameters $\bm{g}_{n, \text{Tx}}$, $\bm{g}_{m, \text{Rx}}$, $\bm{a}_{p, \text{Tx}}$, $\bm{a}_{p, \text{Rx}}$, $\bm{c}_p$, and $\bm{h}_p$ that characterize the array for a single point target at position $p$. Ultimately, we aim to describe a given but unknown array for an arbitrary scene composed of multiple targets at different and unknown positions. According to the superposition principle, a scene with multiple targets can be described as the superposition of multiple, independent single-source cases \cite{paulrajSubspaceMethodsDirectionsofarrival1993}. Thus, we propose a reference-based calibration method where the unknown parameters are estimated from $P$ single-source scenarios with known positions of the calibration target. This provides us with the ability to handle scenarios with multiple unknown targets after calibration.

Using the measurement system in the anechoic chamber, we place a reflector at $P$ different positions and record the respective measurement tensors, each one following \cref{eq:array-output-model}. The collection of tensors $\{ \bm{\mathcal{Y}}_p \}_{p=1}^P$ is then matched jointly to our model in \cref{eq:array-output-model} to determine the parameter estimates $\{ \bm{\hat{g}}_{n, \text{Tx}} \}_{n=1}^N$, $\{ \bm{\hat{g}}_{m, \text{Rx}} \}_{m=1}^M$, and $\{ \bm{\hat{a}}_{p, \text{Tx}}, \, \bm{\hat{a}}_{p, \text{Rx}}, \, \bm{\hat{c}}_p, \, \bm{\hat{h}}_p \}_{p=1}^P$. This parameter estimation process is cast as the optimization problem
\begin{mini!}
	{\substack{\left\{ \bm{g}_{n,\text{Tx}} \in \mathbb{R}_+^{L} \right\}_{n=1}^{N}, \\ \left\{ \bm{g}_{m,\text{Rx}} \in \mathbb{R}_+^{L} \right\}_{m=1}^{M}, \\ \left\{ \bm{a}_{p, \text{Tx}} \in \mathbb{C}^{N}, \, \bm{a}_{p, \text{Rx}} \in \mathbb{C}^{M}, \, \bm{c}_p \in \mathbb{C}^{L}, \, \bm{h}_p \in \mathbb{C}^{T} \right\}_{p=1}^{P}}}{ f \! \left( \left\{\bm{g}_{n,\text{Tx}}\right\}_{n=1}^N, \, \left\{\bm{g}_{m,\text{Rx}}\right\}_{m=1}^M, \, \left\{\bm{a}_{p, \text{Tx}}, \bm{a}_{p, \text{Rx}}, \bm{c}_p, \bm{h}_p\right\}_{p=1}^P \right) \label{eq:scpd-original-problem-objective}}
	{\label{eq:scpd-original-problem}}{}
	\addConstraint{ \imval*{ \elem*{\bm{a}_{p, \text{Tx}}}_1 } }{= 0, \quad p = 1,\, \dots,\, P \label{eq:scpd-original-problem-atx1-constraint}}
	\addConstraint{ \norm{ \bm{a}_{p, \text{Tx}}}_2^2 - N }{= 0, \quad p = 1,\, \dots,\, P \label{eq:scpd-original-problem-atx-constraint}}
	\addConstraint{ \imval*{ \elem*{\bm{a}_{p, \text{Rx}}}_1 } }{= 0, \quad p = 1,\, \dots,\, P \label{eq:scpd-original-problem-arx1-constraint}}
	\addConstraint{ \norm{ \bm{a}_{p, \text{Rx}}}_2^2 - M }{= 0, \quad p = 1,\, \dots,\, P \label{eq:scpd-original-problem-arx-constraint}}
	\addConstraint{ \elem*{\bm{c}_p}_1 - 1 }{= 0, \quad p = 1,\, \dots,\, P \label{eq:scpd-original-problem-c1-constraint}}
	\addConstraint{ \abs{ \elem*{\bm{c}_p}_\ell } - 1}{= 0, \quad p = 1,\, \dots,\, P, \ \ell = 2,\, \dots,\, L \label{eq:scpd-original-problem-c-constraint}}
	\addConstraint{ \bm{g}_{1,\text{Tx}} }{= \bm{1}_L \label{eq:scpd-original-problem-g-tx-constraint-ref}}
	\addConstraint{ -\elem*{\bm{g}_{n,\text{Tx}}}_\ell + \varepsilon}{\leq 0, \quad n = 2,\, \dots,\, N, \ \ell = 1,\, \dots,\, L \label{eq:scpd-original-problem-g-tx-constraint-min}}
	\addConstraint{ \max_\ell \elem*{\bm{g}_{n,\text{Tx}}}_\ell - 1}{= 0, \quad n = 2,\, \dots,\, N, \ \ell = 1,\, \dots,\, L \label{eq:scpd-original-problem-g-tx-constraint-max}}
	\addConstraint{ -\elem*{\bm{g}_{m,\text{Rx}}}_\ell + \varepsilon}{\leq 0, \quad m = 1,\, \dots,\, M, \ \ell = 1,\, \dots,\, L \label{eq:scpd-original-problem-g-rx-constraint-min}}
	\addConstraint{ \max_\ell \elem*{\bm{g}_{m,\text{Rx}}}_\ell - 1}{= 0, \quad m = 1,\, \dots,\, M, \ \ell = 1,\, \dots,\, L, \label{eq:scpd-original-problem-g-rx-constraint-max}}
\end{mini!}
with cost function
\begin{equation}
	\begin{aligned}
		& f \! \left( \left\{\bm{g}_{n,\text{Tx}}\right\}_{n=1}^N, \, \left\{\bm{g}_{m,\text{Rx}}\right\}_{m=1}^M, \, \left\{\bm{a}_{p, \text{Tx}}, \bm{a}_{p, \text{Rx}}, \bm{c}_p, \bm{h}_p\right\}_{p=1}^P \right)  \\
		= & \sum_{p=1}^{P} \norm{ \bm{\mathcal{Y}}_p - \sum_{n=1}^{N} \sum_{m=1}^{M} \left( \bm{e}_n \hadamul \bm{a}_{p, \text{Tx}} \right) \outerprod \left( \bm{e}_m \hadamul \bm{a}_{p, \text{Rx}} \right) \outerprod \left( \bm{g}_{n, \text{Tx}} \hadamul \bm{g}_{m, \text{Rx}} \hadamul \bm{c}_p \right) \outerprod \bm{h}_p }_\mathsf{F}^2,
	\end{aligned}
	\label{eq:costfct}	
\end{equation}
which is designed to match the noise-free array output model corresponding to \cref{eq:array-output-model} to all calibration measurements $\{ \bm{\mathcal{Y}}_p \}_{p=1}^P$ in a \gls{LS} sense.
			
The constraints in \cref{eq:scpd-original-problem-atx1-constraint,eq:scpd-original-problem-atx-constraint,eq:scpd-original-problem-arx1-constraint,eq:scpd-original-problem-arx-constraint,eq:scpd-original-problem-c1-constraint,eq:scpd-original-problem-g-tx-constraint-ref,eq:scpd-original-problem-g-tx-constraint-min,eq:scpd-original-problem-g-tx-constraint-max,eq:scpd-original-problem-g-rx-constraint-min,eq:scpd-original-problem-g-rx-constraint-max} are introduced to address trivial scaling invariances. Generally, if \cref{eq:scpd-original-problem-objective} is minimized without these constraints, then not only $\{ \bm{g}_{n,\text{Tx}}^\star \}_{n=1}^{N}$, $\{ \bm{g}_{m,\text{Rx}}^\star \}_{m=1}^{M}$, and
$\{ \bm{a}_{p, \text{Tx}}^\star, \, \bm{a}_{p, \text{Rx}}^\star, \, \bm{c}_p^\star, \, \bm{h}_p^\star \}_{p=1}^{P}$ are minimizers to \cref{eq:scpd-original-problem-objective}, but also $\{ \bm{\tilde{g}}_{n,\text{Tx}}^\star \}_{n=1}^{N}$, $\{ \bm{\tilde{g}}_{m,\text{Rx}}^\star \}_{m=1}^{M}$, and
$\{ \bm{\tilde{a}}_{p, \text{Tx}}^\star, \, \bm{\tilde{a}}_{p, \text{Rx}}^\star, \, \bm{\tilde{c}}_p^\star, \, \bm{\tilde{h}}_p^\star \}_{p=1}^{P}$ yield the same objective function value in \cref{eq:scpd-original-problem-objective} for $ \bm{\tilde{g}}_{n,\text{Tx}}^\star = \kappa_{n,\bm{g}_{\rm Tx}} \bm{g}_{n,\text{Tx}}^\star$,  $\bm{\tilde{g}}_{m,\text{Rx}}^\star = \kappa_{m,\bm{g}_{\rm Rx}}\bm{g}_{m,\text{Rx}}^\star$, $\bm{\tilde{a}}_{p, \text{Tx}}^\star = \kappa_{p,\bm{a}_{\rm Tx}} \bm{a}_{p, \text{Tx}}^\star$, $\bm{\tilde{a}}_{p, \text{Rx}}^\star =\kappa_{p,\bm{a}_{\rm Rx}} \bm{a}_{p, \text{Rx}}^\star$, $ \bm{\tilde{c}}_p^\star = \kappa_{p,\bm{c}} \bm{c}_p^\star$, and $\bm{\tilde{h}}_p^\star = \kappa_{p,\bm{h}} \bm{h}_p^\star$, where the coefficients $\kappa_{n,\bm{g}_{\rm Tx}}$ and $\kappa_{m,\bm{g}_{\rm Rx}}$ are positive real-valued, $\kappa_{p,\bm{a}_{\rm Tx}}$ and $\kappa_{p,\bm{a}_{\rm Rx}}$ are arbitrary complex, $\kappa_{p,\bm{c}}$ is unit modulus, i.e.\@ $\abs{\kappa_{p,\bm{c}}} = 1$, and $\kappa_{p,\bm{h}} = (\kappa_{n,\bm{g}_{\rm Tx}} \kappa_{m,\bm{g}_{\rm Rx}} \kappa_{p,\bm{a}_{\rm Tx}} \kappa_{p,\bm{a}_{\rm Rx}} \kappa_{p,\bm{c}})^{-1}$. Rotations of the global phase are prevented by keeping the first entry of every optimization variable other than $\bm{h}_p$ real-valued. Scaling ambiguities are further avoided by employing norm constraints on $\bm{a}_{p, \text{Tx}}$ and $\bm{a}_{p, \text{Rx}}$, respectively. Constraint \cref{eq:scpd-original-problem-g-tx-constraint-ref} resolves the scaling ambiguity between the transmit and receive frequency response vectors, $\bm{g}_{n,\text{Tx}}$ and $\bm{g}_{n,\text{Rx}}$, in the objective function \cref{eq:costfct} by fixing the transmit reference sensor $n=1$. On the \gls{DFT} bins that carry relevant signal power, the frequency responses of the $n$-th transmitter and $m$-th receiver, respectively, are furthermore restricted in \cref{eq:scpd-original-problem-g-tx-constraint-min}-\ref{eq:scpd-original-problem-g-rx-constraint-max} to take real-valued and strictly positive values in the interval $\left[ \varepsilon, 1 \right]$ for some small constant $\varepsilon >0$. Lastly, \cref{eq:scpd-original-problem-c-constraint} follows from the unit modulus property of the phase response vector $\bm{c}_p$.

Problem \cref{eq:scpd-original-problem} is jointly nonconvex, with a multivariate objective function \cref{eq:scpd-original-problem-objective}. For each individual variable, however, the problem can be reformulated in an equivalent convex subproblem when the remaining variables are fixed. Hence, \cref{eq:scpd-original-problem} can be solved by various iterative techniques such as gradient projection and the successive convex approximation framework \cite{yangInexactBlockCoordinate2020}. We choose the \gls{BCD} method that approximates the multilinear objective of the optimization problem by a series of approximate problems in which all but one optimization variable are fixed, leading to univariate subproblems \cite{bertsekasNonlinearProgramming2016}.

Applying the \gls{BCD} method straight to problem \cref{eq:scpd-original-problem} results in subproblems that are nonconvex due to the constraints. As nonconvex problems are in general difficult to solve, we relax the nonconvex constraints in each of the subproblems and then show that these constraints can always be satisfied by appropriate scaling. This modification of the \gls{BCD} method convexifies the subproblems such that they exhibit closed-form solutions. Additionally, each approximate problem is separable into further subproblems with independent variables such that the update of any block variable can be done in parallel. This makes our method attractive for implementation on modern parallel hardware architectures such as \glspl{GPU}, \glspl{FPGA}, and multicore \glspl{DSP}. Further parallelization can be achieved with the successive convex approximation framework in \cite{yangInexactBlockCoordinate2020}.
			
Each approximate problem only depends on one set of variables, hence, only one factor in the multilinear product. The following subsections introduce the approximate problems for each block variable and the corresponding solutions. We denote the iterates of the different block variables after the update and before the parameter scaling in iteration $i$ of the modified \gls{BCD} method as $\{ \bm{g}_{n, \text{Tx}}^{(i+1)}\}_{n=1}^N$, $\{\bm{g}_{m, \text{Rx}}^{(i+1)}\}_{m=1}^M$ and $\{ \bm{a}_{p, \text{Tx}}^{(i+1)},\bm{a}_{p, \text{Rx}}^{(i+1)},\bm{c}_{p}^{(i+1)}, \bm{h}_{p}^{(i+1)}\}_{p=1}^P$, respectively.\footnote{We note that the order in which the variables are updated can be modified and does not need to follow the sequence presented here. It is, however, mandatory by the \gls{BCD} method to always use the latest available values of all other parameters when updating one block variable of interest.} The iterates of the different block variables in the $i$-th iteration of the modified \gls{BCD} method after the parameter scaling are denoted as 
\begin{subequations}
	\label{eq:rescaling}
	\begin{align}
		\bm{\tilde{g}}_{n, \text{Tx}}^{(i+1)} &=   \kappa_{n,\bm{g}_{\rm Tx}} {\bm{g}}_{n, \text{Tx}}^{(i+1)}, \label{eq:rescale-g-tx} \\
		\bm{\tilde{g}}_{m, \text{Rx}}^{(i+1)} &=   \kappa_{m,\bm{g}_{\rm Rx}} {\bm{g}}_{m, \text{Rx}}^{(i+1)}, \label{eq:rescale-g-rx} \\
		\bm{\tilde{a}}_{p, \text{Tx}}^{(i+1)} &=   \kappa_{p,\bm{a}_{\rm Tx}} \left( \bm{\kappa}_{\bm{g}_\text{Tx}} \hadamul {\bm{a}}_{p, \text{Tx}}^{(i+1)} \right), \label{eq:rescale-a-tx} \\
		\bm{\tilde{a}}_{p, \text{Rx}}^{(i+1)} &= \kappa_{p,\bm{a}_{\rm Rx}} \left( \bm{\kappa}_{\bm{g}_\text{Rx}} \hadamul {\bm{a}}_{p, \text{Rx}}^{(i+1)} \right), \label{eq:rescale-a-rx} \\
		\bm{\tilde{c}}_{p}^{(i+1)} &= \kappa_{p, \bm{c}}  {\bm{c}}_{p}^{(i+1)}, \label{eq:rescale-c} \\
		\bm{\tilde{h}}_{p}^{(i+1)} &= \kappa_{p, \bm{h}}  {\bm{h}}_{p}^{(i+1)}, \label{eq:rescale-h}
	\end{align}
\end{subequations}
for $n=1,\ldots,N$, $m=1,\ldots,M$, and $p=1,\ldots,P$, respectively, where 
	$\bm{\kappa}_{\bm{g}_\text{Tx}} $ \linebreak
	$= [ \kappa_{1, \bm{g}_\text{Tx}}, \ldots, \kappa_{N, \bm{g}_\text{Tx}} ] 
	= [
	1, (\max_\ell \elem{\bm{g}_{2,\text{Tx}}^{(i+1)}}_\ell)^{-1}, \ldots,
	(\max_\ell \elem{\bm{g}_{N,\text{Tx}}^{(i+1)}}_\ell)^{-1}
	]^\mathsf{T} \in \mathbb{R}_+^N$,
	$\bm{\kappa}_{\bm{g}_\text{Rx}} = [ \kappa_{1, \bm{g}_\text{Rx}}, \ldots, \kappa_{M, \bm{g}_\text{Rx}} ]
	$ \linebreak
	$= [
	(\max_\ell \elem{\bm{g}_{1,\text{Rx}}^{(i+1)}}_\ell)^{-1}, \ldots,
	(\max_\ell \elem{\bm{g}_{M,\text{Rx}}^{(i+1)}}_\ell)^{-1}
	]^\mathsf{T} \in \mathbb{R}_+^M$,
	$\kappa_{p, \bm{c}} = \elem{\bm{c}_p^{(i+1)}}_1^\ast$,
	$\kappa_{p, \bm{h}} =  (\kappa_{p,\bm{a}_{\rm Tx}} \kappa_{p,\bm{a}_{\rm Rx}} \kappa_{p, \bm{c}})^{-1}$,
while $\kappa_{p,\bm{a}_{\rm Tx}}$ and $\kappa_{p,\bm{a}_{\rm Rx}}$ are chosen such that the first entry in $\bm{a}_{p, \text{Tx}}^{(i+1)}$ and $\bm{a}_{p, \text{Rx}}^{(i+1)}$ become real-valued, respectively, and the respective norm constraints \cref{eq:scpd-original-problem-atx-constraint,eq:scpd-original-problem-arx-constraint} are fulfilled.

\subsection{Update of $\{\titlebm{a}_{p, \text{Tx}}\}_{p=1}^P$}
\label{sec:updating-a-tx}

We start our modified \gls{BCD} method by relaxing problem \cref{eq:scpd-original-problem} to remove the constraints \cref{eq:scpd-original-problem-atx1-constraint,eq:scpd-original-problem-atx-constraint}, which can always be satisfied by the proposed rescaling in \cref{eq:rescaling}. Following the definitions \cref{eq:factor-matrix-a-rx,eq:factor-matrix-h} for the factor matrices $\bm{\tilde{A}}_{p, \text{Rx}}^{(i)} = \bm{1}_N^\mathsf{T} \kron \diag{\bm{\tilde{a}}_{p, \text{Rx}}^{(i)}}$ and $\bm{\tilde{H}}_p^{(i)} = \bm{\tilde{h}}_p^{(i)} \bm{1}_{NM}^\mathsf{T}$, respectively, as well as the frequency matrix $\bm{\tilde{B}}_p^{(i)} = ((\bm{\tilde{c}}_p^{(i)})^\mathsf{T} \khatri \, (\bm{\tilde{G}}_\text{Tx}^{(i)})^\mathsf{T} \khatri \, (\bm{\tilde{G}}_\text{Rx}^{(i)})^\mathsf{T} )^\mathsf{T}$ according to \cref{eq:factor-matrix-b}, the first approximate problem at iteration $i$ is given by
\begin{mini}
	{\{ \bm{a}_{p, \text{Tx}} \in \mathbb{C}^{N} \}_{p=1}^{P} }{f_{\bm{a}_{\text{Tx}}} \! \left( \left\{ \bm{a}_{p, \text{Tx}}, \bm{\tilde{a}}_{p, \text{Rx}}^{(i)}, \bm{\tilde{c}}_{p}^{(i)}, \bm{\tilde{h}}_p^{(i)} \right\}_{p=1}^P, \, \left\{ \bm{\tilde{g}}_{n, \text{Tx}}^{(i)} \right\}_{n=1}^N, \, \left\{ \bm{\tilde{g}}_{m, \text{Rx}}^{(i)} \right\}_{m=1}^M \right), }
	{\label{eq:subproblem-block-a-tx}}{}
\end{mini}
where we use the mode-one unfolding on \cref{eq:costfct} as defined in \cref{eq:unfolding-mode-one} to obtain the cost function \linebreak
$f_{\bm{a}_{\text{Tx}}} ( \{ \bm{a}_{p, \text{Tx}}, \bm{\tilde{a}}_{p, \text{Rx}}^{(i)}, \bm{\tilde{c}}_{p}^{(i)}, \bm{\tilde{h}}_p^{(i)} \}_{p=1}^P, \, \{ \bm{\tilde{g}}_{n, \text{Tx}}^{(i)} \}_{n=1}^N, \, \{ \bm{\tilde{g}}_{m, \text{Rx}}^{(i)} \}_{m=1}^M ) $ \linebreak 
$= \sum_{p=1}^{P} \norm*{ {\bm{Y}_p}_{(1)} - ( \diag{ \bm{a}_{p, \text{Tx}}} \kron \bm{1}_M^\mathsf{T} ) (  \bm{\tilde{H}}_p^{(i)} \khatri \bm{\tilde{B}}_p^{(i)} \khatri \bm{\tilde{A}}_{p, \text{Rx}}^{(i)} )^\mathsf{T} }_\mathsf{F}^2$. Eventually, problem \cref{eq:subproblem-block-a-tx} can be separated into $PN$ scalar subproblems
\begin{mini}
	{[\bm{a}_{p, \text{Tx}}]_n \in \mathbb{C}}{\sum_{m=1}^{M} \norm{ [\bm{\mathcal{Y}}_p]_{n,m,:,:} - \elem*{\bm{a}_{p, \text{Tx}}}_n \elem*{\bm{\tilde{a}}_{p, \text{Rx}}^{(i)}}_m \bm{\tilde{b}}_{n,m,p}^{(i)} \left(\bm{\tilde{h}}_p^{(i)}\right)^\mathsf{T} }_2^2 \label{eq:subproblem-decoupled-a-tx}}
	{}{}
\end{mini}
for $ \bm{\tilde{b}}_{n,m,p}^{(i)} = ( \bm{\tilde{c}}_p^{(i)} \hadamul \bm{\tilde{g}}_{n, \text{Tx}}^{(i)} \hadamul \bm{\tilde{g}}_{m, \text{Rx}}^{(i)} ) $ with solutions
\begin{equation}
	\label{eq:subproblem-decoupled-solution-a-tx}
	\begin{aligned}
		\elem*{\bm{a}_{p, \text{Tx}}^{(i+1)}}_n & = \left( \norm{\bm{\tilde{h}}_p^{(i)}}_2^2 \sum\limits_{m = 1}^M \abs{\elem*{\bm{\tilde{a}}_{p, \text{Rx}}^{(i)}}_m}^2 \norm{\bm{\tilde{b}}_{n,m,p}^{(i)}}_2^2 \right)^{-1} \sum\limits_{m = 1}^M  \elem*{\bm{\tilde{a}}_{p, \text{Rx}}^{(i)}}_m^\ast \left(\bm{\tilde{b}}_{n,m,p}^{(i)}\right)^\mathsf{H} \elem*{\bm{\mathcal{Y}}_p}_{n,m,:,:} \left(\bm{\tilde{h}}_p^{(i)}\right)^\ast
	\end{aligned}
\end{equation}
for $n = 1, \ldots, N$ and $p = 1, \ldots, P$, respectively, that can be solved in parallel.

\subsection{Update of $\{ \titlebm{a}_{p, \text{Rx}} \}_{p=1}^P$}
\label{sec:updating-a-rx}

Similar to the update of $\{ \bm{a}_{p, \text{Tx}} \}_{p=1}^P$, learning the receive steering vectors $\{ \bm{a}_{p, \text{Rx}} \}_{p=1}^{P}$ requires solving the approximate problem in which all remaining parameters are fixed. At first, we again relax the constraints \cref{eq:scpd-original-problem-arx1-constraint,eq:scpd-original-problem-arx-constraint}, which can always be satisfied by the proposed rescaling \cref{eq:rescaling}.

Using the mode-two unfolding in \cref{eq:unfolding-mode-two} on \cref{eq:costfct}, the approximate problem at iteration $i$ is given by
\begin{mini}
	{\{\bm{a}_{p, \text{Rx}} \in \mathbb{C}^{M}\}_{p=1}^{P} }{f_{\bm{a}_{\text{Rx}}} \! \left( \left\{ \bm{a}_{p, \text{Rx}}, \bm{a}_{p, \text{Tx}}^{(i+1)}, \bm{\tilde{c}}_{p}^{(i)}, \bm{\tilde{h}}_p^{(i)} \right\}_{p=1}^P, \, \left\{ \bm{\tilde{g}}_{n, \text{Tx}}^{(i)} \right\}_{n=1}^N, \, \left\{ \bm{\tilde{g}}_{m, \text{Rx}}^{(i)} \right\}_{m=1}^M \right) }
	{\label{eq:subproblem-block-a-rx}}{}
\end{mini}
with $f_{\bm{a}_{\text{Rx}}} ( \{ \bm{a}_{p, \text{Rx}}, \bm{a}_{p, \text{Tx}}^{(i+1)}, \bm{\tilde{c}}_{p}^{(i)}, \bm{\tilde{h}}_p^{(i)} \}_{p=1}^P, \, \{ \bm{\tilde{g}}_{n, \text{Tx}}^{(i)} \}_{n=1}^N, \, \{ \bm{\tilde{g}}_{m, \text{Rx}}^{(i)} \}_{m=1}^M ) $ \linebreak
$ = \sum_{p=1}^{P} \norm*{ {\bm{Y}_p}_{(2)} - ( \bm{1}_N^\mathsf{T} \kron \diag{ \bm{a}_{p, \text{Rx}}} ) ( {\bm{A}}_{p, \text{Tx}}^{(i+1)} \khatri \bm{\tilde{H}}_p^{(i)} \khatri \bm{\tilde{B}}_p^{(i)} )^\mathsf{T} }_\mathsf{F}^2$.
Following the same steps as in \Cref{sec:updating-a-tx}, problem \cref{eq:subproblem-block-a-rx} is decoupled into $PM$ scalar subproblems 
\begin{mini}
	{[\bm{a}_{p, \text{Rx}}]_m \in \mathbb{C}}{\sum_{n=1}^{N} \norm{ [\bm{\mathcal{Y}}_p]_{n,m,:,:} - \elem*{\bm{a}_{p, \text{Rx}}}_m \elem*{\bm{a}_{p, \text{Tx}}^{(i+1)}}_n \bm{\tilde{b}}_{n,m,p}^{(i)} \left(\bm{\tilde{h}}_p^{(i)}\right)^\mathsf{T} }_2^2 \label{eq:subproblem-decoupled-a-rx}}
	{}{}
\end{mini}
with solutions
\begin{equation}
	\label{eq:subproblem-decoupled-solution-a-rx}
	\begin{aligned}
		\elem*{\bm{a}_{p, \text{Rx}}^{(i+1)}}_m & = \left( \norm{ \bm{\tilde{h}}_p^{(i)}}_2^2 \sum\limits_{n = 1}^N \abs{\elem*{\bm{a}_{p, \text{Tx}}^{(i+1)}}_n}^2 \norm{\bm{\tilde{b}}_{n,m,p}^{(i)}}_2^2 \right)^{-1} \sum\limits_{n = 1}^N  \elem*{\bm{a}_{p, \text{Tx}}^{(i+1)}}_n^* \left( \bm{\tilde{b}}_{n,m,p}^{(i)} \right)^\mathsf{H} \elem*{\bm{\mathcal{Y}}_p}_{n,m,:,:} \left( \bm{\tilde{h}}_p^{(i)} \right)^*
	\end{aligned}
\end{equation}
for $m = 1, \ldots, M$ and $p = 1, \ldots, P$.

\subsection{Update of $\{\titlebm{c}_{p}\}_{p=1}^P$}

Following the \gls{BCD} procedure as above, we next consider the update of the block variables $\{\bm{c}_{p}\}_{p=1}^P$. First, we make the nonconvex constraint \cref{eq:scpd-original-problem-c-constraint} implicit by introducing the parameterization $\elem{\bm{c}_p}_\ell = \eul^{\ju\phi_{\ell,p}}$ and consider $\phi_{\ell,p}	$ with $\phi_{\ell,p} \in [-\pi, \pi)$ for $\ell=1,\ldots,L$ and $p=1,\ldots,P$ as the optimization variables instead. Second, we relax constraint \cref{eq:scpd-original-problem-c1-constraint}, which can always be satisfied by the rescaling procedure \cref{eq:rescaling}.

Applying the mode-three unfolding in \cref{eq:unfolding-mode-three} on \cref{eq:costfct} yields the modified approximate problem
\begin{mini}
	{\left\{ \bm{\phi}_p \in (-\pi, \pi]^L \right\}_{p=1}^P}{f_{\bm{\phi}} \! \left( \left\{ \bm{\phi}_{p}, \bm{a}_{p, \text{Tx}}^{(i+1)}, \bm{a}_{p, \text{Rx}}^{(i+1)}, \bm{\tilde{h}}_p^{(i)} \right\}_{p=1}^P, \, \left\{ \bm{\tilde{g}}_{n, \text{Tx}}^{(i)} \right\}_{n=1}^N, \, \left\{ \bm{\tilde{g}}_{m, \text{Rx}}^{(i)} \right\}_{m=1}^M \right), }
	{\label{eq:subproblem-block-c}}{}
\end{mini}
where $\bm{\phi}_p = [\phi_{1,p}, \ldots, \phi_{L,p}]^\mathsf{T} \in [-\pi, \pi)^L$ and $f_{\bm{\phi}} ( \{ \bm{\phi}_{p}, \bm{a}_{p, \text{Tx}}^{(i+1)}, \bm{a}_{p, \text{Rx}}^{(i+1)}, \bm{\tilde{h}}_p^{(i)} \}_{p=1}^P, \, \{ \bm{\tilde{g}}_{n, \text{Tx}}^{(i)} \}_{n=1}^N, \, \{ \bm{\tilde{g}}_{m, \text{Rx}}^{(i)} \}_{m=1}^M ) $ \linebreak
$= \sum_{p=1}^{P} \norm*{ {\bm{Y}_p}_{(3)} - \diag{\eul^{\ju\bm{\phi}_{p}}} \, \bm{\tilde{G}}^{(i)} ( \bm{A}_{p, \text{Rx}}^{(i+1)} \khatri \bm{A}_{p, \text{Tx}}^{(i+1)} \khatri \bm{\tilde{H}}_p^{(i)} )^\mathsf{T} }_\mathsf{F}^2$ using
$\bm{\tilde{G}}^{(i)} = ( \bm{\tilde{G}}_\text{Tx}^{(i)\mathsf{T}} \khatri \bm{\tilde{G}}_\text{Rx}^{(i)\mathsf{T}} )^\mathsf{T} $ \linebreak
$= [
\diag{\bm{\tilde{g}}_{1,\text{Tx}}^{(i)}} \, \bm{\tilde{G}}_\text{Rx}^{(i)}, \ldots, \diag{\bm{\tilde{g}}_{N,\text{Tx}}^{(i)}} \, \bm{\tilde{G}}_\text{Rx}^{(i)}
]$
with $\bm{\tilde{G}}_\text{Tx}^{(i)}$ and $\bm{\tilde{G}}_\text{Rx}^{(i)}$ according to \cref{eq:Gtx} and \cref{eq:Grx}, respectively, while $\eul^{(\cdot)}$ denotes the elementwise exponential function.

As in \Cref{sec:updating-a-tx,sec:updating-a-rx}, problem \cref{eq:subproblem-block-c} decouples and can be solved separately for every $p$. Additionally, the diagonal matrix containing the optimization variable, i.e.\@ $\diag{\eul^{\ju\bm{\phi}_{p}}}$, decouples the optimization variable elements $\elem{\bm{\phi}_p}_\ell$, leading to $PL$ parallel surrogate problems
\begin{maxi}
	{\phi_{\ell,p} \in (-\pi, \pi]}{ \reval*{  \elem*{ {\bm{Y}_p}_{(3)} }_{\ell,:}^\mathsf{T} \left( \bm{A}_{p, \text{Rx}}^{(i+1)} \khatri \bm{A}_{p, \text{Tx}}^{(i+1)} \khatri \bm{\tilde{H}}_p^{(i)} \right)^*  \elem*{ \bm{\tilde{G}}^{(i)} }_{\ell,:} \, \eul^{-\ju\phi_{\ell,p}} }}
	{\label{eq:subproblem-decoupled-c}}{}
\end{maxi}
for $\ell = 1, \ldots, L$ and $p = 1,\ldots, P$, respectively.
A maximizer to \cref{eq:subproblem-decoupled-c} is given by $\phi_{\ell,p}^{\rm opt} $ \linebreak 
$= \arg ( (\elem{ {\bm{Y}_p}_{(3)} }_{\ell,:})^\mathsf{T} ( \bm{A}_{p, \text{Rx}}^{(i+1)} \khatri \bm{A}_{p, \text{Tx}}^{(i+1)} \khatri \bm{\tilde{H}}_p^{(i)} )^* (\elem{\bm{\tilde{G}}^{(i)}}_{\ell,:})^* )$ such that the $\ell$-th element of the phase response $\bm{c}_p$ can be estimated as
\begin{equation}
	\label{eq:subproblem-solution-c}
	\begin{aligned}
		\elem*{ \bm{c}_p^{(i+1)} }_\ell & =  \eul^{\ju\phi_{\ell,p}^{\rm opt}} = \eul^{\ju \arg \left( \sum\limits_{n=1}^{N} \sum\limits_{m=1}^M \elem*{\bm{a}_{p, \text{Tx}}^{(i+1)}}_n^* \elem*{\bm{a}_{p, \text{Rx}}^{(i+1)}}_m^* \elem*{\bm{\tilde{g}}_{n,\text{Tx}}^{(i)}}_\ell \elem*{\bm{\tilde{g}}_{m,\text{Rx}}^{(i)}}_\ell \left( \bm{\tilde{h}}_p^{(i)} \right)^\mathsf{H} \elem*{\bm{\mathcal{Y}}_p}_{n,m,\ell,:}  \right)}
	\end{aligned}
\end{equation}
for $\ell = 1, \ldots, L$ and $p = 1,\ldots, P$.

\subsection{Update of $\{ \titlebm{g}_{n, \text{Tx}}\}_{n=1}^N$}
\label{sec:updating-g-tx}

According to the \gls{BCD} procedure, we next consider the block variable update $\{\bm{g}_{n,\text{Tx}}\}_{n=1}^N$. We relax the nonconvex constraint \cref{eq:scpd-original-problem-g-tx-constraint-max} and consider $\elem*{\bm{g}_{n,\text{Tx}}}_\ell \leq 1$ for $n = 2,\ldots,N$ and $\ell = 1, \ldots, L$ instead, because the original constraint can always be satisfied by applying the rescaling \cref{eq:rescaling} to the solution of the relaxed subproblem. The modified relaxed problem with the remaining parameters fixed therefore reduces to
\begin{mini!}
	{\left\{ \bm{g}_{n,\text{Tx}} \in \mathbb{R}_+^{L} \right\}_{n=1}^N}{f_{\bm{g}_{\text{Tx}}} \! \left( \left\{ \bm{g}_{n,\text{Tx}} \right\}_{n=1}^N, \left\{ \bm{a}_{p, \text{Tx}}^{(i+1)}, \bm{a}_{p, \text{Rx}}^{(i+1)} , \bm{c}_p^{(i+1)} ,\bm{\tilde{h}}_p^{(i)}\right\}_{p=1}^P, \left\{ \bm{\tilde{g}}_{m,\text{Rx}}^{(i)} \right\}_{m = 1}^M \right) \label{eq:subproblem-block-g-tx-objective}}
	{\label{eq:subproblem-block-g-tx}}{}
	\addConstraint{ \bm{g}_{1,\text{Tx}}}{= \bm{1}_L \label{eq:subproblem-block-g-tx-constraint-ref}}
	\addConstraint{ -\elem*{\bm{g}_{n,\text{Tx}}}_\ell + \varepsilon}{ \leq 0, \quad n = 2,\, \dots,\, N, \ \ell = 1,\, \dots,\, L \label{eq:subproblem-block-g-tx-constraint-min}}
	\addConstraint{ \elem*{\bm{g}_{n,\text{Tx}}}_\ell - 1 }{\leq 0, \quad n = 2,\, \dots,\, N, \ \ell = 1,\, \dots,\, L \label{eq:subproblem-block-g-tx-constraint-max}},
\end{mini!}
where we remark that $f_{\bm{g}_{\text{Tx}}} ( \{ \bm{g}_{n,\text{Tx}} \}_{n=1}^N, \{ \bm{a}_{p, \text{Tx}}^{(i+1)}, \bm{a}_{p, \text{Rx}}^{(i+1)} , \bm{c}_p^{(i+1)} ,\bm{\tilde{h}}_p^{(i)} \}_{p=1}^P, \{ \bm{\tilde{g}}_{m,\text{Rx}}^{(i)} \}_{m = 1}^M ) $ \linebreak
$= \sum_{p=1}^{P} \sum_{n=1}^{N} \sum_{m=1}^{M} \sum_{\ell=1}^{L} \norm*{ 
	\elem{\bm{c}_p^{(i+1)}}_{\ell}^* \elem{\bm{\mathcal{Y}}_p}_{n,m,\ell,:}
	- \elem{\bm{g}_{n,\text{Tx}}}_{\ell}  \elem{\bm{\tilde{g}}_{m,\text{Rx}}^{(i)}}_{\ell}  \elem{\bm{a}_{p,\text{Tx}}^{(i+1)}}_n \elem{\bm{a}_{p, \text{Rx}}^{(i+1)}}_m \bm{\tilde{h}}_p^{(i)}  }_\mathsf{F}^2$
takes measurements \linebreak
from all $P$ positions into account.

Ignoring the constraints in \cref{eq:subproblem-block-g-tx-constraint-min}-\cref{eq:subproblem-block-g-tx-constraint-max}, the problem reduces to a simple linear \gls{LS} problem with decoupled variables.
The corresponding minimizers are given in closed form as
\begin{equation}
	\label{eq:subproblem-solution-g-tx-decoupled-relaxed}
	\begin{aligned}
		\elem*{ \bm{g}_{n,\text{Tx}}^{\text{LS}} }_\ell & = \frac{
			\reval*{ \sum\limits_{p=1}^{P} \sum\limits_{m=1}^{M} \elem*{\bm{a}_{p,\text{Tx}}^{(i+1)}}_n^* \, \elem*{\bm{a}_{p, \text{Rx}}^{(i+1)}}_m^* \,
				\elem*{\bm{c}_p^{(i+1)}}_{\ell}^* \, \elem*{\bm{\tilde{g}}_{m,\text{Rx}}^{(i)}}_{\ell} \left(\bm{\tilde{h}}_p^{(i)}\right)^\mathsf{H}  \elem*{\bm{\mathcal{Y}}_p}_{n,m,\ell,:} }}{\sum\limits_{p=1}^{P} \sum\limits_{m=1}^{M} \abs{\elem*{\bm{a}_{p, \text{Tx}}^{(i+1)}}_n}^2 \, \abs{\elem*{\bm{a}_{p, \text{Rx}}^{(i+1)}}_m}^2 \left(\elem*{\bm{\tilde{g}}_{m,\text{Rx}}^{(i)}}_{\ell}\right)^2 \, \norm{ \bm{\tilde{h}}_p^{(i)}}_2^2 }
	\end{aligned}
\end{equation}
for $\ell = 1, \ldots, L$ and $n = 2, \ldots, N$. Neglecting the trivial case of $n=1$, there remain $L(N-1)$ terms in \cref{eq:subproblem-block-g-tx-objective} that are quadratic functions in one scalar variable each, such that there exists a respective global minimum that either fulfills \cref{eq:subproblem-block-g-tx-constraint-min}-\cref{eq:subproblem-block-g-tx-constraint-max} or is located outside these boundaries. In the latter case, \cref{eq:subproblem-block-g-tx-objective} is strictly monotonically descending towards \cref{eq:subproblem-solution-g-tx-decoupled-relaxed} within the constrained interval. The solution to \cref{eq:subproblem-block-g-tx} is therefore given by \cref{eq:subproblem-solution-g-tx-decoupled-relaxed} projected onto $[ \varepsilon, 1 ]$, i.e.,
\begin{equation}
	\label{eq:subproblem-solution-g-tx-decoupled}
	\elem*{\bm{g}_{n,\text{Tx}}^{(i+1)}}_\ell = \begin{cases}
		1 & \elem*{\bm{g}_{n,\text{Tx}}^{\text{LS}}}_\ell > 1 \\
		\elem*{\bm{g}_{n,\text{Tx}}^{\text{LS}}}_\ell & \varepsilon \leq \elem*{\bm{g}_{n,\text{Tx}}^{\text{LS}}}_\ell \leq 1 \\
		\varepsilon & \elem*{\bm{g}_{n,\text{Tx}}^{\text{LS}}}_\ell < \varepsilon.
	\end{cases}
\end{equation}

\subsection{Update of $\{\titlebm{g}_{m, \text{Rx}}\}_{m=1}^M$}
\label{sec:updating-g-rx}

Following similar considerations as in the previous subsection, we relax the nonconvex constraint \cref{eq:scpd-original-problem-g-rx-constraint-max} by considering $\elem*{\bm{g}_{m,\text{Rx}}}_\ell \leq 1$ for $m = 1,\ldots, M$ and $\ell = 1, \ldots, L$ instead, and remark that the original constraint can always be satisfied by proper rescaling of the solution of the relaxed subproblem according to \cref{eq:rescaling}.

The resulting relaxed problem is
\begin{mini!}
	{\left\{ \bm{g}_{m,\text{Rx}} \in \mathbb{R}_+^{L} \right\}_{m=1}^M}{f_{\bm{g}_{\text{Rx}}} \! \left( \left\{ \bm{g}_{m,\text{Rx}} \right\}_{m=1}^M, \left\{ \bm{a}_{p, \text{Tx}}^{(i+1)} , \bm{a}_{p, \text{Rx}}^{(i+1)} , \bm{c}_p^{(i+1)} , \bm{\tilde{h}}_p^{(i)} \right\}_{p=1}^P,  \left\{\bm{g}_{n,\text{Tx}}^{(i+1)}\right\}_{n=1}^N \right) \label{eq:subproblem-block-g-rx-objective}}
	{\label{eq:subproblem-block-g-rx}}{}
	\addConstraint{ - \elem*{\bm{g}_{m,\text{Rx}}}_\ell + \varepsilon}{ \leq 0, \quad m = 1, \, \dots,\, M, \ \ell = 1, \, \dots,\, L \label{eq:subproblem-block-g-rx-constraint-min}}
	\addConstraint{ \elem*{\bm{g}_{m,\text{Rx}}}_\ell - 1 }{\leq 0, \quad m = 1, \, \dots,\, M, \ \ell = 1, \, \dots,\, L \label{eq:subproblem-block-g-rx-constraint-max}}
\end{mini!}
with $f_{\bm{g}_{\text{Rx}}} ( \{ \bm{g}_{m,\text{Rx}} \}_{m=1}^M, \{ \bm{a}_{p, \text{Tx}}^{(i+1)} , \bm{a}_{p, \text{Rx}}^{(i+1)} , \bm{c}_p^{(i+1)} , \bm{\tilde{h}}_p^{(i)} \}_{p=1}^P ,\{\bm{g}_{n,\text{Tx}}^{(i+1)} \}_{n=1}^N ) $ \linebreak
$= \sum_{p=1}^{P} \sum_{n=1}^{N} \sum_{m=1}^{M} \sum_{\ell=1}^{L} \norm*{ \elem{\bm{c}_p^{(i+1)}}_{\ell}^* \elem{\bm{\mathcal{Y}}_p}_{n,m,\ell,:} - \elem{\bm{g}_{m,\text{Rx}}}_{\ell}  \elem{\bm{g}_{n,\text{Tx}}^{(i+1)}}_{\ell}  \elem{\bm{a}_{p,\text{Tx}}^{(i+1)}}_n \elem{\bm{a}_{p, \text{Rx}}^{(i+1)}}_m \bm{\tilde{h}}_p^{(i)} }_2^2$.

The corresponding \gls{LS} solution of the relaxed problem is given by
\begin{equation}
	\label{eq:subproblem-solution-g-rx-decoupled-relaxed}
	\begin{aligned}
		\elem*{ \bm{g}_{m,\text{Rx}}^{\text{LS}} }_\ell & = \frac{ \reval*{ \sum\limits_{p=1}^{P} \sum\limits_{n=1}^{N} 	\elem*{\bm{a}_{p,\text{Tx}}^{(i+1)}}_n^* \, \elem*{\bm{a}_{p, \text{Rx}}^{(i+1)}}_m^* \,
		\elem*{\bm{c}_p^{(i+1)}}_{\ell}^* \,  \elem*{\bm{g}_{n,\text{Tx}}^{(i+1)}}_{\ell}  \left( \bm{\tilde{h}}_p^{(i)} \right)^\mathsf{H}  \elem*{\bm{\mathcal{Y}}_p}_{n,m,\ell,:} }}{\sum\limits_{p=1}^{P} \sum\limits_{n=1}^{N} \abs{ \elem*{\bm{a}_{p, \text{Tx}}^{(i+1)}}_n }^2 \, \abs{\elem*{\bm{a}_{p, \text{Rx}}^{(i+1)}}_m}^2 \left(\elem*{\bm{g}_{n,\text{Tx}}^{(i+1)}}_{\ell}\right)^2 \, \norm{ \bm{\tilde{h}}_p^{(i)} }_2^2 }
	\end{aligned}
\end{equation}
for $\ell = 1, \ldots, L$ and $m = 1, \ldots, M$, which are again projected onto $[ \varepsilon, 1 ]$, i.e.,
\begin{equation}
	\label{eq:subproblem-solution-g-rx-decoupled}
	\elem*{\bm{g}_{m,\text{Rx}}^{(i+1)}}_\ell = \begin{cases}
		1 & \elem*{\bm{g}_{m,\text{Rx}}^{\text{LS}}}_\ell > 1 \\
		\elem*{\bm{g}_{m,\text{Rx}}^{\text{LS}}}_\ell & \varepsilon \leq \elem*{\bm{g}_{m,\text{Rx}}^{\text{LS}}}_\ell \leq 1 \\
		\varepsilon & \elem*{\bm{g}_{m,\text{Rx}}^{\text{LS}}}_\ell < \varepsilon
	\end{cases}.
\end{equation}

\subsection{Update of $\{\titlebm{h}_{p}\}_{p=1}^P$}

Making use of the definitions \cref{eq:factor-matrix-a-tx}, \cref{eq:factor-matrix-a-rx}, and  \cref{eq:factor-matrix-b}, the approximate problem in the gain vectors $\{ \bm{h}_{p} \}_{p=1}^P$ for the remaining parameters fixed yields the standard linear \gls{LS} problem 
\begin{mini}
	{\left\{ \bm{h}_{p} \in \mathbb{C}^{T} \right\}_{p=1}^{P}}{f_{\bm{h}} \! \left( \left\{ \bm{h}_{p} , \bm{a}_{p, \text{Tx}}^{(i+1)} , \bm{a}_{p, \text{Rx}}^{(i+1)} , \bm{c}_p^{(i+1)} \right\}_{p = 1}^P, \left\{\bm{g}_{n,\text{Tx}}^{(i+1)}\right\}_{n=1}^N, \left\{\bm{g}_{m,\text{Rx}}^{(i+1)}\right\}_{m=1}^M \right) }
	{\label{eq:subproblem-block-h}}{}
\end{mini}
with $f_{\bm{h}} ( \{ \bm{h}_{p} , \bm{a}_{p, \text{Tx}}^{(i+1)} , \bm{a}_{p, \text{Rx}}^{(i+1)} , \bm{c}_p^{(i+1)} \}_{p = 1}^P, \{ \bm{g}_{n,\text{Tx}}^{(i+1)} \}_{n=1}^N, \{ \bm{g}_{m,\text{Rx}}^{(i+1)} \}_{m=1}^M ) $ \linebreak 
$= \sum_{p=1}^{P} \norm*{ {\bm{Y}_p}_{(4)} - \bm{h}_p \bm{1}_{NM}^\mathsf{T} ( \bm{B}_{p}^{(i+1)} \khatri \bm{A}_{p, \text{Rx}}^{(i+1)} \khatri \bm{A}_{p, \text{Tx}}^{(i+1)} )^\mathsf{T} }_\mathsf{F}^2$. The problem \cref{eq:subproblem-block-h}
decouples into $P$ \gls{LS} problems with individual minimizers given by
\begin{equation}
	\label{eq:scpd-solution-h-decoupled}
	\begin{aligned}
		\bm{h}_{p}^{(i+1)}
		&= \frac{\sum\limits_{n = 1}^N \sum\limits_{m = 1}^M \elem*{\bm{a}_{p, \text{Tx}}^{(i+1)}}_n^* \, \elem*{\bm{a}_{p, \text{Rx}}^{(i+1)}}_m^* \left( \elem*{\bm{\mathcal{Y}}_p]_{n,m,:,:} } \right)^\mathsf{T} \left( \bm{b}_{n,m,p}^{(i+1)} \right)^* }{\sum\limits_{n = 1}^N \sum\limits_{m = 1}^M 
			\abs{\elem*{\bm{a}_{p, \text{Tx}}^{(i+1)}}_n}^2 \, \abs{\elem*{\bm{a}_{p, \text{Rx}}^{(i+1)}}_m}^2 \,
			\norm{\bm{b}_{n,m,p}^{(i+1)}}_2^2 }
	\end{aligned}
\end{equation}
for $p = 1, \ldots, P$. After the update of $\{ \bm{h}_{p} \}_{p=1}^P$, iteration $i$ concludes by initiating the scaling procedure \cref{eq:rescaling}.

With this, we now have introduced the subproblems and corresponding solutions of a modified \gls{BCD} method to learn the array response for $P$ sample calibration points by estimating the corresponding parameters of the tensor model in \cref{eq:array-output-model}, i.e.\@ $\left\{ \bm{g}_{n, \text{Tx}} \right\}_{n=1}^N$, $\left\{ \bm{g}_{m, \text{Rx}} \right\}_{m=1}^M$, and $\{ \bm{a}_{p, \text{Tx}}, \, \bm{a}_{p, \text{Rx}}, \, \bm{c}_p, \, \bm{h}_p \}_{p=1}^P$. Additionally, the solutions to all approximate problems of the modified \gls{BCD} method are given in closed-form and the individual block variables as well as their rescaling \cref{eq:rescaling} can be computed in parallel.

\subsection{Initialization of the BCD}
\label{sec:initialization}

The parameter $\bm{g}_{1,\text{Tx}}$ is fixed according to \cref{eq:scpd-original-problem-g-tx-constraint-ref}, so w.l.o.g., we start our proposed calibration procedure by estimating the transmit magnitude response update $\bm{g}_{n, \text{Tx}}^{(i+1)}$ at iteration $i$. Estimates for the remaining parameters at the initial iteration $i=1$ are obtained 
from the calibration data for snapshots $t = 1, \ldots, T$ using the simple initialization
	$\bm{\tilde{a}}_{p,\text{Tx}}^{(1)} = \bm{1}_N$,
	$\bm{\tilde{a}}_{p,\text{Rx}}^{(1)} = \bm{1}_M$,
	$\bm{\tilde{g}}_{m, \text{Rx}}^{(1)} = \bm{1}_L$,
	$\bm{\tilde{c}}_p^{(1)} = \bm{1}_L$, and
	$\elem{\bm{\tilde{h}}_p^{(1)}}_t = (LMN)^{-1} \sum_{\ell=1}^{L} \sum_{m=1}^{M} \sum_{n=1}^{N} \abs{ \reval*{ \elem{\bm{\mathcal{Y}}_p}_{n,m,\ell,t} }}$ \linebreak 
	$+ \ju (LMN)^{-1} \sum_{\ell=1}^{L} \sum_{m=1}^{M} \sum_{n=1}^{N} \abs{ \imval*{ \elem{\bm{\mathcal{Y}}_p}_{n,m,\ell,t} }}$,
which we denote as the complex-valued \gls{t-av}.
These initial estimates can be obtained from the measurements in $\bm{\mathcal{Y}}_p$ without additional knowledge of the calibration positions $p$.

\subsection{Range scanning}
\label{sec:range-scanning}

Our experiments reveal that the phase responses $\{ \bm{c}_p \}_{p=1}^P \in \mathbb{C}^L$ learned from calibration measurements exhibit approximately linear phases that are distant-dependent. Thus, we conclude that the far field model with constant group delays holds \cite{richardsPrinciplesModernRadar2010}. Hence, we may model the entries of the $p$-th phase response as
\begin{equation}
	\elem*{ \bm{c}_p }_\ell = \eul^{- \ju (\ell-1) \Delta \phi_p}
	\label{eq:phase-response-approximation}
\end{equation}
with phase slope
$\Delta \phi_p = 2 c^{-1} (r_p + r_{0}) \Delta \omega$, where $c$ is the effective speed of sound for the imaging burst, $r_p$ is the adjusted range from the array to the calibration target position indexed by $p$, $r_{0}$ is a system offset, e.g.\@ due to unknown group delays introduced by the waveguide and mechanical inertia of the \glspl{PUT}, and $\Delta \omega = 2 \pi f_\mathrm{s} (L_{\rm DFT})^{-1}$ is the radial frequency separation based on the sampling frequency $f_\mathrm{s}$ as well as the \gls{DFT} length $L_{\rm DFT}$.

While the true range $r_p$ is known for every calibration point $p$, the system offset range $r_{0}$ is generally unknown. However, we can determine $r_0$ using the \gls{ESPRIT} for the simple case of a single source as \cite{royESPRITestimationSignalParameters1989}
\begin{equation}
\hat{r}_0 = \frac{c}{2\Delta \omega} \sum_{p=1}^P \left( \arg \! \left( \elem*{\bm{\hat{c}}_p}_{1:L-1}^\mathsf{H} \elem*{\bm{\hat{c}}_p}_{2:L} \right) - 2 r_p \frac{\Delta \omega}{c}\right).
\label{eq:phase-slope-estimate}
\end{equation}

With the model-based approach described above, the density of dictionary entries in the range dimension can be flexibly increased by choosing different scanning ranges $r_{p_1}', \ldots, r_{p_R}'$ for every calibrated position $p$. Thus, the total number of dictionary entries $P$ is increased to $P = RP_{\rm measured}$. This allows a higher accuracy for the estimation of the range in distances that have not been learned during calibration.

\subsection{Complexity and methods for comparison}
\label{sec:complexity}

Instead of learning from measurements under the common narrow-band assumption \cite{vibergIntroductionArrayProcessing2014}, a normalized and single-snapshot array response $\bm{\mathcal{R}}_p \in \mathbb{C}^{M \times N \times L}$ for a single target at position $p$ can be described in closed form as the third-order tensor
\begin{equation}
	\label{eq:analytic-response}
	\bm{\mathcal{R}}_{p,\text{analytic}} = \bm{a}_{p,\text{Tx,analytic}} \outerprod \bm{a}_{p,\text{Rx,analytic}} \outerprod \bm{b}_{p,\text{analytic}}
\end{equation}
using the analytic steering vectors $\bm{a}_{p,\text{Tx,analytic}} = [ \eul^{-\ju \bm{k}(\vartheta,\varphi)^\mathsf{T} \bm{r}_{1,\text{Tx}}}, \ldots, \eul^{-\ju \bm{k}(\vartheta,\varphi)^\mathsf{T} \bm{r}_{N,\text{Tx}}} ]^\mathsf{T}$ and $\bm{a}_{p,\text{Rx,analytic}} $ \linebreak
$= [ \eul^{-\ju \bm{k}(\vartheta,\varphi)^\mathsf{T} \bm{r}_{1,\text{Rx}}}, \ldots, \eul^{-\ju \bm{k}(\vartheta,\varphi)^\mathsf{T} \bm{r}_{M,\text{Rx}}} ]^\mathsf{T}$, the linear frequency response $\bm{b}_{p,\text{analytic}} = [ 1, \ldots, \eul^{- \ju 2 (L-1) r_p \frac{\Delta \omega}{c}} ]^\mathsf{T}$, the wave vector $\bm{k}(\vartheta,\varphi) = -2 \pi f_0 c^{-1} [\sin(\vartheta) \cos(\varphi), \, \sin(\varphi), \, \cos(\vartheta) \cos(\varphi)]^\mathsf{T}$, the array element positions $\bm{r}_i = [x_i, \, y_i, \, z_i]^\mathsf{T}$, and $\Delta \omega = 2 \pi f_\text{s} (L_\text{DFT})^{-1}$. Note that the analytic frequency response $\bm{b}_{p,\text{analytic}}$ uses the same linear phase model as described in \Cref{sec:range-scanning}.

A simple calibration technique that reduces relative gain and phase offsets in the different transceiver elements is given by broadside compensation \cite{vibergIntroductionArrayProcessing2014,aumannPhasedarrayCalibrationAdaptive1991}. For this, a measurement tensor $\bm{\mathcal{Y}}_{(\ang{0},\ang{0})}$ with a target at broadside direction $(\vartheta,\varphi) = (\ang{0},\ang{0})$ is approximated by the rank-$1$ \gls{CPD} \cref{eq:rank-one-approximation} using a classic \gls{ALS} algorithm \cite{koldaTensorDecompositionsApplications2009} with the initialization in \Cref{sec:initialization}. After appropriate rescaling, unique estimates $\bm{\hat{a}}_{(\ang{0},\ang{0}),\text{Tx,CPD}}$, $\bm{\hat{a}}_{(\ang{0},\ang{0}),\text{Rx,CPD}}$, and $\bm{\hat{b}}_{(\ang{0},\ang{0}),\text{CPD}}$ are used as complex valued gains, i.e.\@ amplitude factors and phase shifts, to correct the components of the analytic array responses $\bm{\mathcal{R}}_p$ of all directions at the same range:
\begin{equation}
	\begin{aligned}
		\bm{a}_{p,\text{Tx,analytic}} &= \bm{\hat{a}}_{(\ang{0},\ang{0}),\text{Tx,CPD}} \hadamul \begin{bmatrix}
			\eul^{-\ju \bm{k}(\vartheta,\varphi)^\mathsf{T} \bm{r}_{1,\text{Tx}}}, & \ldots, & \eul^{-\ju \bm{k}(\vartheta,\varphi)^\mathsf{T} \bm{r}_{N,\text{Tx}}}
		\end{bmatrix}^\mathsf{T}, \\
		\bm{a}_{p,\text{Rx,analytic}} &= \bm{\hat{a}}_{(\ang{0},\ang{0}),\text{Rx,CPD}} \hadamul \begin{bmatrix}
			\eul^{-\ju \bm{k}(\vartheta,\varphi)^\mathsf{T} \bm{r}_{1,\text{Rx}}}, & \ldots, & \eul^{-\ju \bm{k}(\vartheta,\varphi)^\mathsf{T} \bm{r}_{M,\text{Rx}}}
		\end{bmatrix}^\mathsf{T}, \\
		\bm{b}_{p,\text{analytic}} &= \bm{\hat{b}}_{(\ang{0},\ang{0}),\text{CPD}} \hadamul \begin{bmatrix}
			\eul^{- \ju 2 \cdot 0 \cdot r_p \frac{\Delta \omega}{c}}, & \ldots, & \eul^{- \ju 2 (L-1) r_p \frac{\Delta \omega}{c}}
		\end{bmatrix}^\mathsf{T}.
	\end{aligned}
\end{equation}

The analytic model for $\bm{\mathcal{R}}_p$ yields a low computational load with broadside compensation but requires precise knowledge of the array geometry, which may be unavailable if the array is embedded in a system or affected by temperature changes and aging. Our proposed calibration technique and the positionwise rank-$1$ \gls{CPD} approximation \cref{eq:rank-one-approximation} overcome this limitation, albeit with higher computational complexity.

The computational cost of the modified \gls{BCD} procedure for problem \cref{eq:scpd-original-problem} is mainly driven by updating the magnitude responses $\{\bm{g}_{n,\text{Tx}}\}_{n=1}^N$ and $\{\bm{g}_{m,\text{Rx}}\}_{m=1}^M$, which sets our tensor model apart from the rank-$1$ \gls{CPD} approximation. As noted in \Cref{sec:updating-g-tx,sec:updating-g-rx}, estimating $\bm{g}_{n,\text{Tx}}$ and $\bm{g}_{m,\text{Rx}}$ requires measurements from all $P$ positions, respectively. While this makes our method computationally more demanding than a positionwise rank-$1$ \gls{CPD} approximation, the complexity remains in the same order as the decoupling of subproblems allows for elementwise updates. \Cref{table:computational-load} lists the computational load for the three calibration techniques in terms of multiply-accumulate (MAC) operations, where the load for the rank-$1$ \gls{CPD} approximation is based on elementwise updates for a fair comparison.

\begin{table}[t]
	\centering
	\begin{tabular}{ c c }
		\hline
		\textbf{Algorithm} & \textbf{Computational load} \\ 
		\hline
		Broadside compensation & $\mathcal{O}(P(N+M+L))$ \\ 
		\hline
		Rank-$1$ CPD & $\mathcal{O}(PNMLT)$ \\ 
		\hline
		Proposed & $\mathcal{O}(PNMLT)$ \\ 
		\hline
	\end{tabular}
	\caption{Computational load of multiply-accumulate (MAC) operations for different tensor-based calibration algorithms.}
	\label{table:computational-load}
\end{table}

\subsection{Convergence analysis}

In the following, we establish the convergence result for the proposed modified \gls{BCD} algorithm that follows the proof of the conventional \gls{BCD} algorithm provided in~\cite[Proposition 3.7.1]{bertsekasNonlinearProgramming2016}.
Let us define the vector $\bm{\theta}^{(i)}$ that contains all block variables $\{ \bm{a}_{p, \text{Tx}}^{(i+1)}, \bm{a}_{p, \text{Rx}}^{(i+1)},	\bm{\phi}_{p}^{(i+1)},\bm{h}_{p}^{(i+1)} \}_{p = 1}^P$, $\{ \bm{g}_{n, \text{Tx}}^{(i+1)}\}_{n = 1}^N ,\{ \bm{g}_{m, \text{Rx}}^{(i+1)}\}_{m = 1}^M$ at iteration $i$ after performing their respective updates \cref{eq:subproblem-decoupled-solution-a-tx,eq:subproblem-decoupled-solution-a-rx,eq:subproblem-solution-c,eq:subproblem-solution-g-tx-decoupled,eq:subproblem-solution-g-rx-decoupled,eq:scpd-solution-h-decoupled} but prior to the scaling, whereas $\bm{\tilde{\theta}}^{(i)}$ denotes the vector of all block variables $\{ \bm{\tilde{a}}_{p, \text{Tx}}^{(i+1)}, \bm{\tilde{a}}_{p, \text{Rx}}^{(i+1)}, \bm{\tilde{\phi}}_{p}^{(i+1)},\bm{\tilde{h}}_{p}^{(i+1)} \}_{p = 1}^P$, $\{ \bm{\tilde{g}}_{n, \text{Tx}}^{(i+1)}\}_{n = 1}^N$, $\{ \bm{\tilde{g}}_{m, \text{Rx}}^{(i+1)}\}_{m = 1}^M$ at iteration $i$ after the rescaling in \cref{eq:rescaling}. The following convergence result can then be established.

\begin{prop}
	\label{prop-modified-bcd}
	
	Every limit point of the sequence $\{ \bm{\tilde{\theta}}^{(i)} \}$ is a stationary point of problem~\cref{eq:scpd-original-problem}.
\end{prop}

\begin{proof}[\sketchofproof{ Proposition~\ref{prop-modified-bcd}}]
	
	The above convergence statement can be justified by extending the proof of~\cite[Proposition 3.7.1]{bertsekasNonlinearProgramming2016} for the classic \gls{BCD} algorithm.
	We first observe that in each block variable update, as well as in the scaling step, the function value is nonincreasing. As the objective is bounded below, it follows from the Monotone Convergence Theorem that the function values corresponding to the iterates converge. To further show that a limit point $\bar{\bm{\theta}}$ of the sequence $\{\bm{\tilde{\theta}}^{(i)}\}$ is a stationary point of problem \cref{eq:scpd-original-problem}, we follow the proof of~\cite[Proposition 3.7.1]{bertsekasNonlinearProgramming2016}. We first note that all respective subproblems, i.e.\@ \cref{eq:subproblem-decoupled-a-tx,eq:subproblem-decoupled-a-rx,eq:subproblem-decoupled-c,eq:subproblem-block-g-tx,eq:subproblem-block-g-rx,eq:subproblem-block-h} have unique minimizers. The objective functions of the subproblems are continuously differentiable over the constraint sets and monotonically nonincreasing between the current iterates and the minimizers of the subproblems. Furthermore, the constraint sets are convex and bounded. Thus, the conditions in~\cite[Proposition 3.7.1]{bertsekasNonlinearProgramming2016} are satisfied. Performing the exact steps in the proof of~\cite[Proposition 3.7.1]{bertsekasNonlinearProgramming2016} along the chain of \gls{BCD} updates in one iteration, we can demonstrate that every limit point $\bar{\bm{\theta}}$ of the sequence of iterates $\{ \bm{\tilde{\theta}}^{(i)} \}$ satisfies the first-order stationarity condition
	\begin{equation}
		\text{Re} \left( \grad{f}(\bar{\bm{\theta}})^{\mathsf{H}} (\bm{\theta} - \bar{\bm{\theta}}) \right) \geq 0 \quad \forall \, \bm{\theta} \in \Theta,
		\label{eq:stationarity}
	\end{equation}
	where $\grad{f}(\bar{\bm{\theta}})$ and $\Theta$ denote the vector of partial derivatives and the feasible set of the problem without scaling constraints, i.e., the set defined by constraints \cref{eq:subproblem-block-g-tx-constraint-ref}, \cref{eq:subproblem-block-g-tx-constraint-min}, \cref{eq:subproblem-block-g-tx-constraint-max}, \cref{eq:subproblem-block-g-rx-constraint-min}, \cref{eq:subproblem-block-g-rx-constraint-max}, respectively.
	Together with the fact that the sequence of iterates $\{\bm{\tilde{\theta}}^{(i)}\}$ is enforced to satisfy the scaling constraints by the scaling procedure \cref{eq:rescaling}, it can be concluded from \cref{eq:stationarity} that every limit point $\bar{\bm{\theta}}$ is a stationary point of problem \cref{eq:scpd-original-problem}.
\end{proof}

Due to the multilinear nature of its objective, we note that problem \cref{eq:scpd-original-problem} is nonconvex, and the convergence of the modified BCD algorithm to a global minimum cannot generally be guaranteed (c.f.\@ \cite{bertsekasNonlinearProgramming2016}). Therefore, the stationary point to which the proposed modified BCD algorithm converges may be a global minimum, a local minimum, or a saddle point of problem \cref{eq:scpd-original-problem}.

\section{Imaging}
\label{sec:methodology-imaging}

As pointed out in \Cref{sec:introduction}, air-coupled \gls{US} uses the same image formation principles as other in-air sensing modalities. Basic but fast techniques, such as delay-and-sum beamforming in the time domain and conventional beamforming in the frequency domain, have been successfully applied to air-coupled ultrasonic arrays \cite{allevatoRealtime3DImaging2021,kerstensERTISFullyEmbedded2019}. Advanced methods, e.g.\@ \gls{MUSIC} \cite{schmidtMultipleEmitterLocation1986} or \gls{ESPRIT} \cite{royESPRITestimationSignalParameters1989}, promise superresolution, provided the array is well-calibrated. Another class of imaging techniques leverages sparsity in the scene, which can be exploited using sparsity-promoting concepts, e.g.\@ convex $\ell_1$-norm minimization \cite{zhangSurveySparseRepresentation2015c}, including the LASSO problem considered in our previous work \cite{mullerDictionarybasedLearning3Dimaging2020}. Sequential algorithms, such as \gls{MP}, \gls{OMP}, and \gls{OLS} \cite{pesaventoThreeMoreDecades2023}, exploit sparsity using a dictionary, which we obtain from calibration measurements. As a proof-of-concept that array responses can be learned from calibration data with the algorithm provided in \Cref{sec:methodology-calibration}, we employ a sequential multi-target estimation technique that estimates target locations one by one.

Consider the superposition of $K$ sources according to our signal model in \cref{eq:array-output-model}, i.e.,
\begin{equation}
	\begin{aligned}
		\bm{\mathcal{Y}} &= \sum_{k=1}^{K} \bm{\mathcal{Y}}_{p_k} = \sum_{k=1}^{K} \bm{\mathcal{Q}}_{p_k} \outerprod \bm{h}_{p_k} + \bm{\mathcal{Z}},
	\end{aligned}
	\label{eq:multi-source-imaging-model}
\end{equation}
where $\bm{\mathcal{Y}}_{p_k}$ models the single-source tensor \cref{eq:array-output-model} for the $k$-th target at the unknown position indexed by $p_k$. The noiseless signal part contributed by the $k$-th target can be factorized into the signal gain $\bm{h}_{p_k}$ and the third-order tensor
\begin{equation}
	\bm{\mathcal{Q}}_{p}  = 
	\sum_{n=1}^{N} \sum_{m=1}^{M} \left( \bm{e}_n \hadamul \bm{a}_{{p}, \text{Tx}} \right) \outerprod \left( \bm{e}_m \hadamul \bm{a}_{{p}, \text{Rx}} \right) \outerprod \left( \bm{g}_{n, \text{Tx}} \hadamul \bm{g}_{m, \text{Rx}} \hadamul \bm{c}_{p} \right) \in \mathbb{C}^{N \times M \times L}
	\label{eq:TensorQ}
\end{equation}
that is formed from the factors $(\bm{a}_{{p}, \text{Tx}},\bm{a}_{{p}, \text{Rx}}, \bm{c}_{p}, \bm{g}_{n, \text{Tx}},\bm{g}_{m, \text{Rx}})$ for calibration points $p = 1, \ldots, P$. The tensor $\bm{\mathcal{Q}}_{p}$ in \cref{eq:TensorQ} characterizes the array range-angular response of a point reflector at the candidate location indexed by $p$ and is assumed to be fixed and known after calibration. Let
\begin{equation}
	\bm{q}_p = \vectorize*{ \bm{\mathcal{Q}}_{p} } \in \mathbb{C}^{NML}
	\label{eq:range-angle-dict-entry}
\end{equation}
denote the vectorization of the array response tensor $\bm{\mathcal{Q}}_{p}$ at calibration position $p$ and let the set
\begin{equation}
	{\cal Q} = \left\{\bm{q}_1, \bm{q}_2,\ldots, \bm{q}_P  \right\}
	\label{eq:setQ}
\end{equation}
denote the range-angle dictionary for $P$ candidate targets. Our objective is to find the atoms $\bm{q}_{\hat{p}_k}$ in the sparse representation $\unfold{4}{\bm{\mathcal{Y}}} = \bm{Y}_{(4)} \approx \sum_{k = 1}^K \bm{h}_{\hat{p}_k} \bm{q}_{\hat{p}_k}^\mathsf{T}$ under the mode-four unfolding from \cref{eq:unfolding-mode-four}. We remark that the true source number $K$ is generally unknown and needs to be detected \cite{vibergIntroductionArrayProcessing2014}.

\subsection{Sequential imaging with OMP}
\label{sec:imaging-omp}

We select the atoms for the sparse representation of the image measurement tensor $\bm{\mathcal{Y}}$ according to the \gls{OMP} technique \cite{patiOrthogonalMatchingPursuit1993}.
At the beginning of the $i$-th iteration of \gls{OMP}, the location estimate indices $\hat{p}_1, \ldots, \hat{p}_{i-1}$ corresponding to the selected atoms $\bm{q}_{\hat{p}_1}, \ldots,\bm{q}_{\hat{p}_{i-1}} \in {\cal Q}$
along with the corresponding signal gain estimate vectors $\hat{\bm{h}}_{1}^{(i-1)},\ldots, \hat{\bm{h}}_{i-1}^{(i-1)} \in \mathbb{C}^{T}$
of the previously estimated targets are given and the location index estimate $\hat{p}_i$ of the target in the $i$-th iteration is selected as the minimizer 
\begin{equation}
	\hat{p}_i = \argmin_{p \in \left\{ 1,\ldots,P \right\}} \min_{\bm{h} \in \mathbb{C}^T} 
	\norm{ \bm{Y}_{(4)} - \sum_{k = 1}^{i-1}  \hat{\bm{h}}_{k}^{(i-1)} \bm{q}_{\hat{p}_k}^\mathsf{T} - \bm{h} \bm{q}_p^\mathsf{T} }_\mathsf{F}^2.
	\label{eq:imaging-omp-problem}
\end{equation}
The inner optimization problem in \cref{eq:imaging-omp-problem} is a linear \gls{LS} problem in $\bm{h}$ for fixed $\bm{q}_p$ and has the closed form solution 
$\bm{h}^{\text{LS}} = ( \bm{Y}_{(4)} -  \sum_{k = 1}^{i-1} \hat{\bm{h}}_{k}^{(i-1)} \bm{q}_{\hat{p}_k}^\mathsf{T} ) (\norm*{ \bm{q}_p }_2^2)^{-1} \bm{q}_p^*.$ Inserting $\bm{h}^{\text{LS}}$ for $\bm{h}$ in \cref{eq:imaging-omp-problem} yields the simple selection criterion
\begin{equation}
	\text{Step 1:} \qquad \hat{p}_i = 	\argmax_{p \in \left\{ 1,\ldots,P \right\}}
	\norm{ \bm{Y}_{(4),\text{res.}}^{(i-1)} \frac{\bm{q}_p^*}{\norm{\bm{q}_p}_2^2}}_2^2,
	\label{eq:imaging-omp-problem2}
\end{equation}
where
\begin{equation}
	\label{eq:residual}
	\bm{Y}_{(4),\text{res.}}^{(i-1)} = 
	\bm{Y}_{(4)} -  \sum_{k = 1}^{i-1}  \hat{\bm{h}}_{k}^{(i-1)} \bm{q}_{\hat{p}_k}^\mathsf{T}
\end{equation}
is the unfolded residual measurement tensor at the end of the $(i-1)$-th iteration. At the $i$-th iteration, we observe that according to \cref{eq:imaging-omp-problem2}, the atom $\bm{q}_p$ of our learned dictionary ${\cal Q}$ is selected that has the largest correlation to the rows of the residual in \cref{eq:residual}.

In the second step of the $i$-th iteration of the \gls{OMP} method, all gain vectors $\{ \hat{\bm{h}}_k^{(i)} \}_{k=1}^i$ corresponding to the selected target directions $\{ \hat{p}_k \}_{k = 1}^i$ are estimated for fixed atoms $\{ \bm{q}_{\hat{p}_k}^{(i)} \}_{k=1}^i$ as the \gls{LS} minimizers 
\begin{equation}
	\begin{aligned}
		\text{Step 2:} \qquad \hat{\bm{h}}_{k}^{(i)} =
		\argmin_{\bm{h} \in \mathbb{C}^T} \norm{ \bm{Y}_{(4)} - \sum_{k = 1}^{i} \bm{h}_{k}^{(i)} \bm{q}_{\hat{p}_k}^\mathsf{T} }_\mathsf{F}^2
		= \elem*{ \bm{Y}_{(4)} \left(\hat{\bm{Q}}^{(i)}\right)^* \left( \left(\hat{\bm{Q}}^{(i)}\right)^\mathsf{T} \left(\hat{\bm{Q}}^{(i)}\right)^* \right)^{-1} }_{:,k},
	\end{aligned}
	\label{eq:LS-OMP2}
\end{equation}
where $\hat{\bm{Q}}^{(i)} = [\bm{q}_{\hat{p}_1}, \ldots, \bm{q}_{\hat{p}_i}]$. We remark that $\norm*{\hat{\bm{h}}_{k}}_2^2$, which is a measure for the received power of the signal reflected by the target detected at the position indexed by $k$, is later used to create the image.

The two steps of the \gls{OMP} procedure are repeated, e.g., until the Frobenius norm of the updated unfolded residual measurement tensor $\bm{Y}_{(4),\text{res.}}^{(i)}$ according to \cref{eq:residual} falls below a predefined threshold $\eta$, i.e. $\norm*{\bm{Y}_{(4),\text{res.}}^{(i)}}_\mathsf{F}^2 \leq \eta$, or a maximum number of iterations $I_\text{OMP}$ is reached.

\section{Results and Discussion}
\label{sec:results-discussion}

In this section, we carry out numerical experiments to investigate the error performance of the proposed calibration technique and its impact on the imaging quality using both synthetic and real measurements recorded with an air-coupled ultrasonic array \cite{allevatoRealtime3DImaging2021}.

\subsection{Simulations}
\label{sec:simulations}

We first evaluate our proposed modified \gls{BCD} method for tensor-based array calibration with synthetic data from ${P=250}$ different calibration points, ${N=4}$ transmit and ${M=60}$ receive elements, ${L=24}$ \gls{DFT} bins, and ${T=10}$ snapshots. We model the transmit steering vector element $\elem{ \bm{a}_{p,\text{Tx}} }_n$, the receive steering vector element $\elem{ \bm{a}_{p,\text{Rx}} }_m$, the phase response element $\elem{ \bm{c}_{p} }_\ell$, and the pulse gain element
$\elem{ \bm{h}_{p} }_t$ as i.i.d.\@ random variables with unit-magnitude and uniform random phase in $[-\pi,\pi)$, for $p = 1, \dots, P$, $n = 1,\ldots, N$, $m = 1, \ldots, N$, $\ell = 1,\ldots, L$, and $t= 1,\ldots, T$. The transmit magnitude response element $\elem{\bm{g}_{n,\text{Tx}}}_\ell$ and the receive magnitude response element $\elem{\bm{g}_{m,\text{Rx}}}_\ell$ are uniformly drawn from $[1-\delta,1]$, with $0 \leq \delta < 1$ modeling variations among transmitters and receivers and $\delta = 0$ corresponding to identical frequency responses. Simulations at varying SNR levels are performed using the tensor factorization according to \cref{eq:array-output-model}, where we assume i.i.d.\@ zero-mean complex circular Gaussian noise.

We use the \gls{MCNCC} of each estimated range-angle dictionary entry in \cref{eq:range-angle-dict-entry} as a performance metric that is to a large extend independent of existing scaling ambiguities. 
In a Monte Carlo simulation with $I_{\rm MC}$ Monte Carlo trials, the \gls{MCNCC} is defined as
\begin{equation}
	\label{eq:corr-norm}
		\mathrm{MCNCC} = \frac{1}{I_{\rm MC}P} \sum_{i_{\rm MC}=1}^{I_{\rm MC}} \sum_{p=1}^{P} \left( 1 - \frac{\abs{ \bm{q}_{p,i_{\rm MC}}^\mathsf{H} \bm{\hat{q}}_{p,i_{\rm MC}} }}{ \norm{\bm{q}_{p,i_{\rm MC}}}_2 \, \norm{\bm{\hat{q}}_{p,i_{\rm MC}}} }_2 \right),
\end{equation}
where $\bm{q}_{p,i_{\rm MC}}$ and $\bm{\hat{q}}_{p,i_{\rm MC}}$ are the vectorized range-angle responses of the true parameter values $\{\bm{g}_{n, \text{Tx}}\}_{n=1}^N$, \linebreak 
$\{\bm{g}_{m, \text{Rx}}\}_{m=1}^M$, and $\{ \bm{a}_{p, \text{Tx}}, \, \bm{a}_{p, \text{Rx}}, \, \bm{c}_p \}_{p=1}^P$ and the corresponding finite iteration parameter estimates $\{ \bm{\hat{g}}_{n, \text{Tx}} \}_{n=1}^N$, $\{ \bm{\hat{g}}_{m, \text{Rx}} \}_{m=1}^M$, and $\{ \bm{\hat{a}}_{p, \text{Tx}}, \, \bm{\hat{a}}_{p, \text{Rx}}, \, \bm{\hat{c}}_p \}_{p=1}^P$ at the termination of the modified \gls{BCD} algorithm for $p = 1,\ldots, P$, $n = 1,\ldots, N$ and $m=1,\ldots, M$, respectively, according to \cref{eq:TensorQ,eq:range-angle-dict-entry}.  In terms of coherence \cite{donohoUncertaintyPrinciplesIdeal2001}, we remark that the \gls{MCNCC} is inversely proportional to the coherence between $\bm{q}_{p,i_{\rm MC}}$ and $\bm{\hat{q}}_{p,i_{\rm MC}}$, i.e., the \gls{MCNCC} is zero for full coherence and one for full incoherence.

In \Cref{fig:calibration-snr-simulation}, the \gls{MCNCC} metric is averaged over $I_{\rm MC}=100$ Monte Carlo trials after each decomposition reaches $f_\mathrm{rel}^{(i-1)} - f_\mathrm{rel}^{(i)} \leq 10^{-6}$, where
\begin{equation}
	f_\mathrm{rel}^{(i)} = \left( \sum\limits_{p=1}^{P} \norm{\bm{\mathcal{Y}}_p}_\mathsf{F}^2 \right)^{-1} \sum\limits_{p=1}^{P} \norm{\bm{\mathcal{Y}}_p - \bm{\hat{\mathcal{Y}}}_p^{(i)} }_\mathsf{F}^2
	\label{eq:norm-cost-fct}
\end{equation}
is the normalized cost function \cref{eq:costfct} at the $i$-th iteration of our proposed modified \gls{BCD} procedure. We compare the \gls{MCNCC} for $\delta \in \{ 0, 0.1, 0.5 \}$ to illustrate the influence of transmitters and receivers with increasingly different frequency responses. With our proposed method, the \gls{MCNCC} converges to $0$, i.e.\@ full coherence between $\bm{q}_{p}$ and $\bm{\hat{q}}_{p}$, with rising SNR, regardless of the distortion parameter $\delta$. For identical frequency responses, i.e.\@ $\delta=0$, the data can well be approximated with a rank-$1$ \gls{CPD} introduced in \Cref{sec:complexity}. For $\delta > 0$, however, i.e.\@ nonidentical frequency responses of individual transducers, a rank-$1$ \gls{CPD} is unsuitable as the \gls{MCNCC} saturates at a nonzero value depending on $\delta$.

\begin{figure}[t]
	\centering 
	\input{figures/calibration-simulation-mcncc-iter-200-p-250-mc-100.tex}
	\caption{Comparison between our proposed decomposition method and a rank-$1$ \gls{CPD} approximation, respectively, using the mean complementary normalized cross-correlation (\gls{MCNCC}) \cref{eq:corr-norm} of the vectorized range-angle response estimates $\bm{\hat{q}}_p$ and their respective true values $\bm{q}_p$ according to \cref{eq:range-angle-dict-entry}. Unlike the rank-$1$ \gls{CPD}, our proposed method is independent of the frequency response variation $\delta$.}
	\label{fig:calibration-snr-simulation}
\end{figure}

From our simulation results for the considered scenario, we further remark that we observe the parameter estimates approaching the true parameters in the noise-free case, which suggests that in this case, our modified \gls{BCD} algorithm converges to the global minimum of problem \cref{eq:scpd-original-problem}.

\subsection{Real data}
\label{sec:results-real-data}

For experimental validation, we use an ${8 \times 8}$ \gls{URA} with custom electronics based on a Xilinx Zynq 7010 FPGA \cite{allevatoRealtime3DImaging2021}. The URA features \glspl{PUT} of type MA40S4S manufactured by Murata (Nagaokakyo, Kyoto Prefecture, Japan). The MA40S4S operate at $f_0 = \SI{40}{kHz}$ and have an output power level of more than $\SI{120}{dB}$ at $\SI{0.3}{m}$, enabling imaging ranges of several meters. These transducers are widely available to both academic and commercial customers, however, their bulky dimensions prevent a classic \gls{lambda-half}-spacing between neighboring array elements. An individually designed waveguide has been proposed to overcome this limitation and is part of the considered \gls{URA} \cite{jagerAircoupled40KHZUltrasonic2017a}.

All \gls{URA} elements can operate in both transmit and receive mode, with fast switching for simultaneous pulse transmission and reception. However, to avoid ring-out effects, we use each transceiver exclusively as a transmitter or receiver. Specifically, the ${N=4}$ corner transducers sequentially transmit sinusoidal pulses of duration ${T_\text{burst} = \SI{1}{ms}}$ each. This configuration minimizes measurement time, limited by the number of pulses per calibration position.

For both calibration and imaging, we fire $T=10$ pulses from one transmitter before switching to the next, repeating this until all $N=4$ transmitters have emitted in total $NT=40$ bursts. The remaining ${M=60}$ sensor elements in the \gls{URA} operate in receive mode only. For each snapshot $t = 1, \ldots, T$, echoes are recorded at a sampling frequency $f_\text{s} = \SI{195}{kHz}$ and stored for $L=24$ sample frequencies. Unambiguous range detection within the bounds of the anechoic chamber is ensured by setting the pulse repetition rate to ${\Delta T_\text{burst} = \SI{150}{ms}}$, assuming a wave propagation velocity of $c=\SI{343}{ms^{-1}}$.

We select $P=1{,}909$ calibration points distributed on a spherical shell at a constant range of $r = \SI{2}{m}$, covering a \gls{fov} of $\pm \ang{60}$ in both azimuth and elevation. Near broadside, the \gls{fov} is sampled more densely due to a higher sensitivity of the steering vectors $\bm{a}_p$ at lower electrical angles $\arg (\bm{a}_p)$. The exact coordinates of each point $p$ are derived from the rotational axes of the measurement system, which efficiently move the array relative to the calibration target along a meander-shaped path \cite{rutschDuctAcousticsAircoupled2023}.

\begin{figure}[t]
	\centering 
	\hspace*{\fill}
	\begin{subfigure}{.35\textwidth}
		\centering
		\input{figures/calibration-approximation-2d-fov120-mimo_colorinv.tex}
		\caption{2D proj. of relat. pointwise reconstr. error}
		\label{fig:calibration-approximation-2d}
	\end{subfigure}
	\hspace*{\fill}
	\begin{subfigure}{.6\textwidth}
		\centering
		\input{figures/calibration-approximation-convergence-behavior.tex}
		\caption{Normal. cost funct. differ.}
		\label{fig:calibration-approximation-vs-iterations}
	\end{subfigure}
	\caption{Relative pointwise reconstruction error \cref{eq:resonstruction_error} and normalized cost function difference \cref{eq:norm-cost-fct} of the calibration technique proposed in \Cref{sec:methodology-calibration} in a constant range of ${r = \SI{2}{\meter}}$. For ${P=1{,}909}$ calibration points defined by their azimuth $\vartheta$ and elevation $\varphi$ w.r.t.\@ the alignment of the array elements, the reconstruction error $\zeta_\mathrm{rel}(p)$ is computed after $500$ iterations \subref{fig:calibration-approximation-2d}. To show the convergence behavior, the difference $f_\mathrm{rel}^{(i-1)}-f_\mathrm{rel}^{(i)}$ between consecutive normalized cost function values is evaluated  \subref{fig:calibration-approximation-vs-iterations}.}
	\label{fig:calibration-approximation}
\end{figure}

The reconstruction performance of our algorithm for the \gls{US} array measurements described above is shown in \Cref{fig:calibration-approximation}. In \Cref{fig:calibration-approximation}\subref{fig:calibration-approximation-2d}, the relative pointwise reconstruction error 
\begin{equation}
	\zeta_\mathrm{rel}(p) = \left( \norm{\bm{\mathcal{Y}}_p}_\mathsf{F} \right)^{-1} \norm{\bm{\mathcal{Y}}_p - \bm{\hat{\mathcal{Y}}}_p}_\mathsf{F}
	\label{eq:resonstruction_error}
\end{equation}
is displayed for various azimuth and elevation angles at the set range of $r = \SI{2}{m}$. The relative reconstruction error represents the model mismatch for any calibration position $p = 1, \ldots, P$ between the calibration measurement $\bm{\mathcal{Y}}_p$ and the reconstructed model $\bm{\hat{\mathcal{Y}}}_p$ according to \cref{eq:array-output-model} using the parameter estimates learned with our modified \gls{BCD} algorithm.
We observe that our proposed tensor model in \cref{eq:array-output-model} accurately models measured data for different target positions within the \gls{fov}, despite the narrow $\SI{3}{\decibel}$ mainlobe width of $\ang{12}$ \cite{allevatoRealtime3DImaging2021}. The relative pointwise reconstruction error according to \cref{eq:resonstruction_error} is $\SI{20}{\percent}$ near broadside and increases up to $\SI{50}{\percent}$ towards the edge of the \gls{fov}. The reconstruction errors may be due to measurement noise, model errors resulting from the violation of the single-point-source assumption, and reverberation. By construction, the \gls{BCD} method always provides a decrease of the objective function in every variable update. The convergence behavior of our proposed modified \gls{BCD} method is shown by examining the difference between the normalized cost function
$f_\mathrm{rel}^{(i)}$ according to \cref{eq:norm-cost-fct} at each iteration $i$ against $f_\mathrm{rel}^{(i-1)}$ of the previous iteration (\Cref{fig:calibration-approximation}\subref{fig:calibration-approximation-vs-iterations}).

\begin{figure}[t]
	\centering
	\includegraphics[height=0.175\textheight]{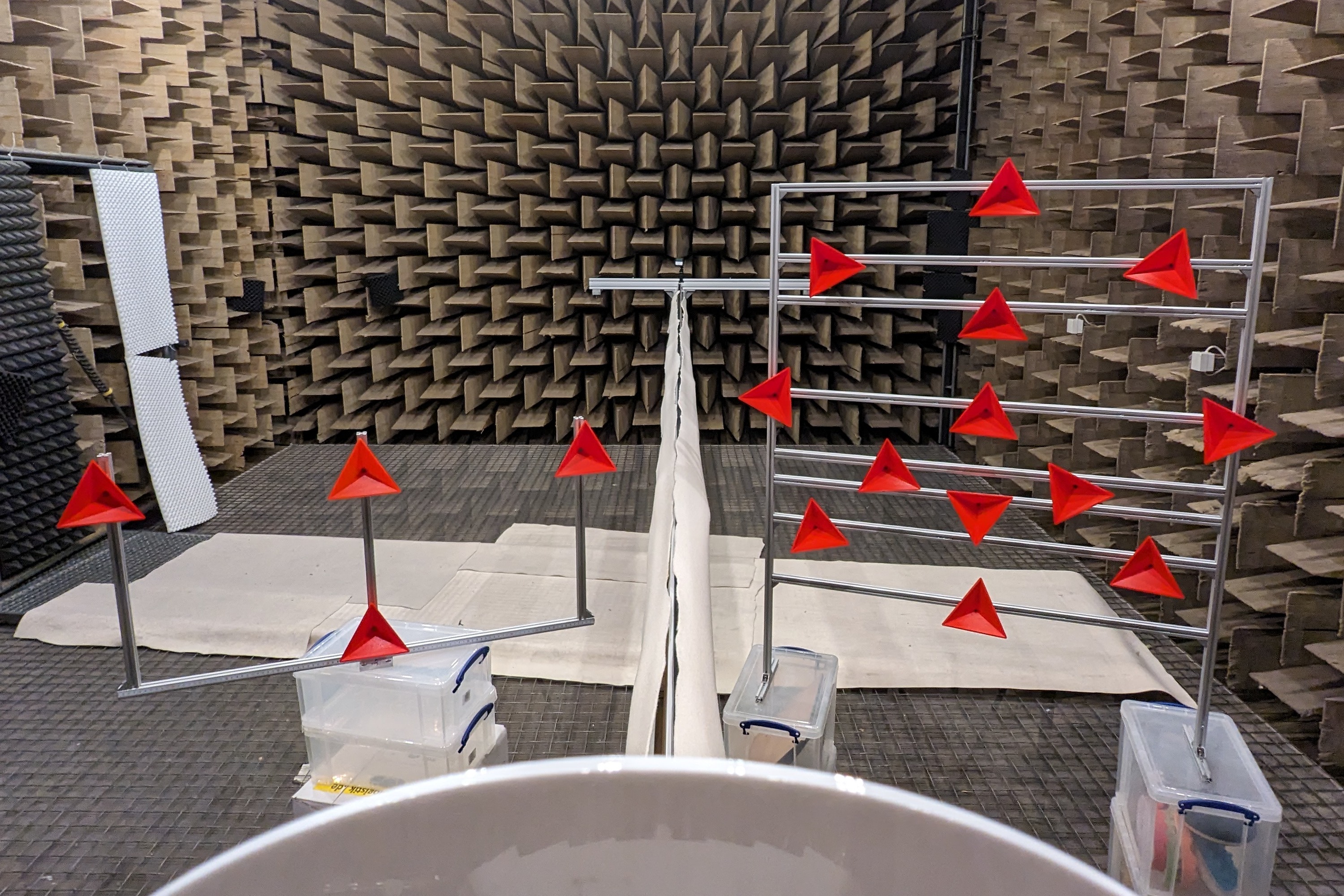}
	\caption{Imaging setup with multiple targets (red triangles) viewed from above and behind the rigid baffle.}
	\label{fig:imaging-setup}
\end{figure}

Finally, projections of scatter plots in Cartesian and spherical space are given as a proof-of-concept for the imaging capabilities of the proposed model and methods. As a test scenario, we arrange multiple tetrahedral corner reflectors of the same kind as used for calibration on two skew planes perpendicular to the $x$-$z$ plane (\Cref{fig:imaging-setup}). A dictionary of $P = R \cdot P_{\rm measured} = 15 \cdot 1{,}909 = 28{,}635$ range-angle response vectors $\{\bm{q}_p\}_{p=1}^P$ is learned from ${P_{\rm measured}=1{,}909}$ measured calibration positions with varying angles at the constant range ${r=\SI{2}{m}}$ and ${R=15}$ different scanning ranges $r'$ including the original calibration range. The scanning ranges are chosen as $r' = r + 0.2 \cdot k \cdot \Delta r $, $k=-7,\ldots,7$, where $\Delta r = 0.5 \cdot c \cdot T_\text{burst} \approx \SI{17}{cm}$ equals half the range resolution given by the Rayleigh bandwidth \cite{richardsFundamentalsradarSignal2022}. For	imaging, we apply the \gls{OMP} algorithm introduced in \Cref{sec:methodology-imaging} with $I_\text{OMP} = 100$ iterations. In each iteration, we select one range-angle response vector $\bm{q}_{\hat{p}}$ from the dictionary according to Step 1 in \cref{eq:imaging-omp-problem2}. The corresponding position estimate $\hat{p}$ is assigned the weighted norm square $T^{-2} \| \bm{\hat{h}}_{\hat{p}} \|_2^2$ according to Step 2 in \cref{eq:LS-OMP2}. A threshold is then applied to exclude position estimates $\hat{p}$ with a weighted power estimate $T^{-2} \| \bm{\hat{h}}_{\hat{p}} \|_2^2$ more than \SI{10}{dB} below the peak estimate. The remaining target estimates are color-coded according to their weighted power estimate.

A \gls{2D} image of the reconstructed model projected onto the $x$-$z$ plane is displayed in \Cref{fig:imaging-example}\subref{fig:imaging-example-3d-proposed} while \Cref{fig:imaging-example}\subref{fig:imaging-example-2d-proposed} shows an angular projection using azimuth and elevation. The latter is obtained by transforming Cartesian coordinates into spherical ones and collapsing the range dimension of the resulting tensor. A positive azimuth angle $\vartheta$ is measured from the $z$-axis to the $x$-axis whereas a positive elevation angle $\varphi$ is measured from the $z$-axis to the $y$-axis. Signal gains associated to targets appearing under the same direction, i.e.\@ same azimuth and elevation but different scanning ranges, are added up.
For comparison, we provide the \gls{2D} projection based on an analytic dictionary that requires exact array geometry knowledge and is used without and with the broadside compensation from \Cref{sec:complexity} (c.f.\@ \Cref{fig:imaging-example}\subref{fig:imaging-example-2d-analytical-no-comp} and \subref{fig:imaging-example-2d-analytical}, respectively).

\begin{figure}[t]
	\centering
	\begin{subfigure}{.45\textwidth}
		\centering
		\input{figures/20240305_imaging_4pleft_peaceright_omp100_thresh10_cart_proposed_normalized_colorinv.tex}
		\vspace{-2.25em}
		\caption{$x$-$z$ projection (proposed)}
		\label{fig:imaging-example-3d-proposed}
	\end{subfigure}
	\hspace*{\fill}
	\begin{subfigure}{0.45\textwidth}
		\centering
		\input{figures/20240305_imaging_4pleft_peaceright_omp100_thresh10_spher_proposed_normalized_colorinv.tex}
		\vspace{-2.25em}
		\caption{Angular proj.\@ (proposed)}
		\label{fig:imaging-example-2d-proposed}
	\end{subfigure} \\
	\vspace{-1em}
	\begin{subfigure}{.45\textwidth}
		\centering
		\input{figures/20240305_imaging_4pleft_peaceright_omp100_thresh10_spher_analytical_nocomp_normalized_colorinv.tex}
		\vspace{-2.25em}
		\caption{Angular proj.\@ (analytic without compensation)}
		\label{fig:imaging-example-2d-analytical-no-comp}
	\end{subfigure}
	\hspace*{\fill}
	\begin{subfigure}{0.45\textwidth}
		\centering
		\input{figures/20240305_imaging_4pleft_peaceright_omp100_thresh10_spher_analytical_normalized_colorinv.tex}
		\vspace{-2.25em}
		\caption{Angular proj.\@ (anal.\@ with compens.)}
		\label{fig:imaging-example-2d-analytical}
	\end{subfigure}
	\caption{From measurements with multiple reflectors (red), scatter plots projected onto the $x$-$z$ plane \subref{fig:imaging-example-3d-proposed} and the azimuth-elevation-plane \subref{fig:imaging-example-2d-proposed}, respectively, are computed using the dictionary of our proposed calibration method. For comparison, the angular projections using an analytic dictionary without \subref{fig:imaging-example-2d-analytical-no-comp} and with \subref{fig:imaging-example-2d-analytical} broadside compensation are also provided.}
	\label{fig:imaging-example}
\end{figure}

Our main observations from \Cref{fig:imaging-example} are threefold:
\begin{enumerate}[label={(\arabic*)}]
	\item In our proof-of-concept experiment with real data, \Cref{fig:imaging-example}\subref{fig:imaging-example-2d-proposed} shows that we achieve a good imaging performance because the proposed dictionary of measured array responses, i.e.\@ without explicit knowledge of the array geometry, is capable of resolving the multi-target scenario.
	\item \Cref{fig:imaging-example}\subref{fig:imaging-example-3d-proposed} shows that the principal target location estimates $\hat{p}$ corresponding to larger signal gain estimates $\bm{\hat{h}}_{\hat{p}}$ are distributed along two planes perpendicular to the $x$-$z$-plane according to the ground truth setup indicated by the red boxes. This proves the correctness of our assumed linear phase response model in \cref{eq:phase-response-approximation}: target ranges can be estimated even if they have not been learned during calibration.
	\item Without and with broadside compensation (c.f.\@ \Cref{fig:imaging-example}\subref{fig:imaging-example-2d-analytical-no-comp} and \subref{fig:imaging-example-2d-analytical}, respectively), the analytic model, which requires precise knowledge of the array geometry, yields similar imaging performance as obtained with the array responses learned from calibration measurements.
\end{enumerate}

\section{Conclusions and Outlook}
\label{sec:conclusions}

In this article, we propose a novel tensor model to describe the measurement data of an air-coupled ultrasonic \gls{MIMO} sensor array for \gls{3D} imaging. A reference-based calibration scheme is suggested in which the tensor model parameters are jointly estimated from all calibration measurements using a modified \gls{BCD} method with proven convergence. The modification exploits scaling invariances between the block variables to convert the originally nonconvex subproblems into equivalent subproblems that admit closed-form solutions. Numerical experiments carried out with synthetic data as well as real measurements recorded with an air-coupled ultrasonic array confirm both, the validity of the model as well as the functionality of the calibration procedure. Additionally, an imaging example with real data verifies that dictionaries with the commonly used, analytically modeled steering and phase response vectors as well as the angular-range response vectors obtained with our calibration procedure are feasible for \gls{3D} imaging with air-coupled ultrasonic sensor arrays. It is assumed that in air-coupled \gls{US} as well as other array processing applications such as radar or sonar, array elements with magnitude responses more distinct than those of our prototype array will benefit from our proposed model as suggested by our simulations.

In the future, we will investigate if more structure enforced onto the tensor signal model and therefore the optimization problem itself, e.g.\@ by exploiting a parametric range model during calibration, will improve the calibration results and imaging performance. Future work also addresses performance bounds of the proposed tensor model and research towards non-reference-based, i.e.\@ blind, calibration methods that perform joint calibration and imaging without the overhead associated with calibration measurements.

\bibliographystyle{elsarticle-num}
\bibliography{references.bib}

\end{CJK*}
\end{document}

%% file: figures/calibration-simulation-mcncc-iter-200-p-250-mc-100.tex
\definecolor{mycolor1}{rgb}{0.00000,0.44531,0.69531}%
\definecolor{mycolor2}{rgb}{0.83203,0.36719,0.00000}%
\begin{tikzpicture}

\begin{axis}[%
width=.65\textwidth,
height=.2\textheight,
at={(0.758in,0.481in)},
scale only axis,
xmin=-20,
xmax=20,
xlabel style={font=\color{white!15!black}},
xlabel={SNR (dB)},
ymode=log,
ymin=1e-07,
ymax=1,
yminorticks=true,
ylabel style={font=\color{white!15!black}},
ylabel={MCNCC},
axis background/.style={fill=white},
xmajorgrids,
ymajorgrids,
yminorgrids,
legend style={at={(0.01,0.02)}, anchor=south west, legend cell align=left, align=left, draw=white!15!black, font=\scriptsize}
]

\addplot [color=mycolor1, dotted, line width=1.5pt, mark=square, mark size=2pt, mark options={solid, mycolor1}, mark phase = 1, mark repeat={3}]
  table[row sep=crcr]{%
-20	0.182054566250665\\
-17.5	0.0840762583128661\\
-15	0.0407626695514663\\
-12.5	0.0169288020166225\\
-10	0.00819428696403325\\
-7.5	0.00371382136078233\\
-5	0.00158132340622362\\
-2.5	0.000591083445468252\\
0	0.000464202070205404\\
2.5	0.000186835138169932\\
5	0.000105301601905623\\
7.5	5.92554641493766e-05\\
10	3.34221432931456e-05\\
12.5	1.88072566643949e-05\\
15	1.06000845355959e-05\\
17.5	5.98650473711874e-06\\
20	3.38352959925955e-06\\
22.5	1.91914448227145e-06\\
25	1.08785505620575e-06\\
27.5	6.20064392290113e-07\\
30	3.52605318102632e-07\\
};
\addlegendentry{Proposed, $\delta = 0$}

\addplot [color=mycolor1, dotted, line width=1.5pt, mark=x, mark size=3pt, mark options={solid, mycolor1}, mark phase = 2, mark repeat={3}]
  table[row sep=crcr]{%
-20	0.170015752513758\\
-17.5	0.079198687601009\\
-15	0.0393011965016128\\
-12.5	0.0153632503796767\\
-10	0.00727187591514305\\
-7.5	0.00319341877638616\\
-5	0.00131839564934437\\
-2.5	0.000727231158955329\\
0	0.000334739908700804\\
2.5	0.000189075362808327\\
5	0.000107026471983613\\
7.5	6.05518574453682e-05\\
10	3.42492275317938e-05\\
12.5	1.96331225240863e-05\\
15	1.11993494652909e-05\\
17.5	6.57574468488861e-06\\
20	3.76553011390757e-06\\
22.5	2.24891081279336e-06\\
25	1.42920026171072e-06\\
27.5	9.05068342417866e-07\\
30	5.68458192555132e-07\\
};
\addlegendentry{Proposed, $\delta = 0.1$}

\addplot [color=mycolor1, dotted, line width=1.5pt, mark=diamond, mark size=2.75pt, mark options={solid, mycolor1}, mark phase = 3, mark repeat={3}]
  table[row sep=crcr]{%
-20	0.11764884146861\\
-17.5	0.046701663202742\\
-15	0.0217464692074844\\
-12.5	0.00870249904959172\\
-10	0.0033512334646954\\
-7.5	0.00188661098813062\\
-5	0.00106574961518206\\
-2.5	0.000603010354646975\\
0	0.000339242246852142\\
2.5	0.000193265374043868\\
5	0.000109908366771465\\
7.5	6.27703493602795e-05\\
10	3.59145337345262e-05\\
12.5	2.07933727577609e-05\\
15	1.20814749248236e-05\\
17.5	7.11191274865368e-06\\
20	4.25142319685129e-06\\
22.5	2.57169780535058e-06\\
25	1.58244896429585e-06\\
27.5	1.00210781144816e-06\\
30	6.41999948062599e-07\\
};
\addlegendentry{Proposed, $\delta = 0.5$}

\addplot [color=mycolor2, dotted, line width=1.5pt, mark=square, mark size=2pt, mark options={solid, mycolor2}, mark phase = 3, mark repeat={3}]
  table[row sep=crcr]{%
-20	0.126695336490162\\
-17.5	0.0474787906202947\\
-15	0.0193992145566083\\
-12.5	0.00907620392104662\\
-10	0.00395106044501838\\
-7.5	0.00207065091183687\\
-5	0.00116582195897978\\
-2.5	0.000656709273435742\\
0	0.000368035628731013\\
2.5	0.000207221195201831\\
5	0.000116672379874353\\
7.5	6.56397460292193e-05\\
10	3.68909873654504e-05\\
12.5	2.07451946985327e-05\\
15	1.16585502055503e-05\\
17.5	6.57082798606866e-06\\
20	3.68673342454774e-06\\
22.5	2.08045381067623e-06\\
25	1.16562242183359e-06\\
27.5	6.56518940926754e-07\\
30	3.6891733860848e-07\\
};
\addlegendentry{Rank-1 CPD, $\delta = 0$}

\addplot [color=mycolor2, dotted, line width=1.5pt, mark=x, mark size=3pt, mark options={solid, mycolor2}, mark phase = 3, mark repeat={3}]
  table[row sep=crcr]{%
-20	0.128881696434926\\
-17.5	0.0507953184492824\\
-15	0.0191969078692841\\
-12.5	0.0090976892368096\\
-10	0.00437019929772778\\
-7.5	0.00276101988729254\\
-5	0.00183578057410721\\
-2.5	0.00132576580059654\\
0	0.00105745587599104\\
2.5	0.000899987021093998\\
5	0.000785898679300359\\
7.5	0.000756965570722679\\
10	0.000715836002821297\\
12.5	0.000696748783909563\\
15	0.000676375130092465\\
17.5	0.000686828337484054\\
20	0.00068059550102552\\
22.5	0.000666817349139749\\
25	0.000687881699803563\\
27.5	0.000677157763414051\\
30	0.000683811198988038\\
};
\addlegendentry{Rank-1 CPD, $\delta = 0.1$}

\addplot [color=mycolor2, dotted, line width=1.5pt, mark=diamond, mark size=2.75pt, mark options={solid, mycolor2}, mark phase = 2, mark repeat={3}]
  table[row sep=crcr]{%
-20	0.145353170486443\\
-17.5	0.0703073357768494\\
-15	0.043130259835505\\
-12.5	0.0343228469253593\\
-10	0.0293737597080755\\
-7.5	0.0279469091758211\\
-5	0.0266430324714825\\
-2.5	0.0258378634776729\\
0	0.0255839433161797\\
2.5	0.0256264062615665\\
5	0.0256685540157415\\
7.5	0.025538351583469\\
10	0.0255613661596579\\
12.5	0.0255470458102308\\
15	0.025258205595593\\
17.5	0.0252674869743657\\
20	0.0254072590825309\\
22.5	0.0253035251735525\\
25	0.025553681341998\\
27.5	0.0257178872702104\\
30	0.0248294129097212\\
};
\addlegendentry{Rank-1 CPD, $\delta = 0.5$}

\end{axis}
\end{tikzpicture}%

%% file: figures/calibration-approximation-convergence-behavior.tex
\definecolor{mycolor1}{rgb}{0.00000,0.44700,0.74100}%
\begin{tikzpicture}

\begin{axis}[%
width = .6\linewidth,
height=.175\textheight,
at={(0.758in,0.481in)},
scale only axis,
xmin=1,
xmax=500,
xlabel style={font=\color{white!15!black}},
xlabel={Iteration $i$},
ymode=log,
ymin=1e-07,
ymax=0.1,
yminorticks=true,
ylabel style={font=\color{white!15!black}},
ylabel={$f_\mathrm{rel}^{(i-1)} - f_\mathrm{rel}^{(i)}$},
axis background/.style={fill=white},
xmajorgrids,
ymajorgrids,
yminorgrids,
legend style={legend cell align=left, align=left, draw=white!15!black}
]
\addplot [color=mycolor1, line width=1.5pt]
  table[row sep=crcr]{%
1	0.0357505144514515\\
2	0.0629104741778892\\
3	0.0776794008057442\\
4	0.0230222021348579\\
5	0.00500865533471662\\
6	0.00190086939667011\\
7	0.000882230389081307\\
8	0.00049196915220806\\
9	0.000269973779961852\\
10	0.000217359017998797\\
11	0.000148362549793335\\
12	0.000113949166191563\\
13	0.000113598312205698\\
14	8.28398391357066e-05\\
15	4.84705803951435e-05\\
16	3.52884727681246e-05\\
17	3.36120472416157e-05\\
18	5.2389526752572e-05\\
19	9.40426442018749e-05\\
20	2.40227931490589e-05\\
21	1.66948811467504e-05\\
22	2.12662631886573e-05\\
23	1.24533163067264e-05\\
24	7.15664334682842e-06\\
25	6.26386210123009e-06\\
26	5.94313886115305e-06\\
27	5.83503377327155e-06\\
28	6.0300300378513e-06\\
29	6.90217153997974e-06\\
30	9.91726627785905e-06\\
31	2.01285971047627e-05\\
32	4.1095278468406e-05\\
33	4.25150684065789e-05\\
34	1.00831452306549e-05\\
35	4.74930004856766e-06\\
36	4.44776193886387e-06\\
37	4.38155660764661e-06\\
38	4.34556853545764e-06\\
39	4.33320480708321e-06\\
40	4.34355032419287e-06\\
41	4.35362037543019e-06\\
42	4.31903479214402e-06\\
43	4.2407089295704e-06\\
44	4.18373366817271e-06\\
45	4.21522161020427e-06\\
46	4.41280401064503e-06\\
47	4.95636415343004e-06\\
48	6.24137753768572e-06\\
49	8.61689946662914e-06\\
50	1.07703545638715e-05\\
51	1.02182238563842e-05\\
52	8.49164836647098e-06\\
53	8.61100708360318e-06\\
54	1.06918668449918e-05\\
55	1.16555514541039e-05\\
56	7.8365918557699e-06\\
57	4.60289786108348e-06\\
58	3.64234571226252e-06\\
59	3.40098391160559e-06\\
60	3.31976144840596e-06\\
61	3.27395478805226e-06\\
62	3.23729533324713e-06\\
63	3.20413791699004e-06\\
64	3.17337908251059e-06\\
65	3.14496758635485e-06\\
66	3.1190163275463e-06\\
67	3.09611169868607e-06\\
68	3.07734156834716e-06\\
69	3.06447181908087e-06\\
70	3.06092282087445e-06\\
71	3.07354360329359e-06\\
72	3.11709819955652e-06\\
73	3.22619975601235e-06\\
74	3.4887312966303e-06\\
75	4.13791328401913e-06\\
76	5.74386159513107e-06\\
77	9.21946449383704e-06\\
78	1.4343203044942e-05\\
79	1.5958540372174e-05\\
80	9.74131900821629e-06\\
81	4.64529039612138e-06\\
82	3.3257043136059e-06\\
83	3.13291199161014e-06\\
84	3.12853021133819e-06\\
85	3.15443332088439e-06\\
86	3.19310401830641e-06\\
87	3.24835443177296e-06\\
88	3.32895121812893e-06\\
89	3.46511772286728e-06\\
90	3.7768930146731e-06\\
91	4.75701412339102e-06\\
92	8.92673236696506e-06\\
93	2.95436474713195e-05\\
94	5.23742069895583e-05\\
95	1.49173312976592e-05\\
96	4.4968301172732e-06\\
97	6.5674473477273e-06\\
98	2.47601741920622e-05\\
99	2.12348895678449e-05\\
100	3.99758724323185e-06\\
101	2.35990016828058e-06\\
102	2.24366540357845e-06\\
103	2.19680048652027e-06\\
104	2.15583735319047e-06\\
105	2.11700547114724e-06\\
106	2.07976870925108e-06\\
107	2.04392419844712e-06\\
108	2.00934746363846e-06\\
109	1.97594713791105e-06\\
110	1.94365021244369e-06\\
111	1.91239529756526e-06\\
112	1.88212903118323e-06\\
113	1.85280395248455e-06\\
114	1.82437718354578e-06\\
115	1.79680977974161e-06\\
116	1.77022131520044e-06\\
117	1.74441577927009e-06\\
118	1.71936323078992e-06\\
119	1.69503370950341e-06\\
120	1.67139935325356e-06\\
121	1.64843460159769e-06\\
122	1.62611572973592e-06\\
123	1.60442060159749e-06\\
124	1.58332855559884e-06\\
125	1.56282036345434e-06\\
126	1.54287825648858e-06\\
127	1.52348601401009e-06\\
128	1.50462912762439e-06\\
129	1.48629506935283e-06\\
130	1.46847369653091e-06\\
131	1.45115786231997e-06\\
132	1.43434433752532e-06\\
133	1.41803519893013e-06\\
134	1.4022399569269e-06\\
135	1.3869788805243e-06\\
136	1.37228827146174e-06\\
137	1.35822905522609e-06\\
138	1.34490112124741e-06\\
139	1.3324679738469e-06\\
140	1.3212004962293e-06\\
141	1.31035763351051e-06\\
142	1.30235960071889e-06\\
143	1.29934958947864e-06\\
144	1.3036340971162e-06\\
145	1.32040723721527e-06\\
146	1.36097289826775e-06\\
147	1.45059200440389e-06\\
148	1.64724905660218e-06\\
149	2.08116514377465e-06\\
150	3.00306091660651e-06\\
151	4.72412638896635e-06\\
152	7.22902776206791e-06\\
153	8.68246642027337e-06\\
154	5.93558807326922e-06\\
155	2.59679879843766e-06\\
156	1.43473220337587e-06\\
157	1.19623752503628e-06\\
158	1.14518385330165e-06\\
159	1.12931823914586e-06\\
160	1.11996459140062e-06\\
161	1.11186649509598e-06\\
162	1.10409243825327e-06\\
163	1.09647378387923e-06\\
164	1.08897520045836e-06\\
165	1.08158551881399e-06\\
166	1.07429863072106e-06\\
167	1.06710994118853e-06\\
168	1.06001558597413e-06\\
169	1.05301219444076e-06\\
170	1.04609675766199e-06\\
171	1.03926656080944e-06\\
172	1.03256524286444e-06\\
173	1.02596239259789e-06\\
174	1.01943451791264e-06\\
175	1.01298042931575e-06\\
176	1.00659818302873e-06\\
177	1.00028590643841e-06\\
178	9.94041923219058e-07\\
179	9.87864711476938e-07\\
180	9.81752863449259e-07\\
181	9.75705055750176e-07\\
182	9.69720038490607e-07\\
183	9.63796618069779e-07\\
184	9.57933656287047e-07\\
185	9.5213005857353e-07\\
186	9.46384772548825e-07\\
187	9.40696785911577e-07\\
188	9.35065116003386e-07\\
189	9.29488817913438e-07\\
190	9.23966971155821e-07\\
191	9.18498687330072e-07\\
192	9.13083101239387e-07\\
193	9.07719372778004e-07\\
194	9.02406685598933e-07\\
195	8.97144242006931e-07\\
196	8.91931271063129e-07\\
197	8.86767015151335e-07\\
198	8.81650740192086e-07\\
199	8.76581727982106e-07\\
200	8.71559276416356e-07\\
201	8.665827068155e-07\\
202	8.61651347161541e-07\\
203	8.56764547751965e-07\\
204	8.51921672762046e-07\\
205	8.47122098801556e-07\\
206	8.42365220243835e-07\\
207	8.37650443119564e-07\\
208	8.32977186115968e-07\\
209	8.28344881909082e-07\\
210	8.23752979051129e-07\\
211	8.1920093275567e-07\\
212	8.14688214001436e-07\\
213	8.10214307311874e-07\\
214	8.05778702872573e-07\\
215	8.01380907966553e-07\\
216	7.97020438758622e-07\\
217	7.92696821516614e-07\\
218	7.88409594276729e-07\\
219	7.8415830684353e-07\\
220	7.79942515460874e-07\\
221	7.75761789251206e-07\\
222	7.7161570699591e-07\\
223	7.67503856469176e-07\\
224	7.63425835548226e-07\\
225	7.59381250547975e-07\\
226	7.55369718108412e-07\\
227	7.51390861974954e-07\\
228	7.47444318327517e-07\\
229	7.43529726787706e-07\\
230	7.39646741965139e-07\\
231	7.35826903852654e-07\\
232	7.32052381158255e-07\\
233	7.28306506325715e-07\\
234	7.24589446998714e-07\\
235	7.20900784956235e-07\\
236	7.17240121961282e-07\\
237	7.13607112734493e-07\\
238	7.10001452741693e-07\\
239	7.0642287142153e-07\\
240	7.02871115643156e-07\\
241	6.9934595114951e-07\\
242	6.95847153120432e-07\\
243	6.92374509170257e-07\\
244	6.88927813352613e-07\\
245	6.85506864384067e-07\\
246	6.82111465866164e-07\\
247	6.78741425064189e-07\\
248	6.75396546467866e-07\\
249	6.72076638785768e-07\\
250	6.68781503843086e-07\\
251	6.65510944353187e-07\\
252	6.62264757589348e-07\\
253	6.5904273627293e-07\\
254	6.55844666574978e-07\\
255	6.52670328893379e-07\\
256	6.49519494300144e-07\\
257	6.4639192831617e-07\\
258	6.43287387247504e-07\\
259	6.40205621071921e-07\\
260	6.37146370774389e-07\\
261	6.3410937467534e-07\\
262	6.31094363878759e-07\\
263	6.28101066268982e-07\\
264	6.25129210840569e-07\\
265	6.22178525810924e-07\\
266	6.19248744615497e-07\\
267	6.16339605574723e-07\\
268	6.13450858222286e-07\\
269	6.10582261417747e-07\\
270	6.0773358590005e-07\\
271	6.04904620948865e-07\\
272	6.0209516439258e-07\\
273	5.99305036264042e-07\\
274	5.96534068031396e-07\\
275	5.93782104374441e-07\\
276	5.9104900418383e-07\\
277	5.88334632567467e-07\\
278	5.85638862404814e-07\\
279	5.82961569572937e-07\\
280	5.80302630392993e-07\\
281	5.77661914635819e-07\\
282	5.75039288741586e-07\\
283	5.7243460693801e-07\\
284	5.69847714460003e-07\\
285	5.67278437335617e-07\\
286	5.64726591600895e-07\\
287	5.62191970088222e-07\\
288	5.59674357525353e-07\\
289	5.57173508775044e-07\\
290	5.54689173259959e-07\\
291	5.52221080751814e-07\\
292	5.49768945035112e-07\\
293	5.47332468014972e-07\\
294	5.44911344491084e-07\\
295	5.42505262157711e-07\\
296	5.40113904490269e-07\\
297	5.37736959627111e-07\\
298	5.35374121146681e-07\\
299	5.33025095617035e-07\\
300	5.30689608924106e-07\\
301	5.28367409158292e-07\\
302	5.26058274163965e-07\\
303	5.2376201709059e-07\\
304	5.21478486836813e-07\\
305	5.19207573934644e-07\\
306	5.16949208662076e-07\\
307	5.14711480548158e-07\\
308	5.12519971351821e-07\\
309	5.10340030235312e-07\\
310	5.0817143293358e-07\\
311	5.06014013579303e-07\\
312	5.03867751078246e-07\\
313	5.01732731250648e-07\\
314	4.99609114634758e-07\\
315	4.97497103180145e-07\\
316	4.95396931809999e-07\\
317	4.9330886320309e-07\\
318	4.91233184352069e-07\\
319	4.89170213113788e-07\\
320	4.8712030842335e-07\\
321	4.85083877177495e-07\\
322	4.83061386669092e-07\\
323	4.81053382461738e-07\\
324	4.79060508595808e-07\\
325	4.77083529903943e-07\\
326	4.75123357324136e-07\\
327	4.73181089866159e-07\\
328	4.71258051026879e-07\\
329	4.69355844301411e-07\\
330	4.6747641702094e-07\\
331	4.65622142176159e-07\\
332	4.63795913008269e-07\\
333	4.62001271239743e-07\\
334	4.60242546962419e-07\\
335	4.58525041269198e-07\\
336	4.56855239638898e-07\\
337	4.55241048080701e-07\\
338	4.5369206036483e-07\\
339	4.52219806157395e-07\\
340	4.50837949639293e-07\\
341	4.49562314375207e-07\\
342	4.48410641729069e-07\\
343	4.47401942160752e-07\\
344	4.4655558828488e-07\\
345	4.45890873335131e-07\\
346	4.45429038098411e-07\\
347	4.45201395193884e-07\\
348	4.45267724247245e-07\\
349	4.45746946575731e-07\\
350	4.46857350788754e-07\\
351	4.48962805710273e-07\\
352	4.52631890968114e-07\\
353	4.58738740771736e-07\\
354	4.6866162650705e-07\\
355	4.84674749778868e-07\\
356	5.1071671502978e-07\\
357	5.53940059377922e-07\\
358	6.28018990478374e-07\\
359	7.60859146908821e-07\\
360	1.01539552643803e-06\\
361	1.56227389458419e-06\\
362	3.05060619543163e-06\\
363	8.47325720998438e-06\\
364	1.69879564576592e-05\\
365	1.07783090618208e-05\\
366	2.80749976144801e-06\\
367	8.85122114158143e-07\\
368	5.25476720025786e-07\\
369	4.36982296347921e-07\\
370	4.13857738301715e-07\\
371	4.06869452906911e-07\\
372	4.03762718925726e-07\\
373	4.01611306766192e-07\\
374	3.99723354527204e-07\\
375	3.97931370521576e-07\\
376	3.96190914186967e-07\\
377	3.9448719935109e-07\\
378	3.92813275573722e-07\\
379	3.91164764246277e-07\\
380	3.89538464262706e-07\\
381	3.87931907930295e-07\\
382	3.86343146474566e-07\\
383	3.84770622363639e-07\\
384	3.83213087373768e-07\\
385	3.81669526761108e-07\\
386	3.80139113520528e-07\\
387	3.78621170527005e-07\\
388	3.77115134786443e-07\\
389	3.75620545334243e-07\\
390	3.74137010039632e-07\\
391	3.72664209824514e-07\\
392	3.71201869131532e-07\\
393	3.69749768025507e-07\\
394	3.68307717657501e-07\\
395	3.66875564594693e-07\\
396	3.65453188821974e-07\\
397	3.64040497302653e-07\\
398	3.62637420314726e-07\\
399	3.61243916335852e-07\\
400	3.59859965048948e-07\\
401	3.58485575446821e-07\\
402	3.5712077495198e-07\\
403	3.55765623738513e-07\\
404	3.54420204295991e-07\\
405	3.53084631976586e-07\\
406	3.51759053773826e-07\\
407	3.50443651986332e-07\\
408	3.49138650324043e-07\\
409	3.47844313242085e-07\\
410	3.46560959929576e-07\\
411	3.45288962200208e-07\\
412	3.44028755816517e-07\\
413	3.42780849149626e-07\\
414	3.41545831616941e-07\\
415	3.40324380898593e-07\\
416	3.39117282366352e-07\\
417	3.37925430748953e-07\\
418	3.36749849116913e-07\\
419	3.35591700761917e-07\\
420	3.34452297745536e-07\\
421	3.33333111335321e-07\\
422	3.32235775113432e-07\\
423	3.31162085531744e-07\\
424	3.30113980706592e-07\\
425	3.29093517104084e-07\\
426	3.28102814028952e-07\\
427	3.27143959255594e-07\\
428	3.26218897006569e-07\\
429	3.25329229222682e-07\\
430	3.24475980306715e-07\\
431	3.23659273382404e-07\\
432	3.22877925951204e-07\\
433	3.22128990815074e-07\\
434	3.21407238601523e-07\\
435	3.20704662160054e-07\\
436	3.20010069665422e-07\\
437	3.19308895546655e-07\\
438	3.18583369351977e-07\\
439	3.17813177885817e-07\\
440	3.16976737968488e-07\\
441	3.16053058169175e-07\\
442	3.15024061281477e-07\\
443	3.1387704713115e-07\\
444	3.12606848185126e-07\\
445	3.1121717514182e-07\\
446	3.09720790681034e-07\\
447	3.08138342397513e-07\\
448	3.06496038104953e-07\\
449	3.04822608709898e-07\\
450	3.0314616594751e-07\\
451	3.01491517640251e-07\\
452	2.9987833605194e-07\\
453	2.98320325442525e-07\\
454	2.96825303558457e-07\\
455	2.9539594148531e-07\\
456	2.94030881420326e-07\\
457	2.92725964801122e-07\\
458	2.91475353408899e-07\\
459	2.90272466951791e-07\\
460	2.89110657525349e-07\\
461	2.87983673574743e-07\\
462	2.86885907807566e-07\\
463	2.85812517542006e-07\\
464	2.84759429702852e-07\\
465	2.83723299854266e-07\\
466	2.82701422826825e-07\\
467	2.81691646342175e-07\\
468	2.8069227142602e-07\\
469	2.79701968364243e-07\\
470	2.78719704427388e-07\\
471	2.77744676480118e-07\\
472	2.76776265240031e-07\\
473	2.75813987427043e-07\\
474	2.74857467896794e-07\\
475	2.73906413439384e-07\\
476	2.72960593461491e-07\\
477	2.72019820002356e-07\\
478	2.71083946623563e-07\\
479	2.70152851089556e-07\\
480	2.69226432592085e-07\\
481	2.68304607087266e-07\\
482	2.67387302854694e-07\\
483	2.66474456944721e-07\\
484	2.65566015178464e-07\\
485	2.64663273852328e-07\\
486	2.63765816632677e-07\\
487	2.62871837297673e-07\\
488	2.61981761062735e-07\\
489	2.61095722153826e-07\\
490	2.6021377286245e-07\\
491	2.59335932284444e-07\\
492	2.58462192648246e-07\\
493	2.57592541519358e-07\\
494	2.56726960579101e-07\\
495	2.55865430176527e-07\\
496	2.5500792877331e-07\\
497	2.54154438827925e-07\\
498	2.53304939579202e-07\\
499	2.52459415706063e-07\\
};

\end{axis}

\end{tikzpicture}%

%% file: figures/20240305_imaging_4pleft_peaceright_omp100_thresh10_cart_proposed_normalized_colorinv.tex
\definecolor{targetcolor}{rgb}{0.83203,0.36719,0.00000}%

\begin{tikzpicture}

\begin{axis}[%
height=.14\textheight,
at={(0.679in,0.638in)},
scale only axis,
point meta min=-28,
point meta max=-14,
colormap={mymap}{[1pt] rgb(0pt)=(0.993248,0.906157,0.143936); rgb(1pt)=(0.983868,0.904867,0.136897); rgb(2pt)=(0.974417,0.90359,0.130215); rgb(3pt)=(0.964894,0.902323,0.123941); rgb(4pt)=(0.9553,0.901065,0.118128); rgb(5pt)=(0.945636,0.899815,0.112838); rgb(6pt)=(0.935904,0.89857,0.108131); rgb(7pt)=(0.926106,0.89733,0.104071); rgb(8pt)=(0.916242,0.896091,0.100717); rgb(9pt)=(0.906311,0.894855,0.098125); rgb(10pt)=(0.89632,0.893616,0.0963354); rgb(11pt)=(0.886271,0.892374,0.0953744); rgb(12pt)=(0.876168,0.891125,0.0952505); rgb(13pt)=(0.866013,0.889868,0.0959528); rgb(14pt)=(0.85581,0.888601,0.0974519); rgb(15pt)=(0.845561,0.887322,0.0997022); rgb(16pt)=(0.83527,0.886029,0.102646); rgb(17pt)=(0.82494,0.88472,0.106217); rgb(18pt)=(0.814576,0.883393,0.110347); rgb(19pt)=(0.804182,0.882046,0.114965); rgb(20pt)=(0.79376,0.880678,0.120005); rgb(21pt)=(0.783315,0.879285,0.125405); rgb(22pt)=(0.772852,0.877868,0.131109); rgb(23pt)=(0.762373,0.876424,0.137064); rgb(24pt)=(0.751884,0.874951,0.143228); rgb(25pt)=(0.741388,0.873449,0.149561); rgb(26pt)=(0.730889,0.871916,0.156029); rgb(27pt)=(0.720391,0.87035,0.162603); rgb(28pt)=(0.709898,0.868751,0.169257); rgb(29pt)=(0.699415,0.867117,0.175971); rgb(30pt)=(0.688944,0.865448,0.182725); rgb(31pt)=(0.678489,0.863742,0.189503); rgb(32pt)=(0.668054,0.861999,0.196293); rgb(33pt)=(0.657642,0.860219,0.203082); rgb(34pt)=(0.647257,0.8584,0.209861); rgb(35pt)=(0.636902,0.856542,0.21662); rgb(36pt)=(0.626579,0.854645,0.223353); rgb(37pt)=(0.616293,0.852709,0.230052); rgb(38pt)=(0.606045,0.850733,0.236712); rgb(39pt)=(0.595839,0.848717,0.243329); rgb(40pt)=(0.585678,0.846661,0.249897); rgb(41pt)=(0.575563,0.844566,0.256415); rgb(42pt)=(0.565498,0.84243,0.262877); rgb(43pt)=(0.555484,0.840254,0.269281); rgb(44pt)=(0.545524,0.838039,0.275626); rgb(45pt)=(0.535621,0.835785,0.281908); rgb(46pt)=(0.525776,0.833491,0.288127); rgb(47pt)=(0.515992,0.831158,0.294279); rgb(48pt)=(0.506271,0.828786,0.300362); rgb(49pt)=(0.496615,0.826376,0.306377); rgb(50pt)=(0.487026,0.823929,0.312321); rgb(51pt)=(0.477504,0.821444,0.318195); rgb(52pt)=(0.468053,0.818921,0.323998); rgb(53pt)=(0.458674,0.816363,0.329727); rgb(54pt)=(0.449368,0.813768,0.335384); rgb(55pt)=(0.440137,0.811138,0.340967); rgb(56pt)=(0.430983,0.808473,0.346476); rgb(57pt)=(0.421908,0.805774,0.35191); rgb(58pt)=(0.412913,0.803041,0.357269); rgb(59pt)=(0.404001,0.800275,0.362552); rgb(60pt)=(0.395174,0.797475,0.367757); rgb(61pt)=(0.386433,0.794644,0.372886); rgb(62pt)=(0.377779,0.791781,0.377939); rgb(63pt)=(0.369214,0.788888,0.382914); rgb(64pt)=(0.360741,0.785964,0.387814); rgb(65pt)=(0.35236,0.783011,0.392636); rgb(66pt)=(0.344074,0.780029,0.397381); rgb(67pt)=(0.335885,0.777018,0.402049); rgb(68pt)=(0.327796,0.77398,0.40664); rgb(69pt)=(0.319809,0.770914,0.411152); rgb(70pt)=(0.311925,0.767822,0.415586); rgb(71pt)=(0.304148,0.764704,0.419943); rgb(72pt)=(0.296479,0.761561,0.424223); rgb(73pt)=(0.288921,0.758394,0.428426); rgb(74pt)=(0.281477,0.755203,0.432552); rgb(75pt)=(0.274149,0.751988,0.436601); rgb(76pt)=(0.266941,0.748751,0.440573); rgb(77pt)=(0.259857,0.745492,0.444467); rgb(78pt)=(0.252899,0.742211,0.448284); rgb(79pt)=(0.24607,0.73891,0.452024); rgb(80pt)=(0.239374,0.735588,0.455688); rgb(81pt)=(0.232815,0.732247,0.459277); rgb(82pt)=(0.226397,0.728888,0.462789); rgb(83pt)=(0.220124,0.725509,0.466226); rgb(84pt)=(0.214,0.722114,0.469588); rgb(85pt)=(0.20803,0.718701,0.472873); rgb(86pt)=(0.202219,0.715272,0.476084); rgb(87pt)=(0.196571,0.711827,0.479221); rgb(88pt)=(0.19109,0.708366,0.482284); rgb(89pt)=(0.185783,0.704891,0.485273); rgb(90pt)=(0.180653,0.701402,0.488189); rgb(91pt)=(0.175707,0.6979,0.491033); rgb(92pt)=(0.170948,0.694384,0.493803); rgb(93pt)=(0.166383,0.690856,0.496502); rgb(94pt)=(0.162016,0.687316,0.499129); rgb(95pt)=(0.157851,0.683765,0.501686); rgb(96pt)=(0.153894,0.680203,0.504172); rgb(97pt)=(0.150148,0.676631,0.506589); rgb(98pt)=(0.146616,0.67305,0.508936); rgb(99pt)=(0.143303,0.669459,0.511215); rgb(100pt)=(0.14021,0.665859,0.513427); rgb(101pt)=(0.137339,0.662252,0.515571); rgb(102pt)=(0.134692,0.658636,0.517649); rgb(103pt)=(0.132268,0.655014,0.519661); rgb(104pt)=(0.130067,0.651384,0.521608); rgb(105pt)=(0.128087,0.647749,0.523491); rgb(106pt)=(0.126326,0.644107,0.525311); rgb(107pt)=(0.12478,0.640461,0.527068); rgb(108pt)=(0.123444,0.636809,0.528763); rgb(109pt)=(0.122312,0.633153,0.530398); rgb(110pt)=(0.12138,0.629492,0.531973); rgb(111pt)=(0.120638,0.625828,0.533488); rgb(112pt)=(0.120081,0.622161,0.534946); rgb(113pt)=(0.119699,0.61849,0.536347); rgb(114pt)=(0.119483,0.614817,0.537692); rgb(115pt)=(0.119423,0.611141,0.538982); rgb(116pt)=(0.119512,0.607464,0.540218); rgb(117pt)=(0.119738,0.603785,0.5414); rgb(118pt)=(0.120092,0.600104,0.54253); rgb(119pt)=(0.120565,0.596422,0.543611); rgb(120pt)=(0.121148,0.592739,0.544641); rgb(121pt)=(0.121831,0.589055,0.545623); rgb(122pt)=(0.122606,0.585371,0.546557); rgb(123pt)=(0.123463,0.581687,0.547445); rgb(124pt)=(0.124395,0.578002,0.548287); rgb(125pt)=(0.125394,0.574318,0.549086); rgb(126pt)=(0.126453,0.570633,0.549841); rgb(127pt)=(0.127568,0.566949,0.550556); rgb(128pt)=(0.128729,0.563265,0.551229); rgb(129pt)=(0.129933,0.559582,0.551864); rgb(130pt)=(0.131172,0.555899,0.552459); rgb(131pt)=(0.132444,0.552216,0.553018); rgb(132pt)=(0.133743,0.548535,0.553541); rgb(133pt)=(0.135066,0.544853,0.554029); rgb(134pt)=(0.136408,0.541173,0.554483); rgb(135pt)=(0.13777,0.537492,0.554906); rgb(136pt)=(0.139147,0.533812,0.555298); rgb(137pt)=(0.140536,0.530132,0.555659); rgb(138pt)=(0.141935,0.526453,0.555991); rgb(139pt)=(0.143343,0.522773,0.556295); rgb(140pt)=(0.144759,0.519093,0.556572); rgb(141pt)=(0.14618,0.515413,0.556823); rgb(142pt)=(0.147607,0.511733,0.557049); rgb(143pt)=(0.149039,0.508051,0.55725); rgb(144pt)=(0.150476,0.504369,0.55743); rgb(145pt)=(0.151918,0.500685,0.557587); rgb(146pt)=(0.153364,0.497,0.557724); rgb(147pt)=(0.154815,0.493313,0.55784); rgb(148pt)=(0.15627,0.489624,0.557936); rgb(149pt)=(0.157729,0.485932,0.558013); rgb(150pt)=(0.159194,0.482237,0.558073); rgb(151pt)=(0.160665,0.47854,0.558115); rgb(152pt)=(0.162142,0.474838,0.55814); rgb(153pt)=(0.163625,0.471133,0.558148); rgb(154pt)=(0.165117,0.467423,0.558141); rgb(155pt)=(0.166617,0.463708,0.558119); rgb(156pt)=(0.168126,0.459988,0.558082); rgb(157pt)=(0.169646,0.456262,0.55803); rgb(158pt)=(0.171176,0.45253,0.557965); rgb(159pt)=(0.172719,0.448791,0.557885); rgb(160pt)=(0.174274,0.445044,0.557792); rgb(161pt)=(0.175841,0.44129,0.557685); rgb(162pt)=(0.177423,0.437527,0.557565); rgb(163pt)=(0.179019,0.433756,0.55743); rgb(164pt)=(0.180629,0.429975,0.557282); rgb(165pt)=(0.182256,0.426184,0.55712); rgb(166pt)=(0.183898,0.422383,0.556944); rgb(167pt)=(0.185556,0.41857,0.556753); rgb(168pt)=(0.187231,0.414746,0.556547); rgb(169pt)=(0.188923,0.41091,0.556326); rgb(170pt)=(0.190631,0.407061,0.556089); rgb(171pt)=(0.192357,0.403199,0.555836); rgb(172pt)=(0.1941,0.399323,0.555565); rgb(173pt)=(0.19586,0.395433,0.555276); rgb(174pt)=(0.197636,0.391528,0.554969); rgb(175pt)=(0.19943,0.387607,0.554642); rgb(176pt)=(0.201239,0.38367,0.554294); rgb(177pt)=(0.203063,0.379716,0.553925); rgb(178pt)=(0.204903,0.375746,0.553533); rgb(179pt)=(0.206756,0.371758,0.553117); rgb(180pt)=(0.208623,0.367752,0.552675); rgb(181pt)=(0.210503,0.363727,0.552206); rgb(182pt)=(0.212395,0.359683,0.55171); rgb(183pt)=(0.214298,0.355619,0.551184); rgb(184pt)=(0.21621,0.351535,0.550627); rgb(185pt)=(0.21813,0.347432,0.550038); rgb(186pt)=(0.220057,0.343307,0.549413); rgb(187pt)=(0.221989,0.339161,0.548752); rgb(188pt)=(0.223925,0.334994,0.548053); rgb(189pt)=(0.225863,0.330805,0.547314); rgb(190pt)=(0.227802,0.326594,0.546532); rgb(191pt)=(0.229739,0.322361,0.545706); rgb(192pt)=(0.231674,0.318106,0.544834); rgb(193pt)=(0.233603,0.313828,0.543914); rgb(194pt)=(0.235526,0.309527,0.542944); rgb(195pt)=(0.237441,0.305202,0.541921); rgb(196pt)=(0.239346,0.300855,0.540844); rgb(197pt)=(0.241237,0.296485,0.539709); rgb(198pt)=(0.243113,0.292092,0.538516); rgb(199pt)=(0.244972,0.287675,0.53726); rgb(200pt)=(0.246811,0.283237,0.535941); rgb(201pt)=(0.248629,0.278775,0.534556); rgb(202pt)=(0.250425,0.27429,0.533103); rgb(203pt)=(0.252194,0.269783,0.531579); rgb(204pt)=(0.253935,0.265254,0.529983); rgb(205pt)=(0.255645,0.260703,0.528312); rgb(206pt)=(0.257322,0.25613,0.526563); rgb(207pt)=(0.258965,0.251537,0.524736); rgb(208pt)=(0.260571,0.246922,0.522828); rgb(209pt)=(0.262138,0.242286,0.520837); rgb(210pt)=(0.263663,0.237631,0.518762); rgb(211pt)=(0.265145,0.232956,0.516599); rgb(212pt)=(0.26658,0.228262,0.514349); rgb(213pt)=(0.267968,0.223549,0.512008); rgb(214pt)=(0.269308,0.218818,0.509577); rgb(215pt)=(0.270595,0.214069,0.507052); rgb(216pt)=(0.271828,0.209303,0.504434); rgb(217pt)=(0.273006,0.20452,0.501721); rgb(218pt)=(0.274128,0.199721,0.498911); rgb(219pt)=(0.275191,0.194905,0.496005); rgb(220pt)=(0.276194,0.190074,0.493001); rgb(221pt)=(0.277134,0.185228,0.489898); rgb(222pt)=(0.278012,0.180367,0.486697); rgb(223pt)=(0.278826,0.17549,0.483397); rgb(224pt)=(0.279574,0.170599,0.479997); rgb(225pt)=(0.280255,0.165693,0.476498); rgb(226pt)=(0.280868,0.160771,0.472899); rgb(227pt)=(0.281412,0.155834,0.469201); rgb(228pt)=(0.281887,0.150881,0.465405); rgb(229pt)=(0.28229,0.145912,0.46151); rgb(230pt)=(0.282623,0.140926,0.457517); rgb(231pt)=(0.282884,0.13592,0.453427); rgb(232pt)=(0.283072,0.130895,0.449241); rgb(233pt)=(0.283187,0.125848,0.44496); rgb(234pt)=(0.283229,0.120777,0.440584); rgb(235pt)=(0.283197,0.11568,0.436115); rgb(236pt)=(0.283091,0.110553,0.431554); rgb(237pt)=(0.28291,0.105393,0.426902); rgb(238pt)=(0.282656,0.100196,0.42216); rgb(239pt)=(0.282327,0.0949554,0.417331); rgb(240pt)=(0.281924,0.0896662,0.412415); rgb(241pt)=(0.281446,0.0843197,0.407414); rgb(242pt)=(0.280894,0.078907,0.402329); rgb(243pt)=(0.280267,0.0734172,0.397163); rgb(244pt)=(0.279566,0.0678359,0.391917); rgb(245pt)=(0.278791,0.0621454,0.386592); rgb(246pt)=(0.277941,0.0563244,0.381191); rgb(247pt)=(0.277018,0.0503444,0.375715); rgb(248pt)=(0.276022,0.0441672,0.370164); rgb(249pt)=(0.274952,0.0377518,0.364543); rgb(250pt)=(0.273809,0.0314975,0.358853); rgb(251pt)=(0.272594,0.0255631,0.353093); rgb(252pt)=(0.271305,0.0199419,0.347269); rgb(253pt)=(0.269944,0.0146249,0.341379); rgb(254pt)=(0.26851,0.00960483,0.335427); rgb(255pt)=(0,0,0)},
x dir=reverse,
xmin=-1.4,
xmax=1.4,
xlabel style={font=\color{white!15!black},yshift=0.25em},
xlabel={$x$ / m},
ymin=0,
ymax=2.25,
ylabel style={font=\color{white!15!black},  yshift=-0.25em},
ylabel={$z$ / m},
axis background/.style={fill=white},
axis x line*=bottom,
axis y line*=left,
ytick={0.5,1,1.5,2},
axis equal,
xmajorgrids,
ymajorgrids,
legend style={legend cell align=left, align=left, draw=white!15!black},
colorbar,
colorbar style={title={$\norm*{\bm{\hat{h}}_{\hat{p}}}_2^2 \, / \, T^2$ (dB)},font=\footnotesize,xshift=3.4em}
]

\draw[targetcolor, semithick, rotate around={15:(axis cs:0.67,1.64)}] (axis cs:1.21,1.59) rectangle (axis cs:0.13,1.66);

\draw[targetcolor, semithick, rotate around={-19:(axis cs:-0.75,1.83)}] (axis cs:-0.18,1.71) rectangle (axis cs:-1.3,1.79);

\addplot[scatter, only marks, mark=*, mark size=.95pt, scatter src=explicit, scatter/use mapped color={mark options={}, draw=mapped color, fill=mapped color}] table[row sep=crcr, meta=color]{%
x	y	color\\
0.631887663848461	1.67181761507261	-23.3514628690322\\
0.165625046930622	1.78724820822149	-22.3229238127977\\
1.08104399461463	1.51556923076923	-19.9132951495632\\
0.682231127243267	1.67111814838979	-18.1916650427729\\
-0.707535042735043	1.79524612550449	-17.2277611533342\\
-0.452596231112416	1.84076021902138	-20.0629644229602\\
-0.510819138660671	1.85976644645734	-19.6121479445526\\
-1.10387301387606	1.61890824140625	-22.5532693327569\\
0.470819119928007	1.70158447797933	-15.3327582691345\\
-0.71229285235943	1.80078774540039	-16.2221625640815\\
-0.738530934826488	1.80078774540039	-17.1605695293529\\
-0.231556318071636	1.93733915221948	-23.070706908159\\
-0.236032281742313	1.94727712131542	-18.3950525613782\\
-1.26783393660681	1.57498290600053	-20.7064011738922\\
-1.07006482782218	1.62814358974359	-18.9621582746252\\
-1.09206457269304	1.67688031707596	-18.4673237834963\\
-0.319792323037206	1.91878562175517	-17.1535867118903\\
-0.806982825294672	1.75132735884796	-18.6691310395529\\
};

\end{axis}
\end{tikzpicture}%

%% file: figures/20240305_imaging_4pleft_peaceright_omp100_thresh10_spher_proposed_normalized_colorinv.tex
\definecolor{targetcolor}{rgb}{0.83203,0.36719,0.00000}%

\begin{tikzpicture}

\begin{axis}[%
height=.14\textheight,
at={(0.679in,0.668in)},
scale only axis,
point meta min=-28,
point meta max=-14,
colormap={mymap}{[1pt] rgb(0pt)=(0.993248,0.906157,0.143936); rgb(1pt)=(0.983868,0.904867,0.136897); rgb(2pt)=(0.974417,0.90359,0.130215); rgb(3pt)=(0.964894,0.902323,0.123941); rgb(4pt)=(0.9553,0.901065,0.118128); rgb(5pt)=(0.945636,0.899815,0.112838); rgb(6pt)=(0.935904,0.89857,0.108131); rgb(7pt)=(0.926106,0.89733,0.104071); rgb(8pt)=(0.916242,0.896091,0.100717); rgb(9pt)=(0.906311,0.894855,0.098125); rgb(10pt)=(0.89632,0.893616,0.0963354); rgb(11pt)=(0.886271,0.892374,0.0953744); rgb(12pt)=(0.876168,0.891125,0.0952505); rgb(13pt)=(0.866013,0.889868,0.0959528); rgb(14pt)=(0.85581,0.888601,0.0974519); rgb(15pt)=(0.845561,0.887322,0.0997022); rgb(16pt)=(0.83527,0.886029,0.102646); rgb(17pt)=(0.82494,0.88472,0.106217); rgb(18pt)=(0.814576,0.883393,0.110347); rgb(19pt)=(0.804182,0.882046,0.114965); rgb(20pt)=(0.79376,0.880678,0.120005); rgb(21pt)=(0.783315,0.879285,0.125405); rgb(22pt)=(0.772852,0.877868,0.131109); rgb(23pt)=(0.762373,0.876424,0.137064); rgb(24pt)=(0.751884,0.874951,0.143228); rgb(25pt)=(0.741388,0.873449,0.149561); rgb(26pt)=(0.730889,0.871916,0.156029); rgb(27pt)=(0.720391,0.87035,0.162603); rgb(28pt)=(0.709898,0.868751,0.169257); rgb(29pt)=(0.699415,0.867117,0.175971); rgb(30pt)=(0.688944,0.865448,0.182725); rgb(31pt)=(0.678489,0.863742,0.189503); rgb(32pt)=(0.668054,0.861999,0.196293); rgb(33pt)=(0.657642,0.860219,0.203082); rgb(34pt)=(0.647257,0.8584,0.209861); rgb(35pt)=(0.636902,0.856542,0.21662); rgb(36pt)=(0.626579,0.854645,0.223353); rgb(37pt)=(0.616293,0.852709,0.230052); rgb(38pt)=(0.606045,0.850733,0.236712); rgb(39pt)=(0.595839,0.848717,0.243329); rgb(40pt)=(0.585678,0.846661,0.249897); rgb(41pt)=(0.575563,0.844566,0.256415); rgb(42pt)=(0.565498,0.84243,0.262877); rgb(43pt)=(0.555484,0.840254,0.269281); rgb(44pt)=(0.545524,0.838039,0.275626); rgb(45pt)=(0.535621,0.835785,0.281908); rgb(46pt)=(0.525776,0.833491,0.288127); rgb(47pt)=(0.515992,0.831158,0.294279); rgb(48pt)=(0.506271,0.828786,0.300362); rgb(49pt)=(0.496615,0.826376,0.306377); rgb(50pt)=(0.487026,0.823929,0.312321); rgb(51pt)=(0.477504,0.821444,0.318195); rgb(52pt)=(0.468053,0.818921,0.323998); rgb(53pt)=(0.458674,0.816363,0.329727); rgb(54pt)=(0.449368,0.813768,0.335384); rgb(55pt)=(0.440137,0.811138,0.340967); rgb(56pt)=(0.430983,0.808473,0.346476); rgb(57pt)=(0.421908,0.805774,0.35191); rgb(58pt)=(0.412913,0.803041,0.357269); rgb(59pt)=(0.404001,0.800275,0.362552); rgb(60pt)=(0.395174,0.797475,0.367757); rgb(61pt)=(0.386433,0.794644,0.372886); rgb(62pt)=(0.377779,0.791781,0.377939); rgb(63pt)=(0.369214,0.788888,0.382914); rgb(64pt)=(0.360741,0.785964,0.387814); rgb(65pt)=(0.35236,0.783011,0.392636); rgb(66pt)=(0.344074,0.780029,0.397381); rgb(67pt)=(0.335885,0.777018,0.402049); rgb(68pt)=(0.327796,0.77398,0.40664); rgb(69pt)=(0.319809,0.770914,0.411152); rgb(70pt)=(0.311925,0.767822,0.415586); rgb(71pt)=(0.304148,0.764704,0.419943); rgb(72pt)=(0.296479,0.761561,0.424223); rgb(73pt)=(0.288921,0.758394,0.428426); rgb(74pt)=(0.281477,0.755203,0.432552); rgb(75pt)=(0.274149,0.751988,0.436601); rgb(76pt)=(0.266941,0.748751,0.440573); rgb(77pt)=(0.259857,0.745492,0.444467); rgb(78pt)=(0.252899,0.742211,0.448284); rgb(79pt)=(0.24607,0.73891,0.452024); rgb(80pt)=(0.239374,0.735588,0.455688); rgb(81pt)=(0.232815,0.732247,0.459277); rgb(82pt)=(0.226397,0.728888,0.462789); rgb(83pt)=(0.220124,0.725509,0.466226); rgb(84pt)=(0.214,0.722114,0.469588); rgb(85pt)=(0.20803,0.718701,0.472873); rgb(86pt)=(0.202219,0.715272,0.476084); rgb(87pt)=(0.196571,0.711827,0.479221); rgb(88pt)=(0.19109,0.708366,0.482284); rgb(89pt)=(0.185783,0.704891,0.485273); rgb(90pt)=(0.180653,0.701402,0.488189); rgb(91pt)=(0.175707,0.6979,0.491033); rgb(92pt)=(0.170948,0.694384,0.493803); rgb(93pt)=(0.166383,0.690856,0.496502); rgb(94pt)=(0.162016,0.687316,0.499129); rgb(95pt)=(0.157851,0.683765,0.501686); rgb(96pt)=(0.153894,0.680203,0.504172); rgb(97pt)=(0.150148,0.676631,0.506589); rgb(98pt)=(0.146616,0.67305,0.508936); rgb(99pt)=(0.143303,0.669459,0.511215); rgb(100pt)=(0.14021,0.665859,0.513427); rgb(101pt)=(0.137339,0.662252,0.515571); rgb(102pt)=(0.134692,0.658636,0.517649); rgb(103pt)=(0.132268,0.655014,0.519661); rgb(104pt)=(0.130067,0.651384,0.521608); rgb(105pt)=(0.128087,0.647749,0.523491); rgb(106pt)=(0.126326,0.644107,0.525311); rgb(107pt)=(0.12478,0.640461,0.527068); rgb(108pt)=(0.123444,0.636809,0.528763); rgb(109pt)=(0.122312,0.633153,0.530398); rgb(110pt)=(0.12138,0.629492,0.531973); rgb(111pt)=(0.120638,0.625828,0.533488); rgb(112pt)=(0.120081,0.622161,0.534946); rgb(113pt)=(0.119699,0.61849,0.536347); rgb(114pt)=(0.119483,0.614817,0.537692); rgb(115pt)=(0.119423,0.611141,0.538982); rgb(116pt)=(0.119512,0.607464,0.540218); rgb(117pt)=(0.119738,0.603785,0.5414); rgb(118pt)=(0.120092,0.600104,0.54253); rgb(119pt)=(0.120565,0.596422,0.543611); rgb(120pt)=(0.121148,0.592739,0.544641); rgb(121pt)=(0.121831,0.589055,0.545623); rgb(122pt)=(0.122606,0.585371,0.546557); rgb(123pt)=(0.123463,0.581687,0.547445); rgb(124pt)=(0.124395,0.578002,0.548287); rgb(125pt)=(0.125394,0.574318,0.549086); rgb(126pt)=(0.126453,0.570633,0.549841); rgb(127pt)=(0.127568,0.566949,0.550556); rgb(128pt)=(0.128729,0.563265,0.551229); rgb(129pt)=(0.129933,0.559582,0.551864); rgb(130pt)=(0.131172,0.555899,0.552459); rgb(131pt)=(0.132444,0.552216,0.553018); rgb(132pt)=(0.133743,0.548535,0.553541); rgb(133pt)=(0.135066,0.544853,0.554029); rgb(134pt)=(0.136408,0.541173,0.554483); rgb(135pt)=(0.13777,0.537492,0.554906); rgb(136pt)=(0.139147,0.533812,0.555298); rgb(137pt)=(0.140536,0.530132,0.555659); rgb(138pt)=(0.141935,0.526453,0.555991); rgb(139pt)=(0.143343,0.522773,0.556295); rgb(140pt)=(0.144759,0.519093,0.556572); rgb(141pt)=(0.14618,0.515413,0.556823); rgb(142pt)=(0.147607,0.511733,0.557049); rgb(143pt)=(0.149039,0.508051,0.55725); rgb(144pt)=(0.150476,0.504369,0.55743); rgb(145pt)=(0.151918,0.500685,0.557587); rgb(146pt)=(0.153364,0.497,0.557724); rgb(147pt)=(0.154815,0.493313,0.55784); rgb(148pt)=(0.15627,0.489624,0.557936); rgb(149pt)=(0.157729,0.485932,0.558013); rgb(150pt)=(0.159194,0.482237,0.558073); rgb(151pt)=(0.160665,0.47854,0.558115); rgb(152pt)=(0.162142,0.474838,0.55814); rgb(153pt)=(0.163625,0.471133,0.558148); rgb(154pt)=(0.165117,0.467423,0.558141); rgb(155pt)=(0.166617,0.463708,0.558119); rgb(156pt)=(0.168126,0.459988,0.558082); rgb(157pt)=(0.169646,0.456262,0.55803); rgb(158pt)=(0.171176,0.45253,0.557965); rgb(159pt)=(0.172719,0.448791,0.557885); rgb(160pt)=(0.174274,0.445044,0.557792); rgb(161pt)=(0.175841,0.44129,0.557685); rgb(162pt)=(0.177423,0.437527,0.557565); rgb(163pt)=(0.179019,0.433756,0.55743); rgb(164pt)=(0.180629,0.429975,0.557282); rgb(165pt)=(0.182256,0.426184,0.55712); rgb(166pt)=(0.183898,0.422383,0.556944); rgb(167pt)=(0.185556,0.41857,0.556753); rgb(168pt)=(0.187231,0.414746,0.556547); rgb(169pt)=(0.188923,0.41091,0.556326); rgb(170pt)=(0.190631,0.407061,0.556089); rgb(171pt)=(0.192357,0.403199,0.555836); rgb(172pt)=(0.1941,0.399323,0.555565); rgb(173pt)=(0.19586,0.395433,0.555276); rgb(174pt)=(0.197636,0.391528,0.554969); rgb(175pt)=(0.19943,0.387607,0.554642); rgb(176pt)=(0.201239,0.38367,0.554294); rgb(177pt)=(0.203063,0.379716,0.553925); rgb(178pt)=(0.204903,0.375746,0.553533); rgb(179pt)=(0.206756,0.371758,0.553117); rgb(180pt)=(0.208623,0.367752,0.552675); rgb(181pt)=(0.210503,0.363727,0.552206); rgb(182pt)=(0.212395,0.359683,0.55171); rgb(183pt)=(0.214298,0.355619,0.551184); rgb(184pt)=(0.21621,0.351535,0.550627); rgb(185pt)=(0.21813,0.347432,0.550038); rgb(186pt)=(0.220057,0.343307,0.549413); rgb(187pt)=(0.221989,0.339161,0.548752); rgb(188pt)=(0.223925,0.334994,0.548053); rgb(189pt)=(0.225863,0.330805,0.547314); rgb(190pt)=(0.227802,0.326594,0.546532); rgb(191pt)=(0.229739,0.322361,0.545706); rgb(192pt)=(0.231674,0.318106,0.544834); rgb(193pt)=(0.233603,0.313828,0.543914); rgb(194pt)=(0.235526,0.309527,0.542944); rgb(195pt)=(0.237441,0.305202,0.541921); rgb(196pt)=(0.239346,0.300855,0.540844); rgb(197pt)=(0.241237,0.296485,0.539709); rgb(198pt)=(0.243113,0.292092,0.538516); rgb(199pt)=(0.244972,0.287675,0.53726); rgb(200pt)=(0.246811,0.283237,0.535941); rgb(201pt)=(0.248629,0.278775,0.534556); rgb(202pt)=(0.250425,0.27429,0.533103); rgb(203pt)=(0.252194,0.269783,0.531579); rgb(204pt)=(0.253935,0.265254,0.529983); rgb(205pt)=(0.255645,0.260703,0.528312); rgb(206pt)=(0.257322,0.25613,0.526563); rgb(207pt)=(0.258965,0.251537,0.524736); rgb(208pt)=(0.260571,0.246922,0.522828); rgb(209pt)=(0.262138,0.242286,0.520837); rgb(210pt)=(0.263663,0.237631,0.518762); rgb(211pt)=(0.265145,0.232956,0.516599); rgb(212pt)=(0.26658,0.228262,0.514349); rgb(213pt)=(0.267968,0.223549,0.512008); rgb(214pt)=(0.269308,0.218818,0.509577); rgb(215pt)=(0.270595,0.214069,0.507052); rgb(216pt)=(0.271828,0.209303,0.504434); rgb(217pt)=(0.273006,0.20452,0.501721); rgb(218pt)=(0.274128,0.199721,0.498911); rgb(219pt)=(0.275191,0.194905,0.496005); rgb(220pt)=(0.276194,0.190074,0.493001); rgb(221pt)=(0.277134,0.185228,0.489898); rgb(222pt)=(0.278012,0.180367,0.486697); rgb(223pt)=(0.278826,0.17549,0.483397); rgb(224pt)=(0.279574,0.170599,0.479997); rgb(225pt)=(0.280255,0.165693,0.476498); rgb(226pt)=(0.280868,0.160771,0.472899); rgb(227pt)=(0.281412,0.155834,0.469201); rgb(228pt)=(0.281887,0.150881,0.465405); rgb(229pt)=(0.28229,0.145912,0.46151); rgb(230pt)=(0.282623,0.140926,0.457517); rgb(231pt)=(0.282884,0.13592,0.453427); rgb(232pt)=(0.283072,0.130895,0.449241); rgb(233pt)=(0.283187,0.125848,0.44496); rgb(234pt)=(0.283229,0.120777,0.440584); rgb(235pt)=(0.283197,0.11568,0.436115); rgb(236pt)=(0.283091,0.110553,0.431554); rgb(237pt)=(0.28291,0.105393,0.426902); rgb(238pt)=(0.282656,0.100196,0.42216); rgb(239pt)=(0.282327,0.0949554,0.417331); rgb(240pt)=(0.281924,0.0896662,0.412415); rgb(241pt)=(0.281446,0.0843197,0.407414); rgb(242pt)=(0.280894,0.078907,0.402329); rgb(243pt)=(0.280267,0.0734172,0.397163); rgb(244pt)=(0.279566,0.0678359,0.391917); rgb(245pt)=(0.278791,0.0621454,0.386592); rgb(246pt)=(0.277941,0.0563244,0.381191); rgb(247pt)=(0.277018,0.0503444,0.375715); rgb(248pt)=(0.276022,0.0441672,0.370164); rgb(249pt)=(0.274952,0.0377518,0.364543); rgb(250pt)=(0.273809,0.0314975,0.358853); rgb(251pt)=(0.272594,0.0255631,0.353093); rgb(252pt)=(0.271305,0.0199419,0.347269); rgb(253pt)=(0.269944,0.0146249,0.341379); rgb(254pt)=(0.26851,0.00960483,0.335427); rgb(255pt)=(0,0,0)},
x dir=reverse,
xmin=-45,
xmax=45,
xlabel style={font=\color{white!15!black},yshift=0.25em},
xlabel={Azimuth $\vartheta \ (^\circ)$},
ymin=-32,
ymax=32,
axis equal,
ylabel style={font=\color{white!15!black}, yshift=-0.25em},
ylabel={Elevation $\varphi \ (^\circ)$},
axis background/.style={fill=white},
axis x line*=bottom,
axis y line*=left,
xmajorgrids,
ymajorgrids,
legend style={legend cell align=left, align=left, draw=white!15!black},
colorbar,
colorbar style={title={$\norm*{\bm{\hat{h}}_{\hat{p}}}_2^2 \, / \, T^2$ (dB)},font=\footnotesize,xshift=2.75em}
]

\addplot[scatter, only marks, mark=triangle*, mark size=2.25pt, semithick, scatter src=explicit, scatter/use mapped color={mark options={}, draw=targetcolor, fill=white}] table[row sep=crcr, meta=color]{%
	x	y	color\\
	37	-10.5	-24\\
	22	-11.0	-24\\
	6	-11		-24\\
	22 	-22		-24\\
	-7.4437   -6.5378  -24.0000\\
	-22.2474   -6.6584  -24.0000\\
	-36.3996   -6.3817  -24.0000\\
	-15.5726  -10.7206  -24.0000\\
	-28.2202  -11.0848  -24.0000\\
	-22.2474    7.9459  -24.0000\\
	-22.2474    0.1454  -24.0000\\
	-22.2474  -13.2796  -24.0000\\
	-22.2474  -20.3304  -24.0000\\
	-10.8473    3.1690  -24.0000\\
	-32.7147    3.1225  -24.0000\\
	-10.4734  -15.7311  -24.0000\\
	-33.2373  -15.4606  -24.0000\\
};

\addplot[scatter, only marks, mark=*, mark size=.95pt, scatter src=explicit, scatter/use mapped color={mark options={}, draw=mapped color, fill=mapped color}] table[row sep=crcr, meta=color]{%
x	y	color\\
-38.8334622656322	-6.55526540862824	-23.3514628690322\\
-33.314118158639	-16.8012213982218	-22.3229238127977\\
35.4999741515178	-10.6850486877471	-19.9132951495632\\
-33.0740970042897	-10.4949598351109	-18.1916650427729\\
-34.2886150266354	4.24174692905709	-17.2277611533342\\
-24.7394724425013	-21.347797507179	-20.0629644229602\\
15.4664754239383	-26.0298597610893	-19.6121479445526\\
22.2076542985965	-20.7048110546354	-22.5532693327569\\
-21.5810338411767	-9.73243196820865	-15.3327582691345\\
20.7048110546354	-11.5369590328155	-16.2221625640815\\
-22.2992933865783	7.86318410398364	-17.1605695293529\\
-21.5101882668875	-0	-23.070706908159\\
-9.46216878890615	-17.096009423715	-18.3950525613782\\
-13.8135658545711	-10.7804969483605	-20.7064011738922\\
-15.3585881179105	1.84198209072813	-18.9621582746252\\
5.2945026039643	-10.2650083971775	-18.4673237834963\\
-6.81582191825822	-6.76810136869058	-17.1535867118903\\
-6.9111891694161	3.31647927251333	-18.6691310395529\\
};

\end{axis}
\end{tikzpicture}%

%% file: figures/20240305_imaging_4pleft_peaceright_omp100_thresh10_spher_analytical_nocomp_normalized_colorinv.tex
\definecolor{targetcolor}{rgb}{0.83203,0.36719,0.00000}%

\begin{tikzpicture}

\begin{axis}[%
height=.14\textheight,
at={(0.67in,0.69in)},
scale only axis,
point meta min=-28,
point meta max=-14,
colormap={mymap}{[1pt] rgb(0pt)=(0.993248,0.906157,0.143936); rgb(1pt)=(0.983868,0.904867,0.136897); rgb(2pt)=(0.974417,0.90359,0.130215); rgb(3pt)=(0.964894,0.902323,0.123941); rgb(4pt)=(0.9553,0.901065,0.118128); rgb(5pt)=(0.945636,0.899815,0.112838); rgb(6pt)=(0.935904,0.89857,0.108131); rgb(7pt)=(0.926106,0.89733,0.104071); rgb(8pt)=(0.916242,0.896091,0.100717); rgb(9pt)=(0.906311,0.894855,0.098125); rgb(10pt)=(0.89632,0.893616,0.0963354); rgb(11pt)=(0.886271,0.892374,0.0953744); rgb(12pt)=(0.876168,0.891125,0.0952505); rgb(13pt)=(0.866013,0.889868,0.0959528); rgb(14pt)=(0.85581,0.888601,0.0974519); rgb(15pt)=(0.845561,0.887322,0.0997022); rgb(16pt)=(0.83527,0.886029,0.102646); rgb(17pt)=(0.82494,0.88472,0.106217); rgb(18pt)=(0.814576,0.883393,0.110347); rgb(19pt)=(0.804182,0.882046,0.114965); rgb(20pt)=(0.79376,0.880678,0.120005); rgb(21pt)=(0.783315,0.879285,0.125405); rgb(22pt)=(0.772852,0.877868,0.131109); rgb(23pt)=(0.762373,0.876424,0.137064); rgb(24pt)=(0.751884,0.874951,0.143228); rgb(25pt)=(0.741388,0.873449,0.149561); rgb(26pt)=(0.730889,0.871916,0.156029); rgb(27pt)=(0.720391,0.87035,0.162603); rgb(28pt)=(0.709898,0.868751,0.169257); rgb(29pt)=(0.699415,0.867117,0.175971); rgb(30pt)=(0.688944,0.865448,0.182725); rgb(31pt)=(0.678489,0.863742,0.189503); rgb(32pt)=(0.668054,0.861999,0.196293); rgb(33pt)=(0.657642,0.860219,0.203082); rgb(34pt)=(0.647257,0.8584,0.209861); rgb(35pt)=(0.636902,0.856542,0.21662); rgb(36pt)=(0.626579,0.854645,0.223353); rgb(37pt)=(0.616293,0.852709,0.230052); rgb(38pt)=(0.606045,0.850733,0.236712); rgb(39pt)=(0.595839,0.848717,0.243329); rgb(40pt)=(0.585678,0.846661,0.249897); rgb(41pt)=(0.575563,0.844566,0.256415); rgb(42pt)=(0.565498,0.84243,0.262877); rgb(43pt)=(0.555484,0.840254,0.269281); rgb(44pt)=(0.545524,0.838039,0.275626); rgb(45pt)=(0.535621,0.835785,0.281908); rgb(46pt)=(0.525776,0.833491,0.288127); rgb(47pt)=(0.515992,0.831158,0.294279); rgb(48pt)=(0.506271,0.828786,0.300362); rgb(49pt)=(0.496615,0.826376,0.306377); rgb(50pt)=(0.487026,0.823929,0.312321); rgb(51pt)=(0.477504,0.821444,0.318195); rgb(52pt)=(0.468053,0.818921,0.323998); rgb(53pt)=(0.458674,0.816363,0.329727); rgb(54pt)=(0.449368,0.813768,0.335384); rgb(55pt)=(0.440137,0.811138,0.340967); rgb(56pt)=(0.430983,0.808473,0.346476); rgb(57pt)=(0.421908,0.805774,0.35191); rgb(58pt)=(0.412913,0.803041,0.357269); rgb(59pt)=(0.404001,0.800275,0.362552); rgb(60pt)=(0.395174,0.797475,0.367757); rgb(61pt)=(0.386433,0.794644,0.372886); rgb(62pt)=(0.377779,0.791781,0.377939); rgb(63pt)=(0.369214,0.788888,0.382914); rgb(64pt)=(0.360741,0.785964,0.387814); rgb(65pt)=(0.35236,0.783011,0.392636); rgb(66pt)=(0.344074,0.780029,0.397381); rgb(67pt)=(0.335885,0.777018,0.402049); rgb(68pt)=(0.327796,0.77398,0.40664); rgb(69pt)=(0.319809,0.770914,0.411152); rgb(70pt)=(0.311925,0.767822,0.415586); rgb(71pt)=(0.304148,0.764704,0.419943); rgb(72pt)=(0.296479,0.761561,0.424223); rgb(73pt)=(0.288921,0.758394,0.428426); rgb(74pt)=(0.281477,0.755203,0.432552); rgb(75pt)=(0.274149,0.751988,0.436601); rgb(76pt)=(0.266941,0.748751,0.440573); rgb(77pt)=(0.259857,0.745492,0.444467); rgb(78pt)=(0.252899,0.742211,0.448284); rgb(79pt)=(0.24607,0.73891,0.452024); rgb(80pt)=(0.239374,0.735588,0.455688); rgb(81pt)=(0.232815,0.732247,0.459277); rgb(82pt)=(0.226397,0.728888,0.462789); rgb(83pt)=(0.220124,0.725509,0.466226); rgb(84pt)=(0.214,0.722114,0.469588); rgb(85pt)=(0.20803,0.718701,0.472873); rgb(86pt)=(0.202219,0.715272,0.476084); rgb(87pt)=(0.196571,0.711827,0.479221); rgb(88pt)=(0.19109,0.708366,0.482284); rgb(89pt)=(0.185783,0.704891,0.485273); rgb(90pt)=(0.180653,0.701402,0.488189); rgb(91pt)=(0.175707,0.6979,0.491033); rgb(92pt)=(0.170948,0.694384,0.493803); rgb(93pt)=(0.166383,0.690856,0.496502); rgb(94pt)=(0.162016,0.687316,0.499129); rgb(95pt)=(0.157851,0.683765,0.501686); rgb(96pt)=(0.153894,0.680203,0.504172); rgb(97pt)=(0.150148,0.676631,0.506589); rgb(98pt)=(0.146616,0.67305,0.508936); rgb(99pt)=(0.143303,0.669459,0.511215); rgb(100pt)=(0.14021,0.665859,0.513427); rgb(101pt)=(0.137339,0.662252,0.515571); rgb(102pt)=(0.134692,0.658636,0.517649); rgb(103pt)=(0.132268,0.655014,0.519661); rgb(104pt)=(0.130067,0.651384,0.521608); rgb(105pt)=(0.128087,0.647749,0.523491); rgb(106pt)=(0.126326,0.644107,0.525311); rgb(107pt)=(0.12478,0.640461,0.527068); rgb(108pt)=(0.123444,0.636809,0.528763); rgb(109pt)=(0.122312,0.633153,0.530398); rgb(110pt)=(0.12138,0.629492,0.531973); rgb(111pt)=(0.120638,0.625828,0.533488); rgb(112pt)=(0.120081,0.622161,0.534946); rgb(113pt)=(0.119699,0.61849,0.536347); rgb(114pt)=(0.119483,0.614817,0.537692); rgb(115pt)=(0.119423,0.611141,0.538982); rgb(116pt)=(0.119512,0.607464,0.540218); rgb(117pt)=(0.119738,0.603785,0.5414); rgb(118pt)=(0.120092,0.600104,0.54253); rgb(119pt)=(0.120565,0.596422,0.543611); rgb(120pt)=(0.121148,0.592739,0.544641); rgb(121pt)=(0.121831,0.589055,0.545623); rgb(122pt)=(0.122606,0.585371,0.546557); rgb(123pt)=(0.123463,0.581687,0.547445); rgb(124pt)=(0.124395,0.578002,0.548287); rgb(125pt)=(0.125394,0.574318,0.549086); rgb(126pt)=(0.126453,0.570633,0.549841); rgb(127pt)=(0.127568,0.566949,0.550556); rgb(128pt)=(0.128729,0.563265,0.551229); rgb(129pt)=(0.129933,0.559582,0.551864); rgb(130pt)=(0.131172,0.555899,0.552459); rgb(131pt)=(0.132444,0.552216,0.553018); rgb(132pt)=(0.133743,0.548535,0.553541); rgb(133pt)=(0.135066,0.544853,0.554029); rgb(134pt)=(0.136408,0.541173,0.554483); rgb(135pt)=(0.13777,0.537492,0.554906); rgb(136pt)=(0.139147,0.533812,0.555298); rgb(137pt)=(0.140536,0.530132,0.555659); rgb(138pt)=(0.141935,0.526453,0.555991); rgb(139pt)=(0.143343,0.522773,0.556295); rgb(140pt)=(0.144759,0.519093,0.556572); rgb(141pt)=(0.14618,0.515413,0.556823); rgb(142pt)=(0.147607,0.511733,0.557049); rgb(143pt)=(0.149039,0.508051,0.55725); rgb(144pt)=(0.150476,0.504369,0.55743); rgb(145pt)=(0.151918,0.500685,0.557587); rgb(146pt)=(0.153364,0.497,0.557724); rgb(147pt)=(0.154815,0.493313,0.55784); rgb(148pt)=(0.15627,0.489624,0.557936); rgb(149pt)=(0.157729,0.485932,0.558013); rgb(150pt)=(0.159194,0.482237,0.558073); rgb(151pt)=(0.160665,0.47854,0.558115); rgb(152pt)=(0.162142,0.474838,0.55814); rgb(153pt)=(0.163625,0.471133,0.558148); rgb(154pt)=(0.165117,0.467423,0.558141); rgb(155pt)=(0.166617,0.463708,0.558119); rgb(156pt)=(0.168126,0.459988,0.558082); rgb(157pt)=(0.169646,0.456262,0.55803); rgb(158pt)=(0.171176,0.45253,0.557965); rgb(159pt)=(0.172719,0.448791,0.557885); rgb(160pt)=(0.174274,0.445044,0.557792); rgb(161pt)=(0.175841,0.44129,0.557685); rgb(162pt)=(0.177423,0.437527,0.557565); rgb(163pt)=(0.179019,0.433756,0.55743); rgb(164pt)=(0.180629,0.429975,0.557282); rgb(165pt)=(0.182256,0.426184,0.55712); rgb(166pt)=(0.183898,0.422383,0.556944); rgb(167pt)=(0.185556,0.41857,0.556753); rgb(168pt)=(0.187231,0.414746,0.556547); rgb(169pt)=(0.188923,0.41091,0.556326); rgb(170pt)=(0.190631,0.407061,0.556089); rgb(171pt)=(0.192357,0.403199,0.555836); rgb(172pt)=(0.1941,0.399323,0.555565); rgb(173pt)=(0.19586,0.395433,0.555276); rgb(174pt)=(0.197636,0.391528,0.554969); rgb(175pt)=(0.19943,0.387607,0.554642); rgb(176pt)=(0.201239,0.38367,0.554294); rgb(177pt)=(0.203063,0.379716,0.553925); rgb(178pt)=(0.204903,0.375746,0.553533); rgb(179pt)=(0.206756,0.371758,0.553117); rgb(180pt)=(0.208623,0.367752,0.552675); rgb(181pt)=(0.210503,0.363727,0.552206); rgb(182pt)=(0.212395,0.359683,0.55171); rgb(183pt)=(0.214298,0.355619,0.551184); rgb(184pt)=(0.21621,0.351535,0.550627); rgb(185pt)=(0.21813,0.347432,0.550038); rgb(186pt)=(0.220057,0.343307,0.549413); rgb(187pt)=(0.221989,0.339161,0.548752); rgb(188pt)=(0.223925,0.334994,0.548053); rgb(189pt)=(0.225863,0.330805,0.547314); rgb(190pt)=(0.227802,0.326594,0.546532); rgb(191pt)=(0.229739,0.322361,0.545706); rgb(192pt)=(0.231674,0.318106,0.544834); rgb(193pt)=(0.233603,0.313828,0.543914); rgb(194pt)=(0.235526,0.309527,0.542944); rgb(195pt)=(0.237441,0.305202,0.541921); rgb(196pt)=(0.239346,0.300855,0.540844); rgb(197pt)=(0.241237,0.296485,0.539709); rgb(198pt)=(0.243113,0.292092,0.538516); rgb(199pt)=(0.244972,0.287675,0.53726); rgb(200pt)=(0.246811,0.283237,0.535941); rgb(201pt)=(0.248629,0.278775,0.534556); rgb(202pt)=(0.250425,0.27429,0.533103); rgb(203pt)=(0.252194,0.269783,0.531579); rgb(204pt)=(0.253935,0.265254,0.529983); rgb(205pt)=(0.255645,0.260703,0.528312); rgb(206pt)=(0.257322,0.25613,0.526563); rgb(207pt)=(0.258965,0.251537,0.524736); rgb(208pt)=(0.260571,0.246922,0.522828); rgb(209pt)=(0.262138,0.242286,0.520837); rgb(210pt)=(0.263663,0.237631,0.518762); rgb(211pt)=(0.265145,0.232956,0.516599); rgb(212pt)=(0.26658,0.228262,0.514349); rgb(213pt)=(0.267968,0.223549,0.512008); rgb(214pt)=(0.269308,0.218818,0.509577); rgb(215pt)=(0.270595,0.214069,0.507052); rgb(216pt)=(0.271828,0.209303,0.504434); rgb(217pt)=(0.273006,0.20452,0.501721); rgb(218pt)=(0.274128,0.199721,0.498911); rgb(219pt)=(0.275191,0.194905,0.496005); rgb(220pt)=(0.276194,0.190074,0.493001); rgb(221pt)=(0.277134,0.185228,0.489898); rgb(222pt)=(0.278012,0.180367,0.486697); rgb(223pt)=(0.278826,0.17549,0.483397); rgb(224pt)=(0.279574,0.170599,0.479997); rgb(225pt)=(0.280255,0.165693,0.476498); rgb(226pt)=(0.280868,0.160771,0.472899); rgb(227pt)=(0.281412,0.155834,0.469201); rgb(228pt)=(0.281887,0.150881,0.465405); rgb(229pt)=(0.28229,0.145912,0.46151); rgb(230pt)=(0.282623,0.140926,0.457517); rgb(231pt)=(0.282884,0.13592,0.453427); rgb(232pt)=(0.283072,0.130895,0.449241); rgb(233pt)=(0.283187,0.125848,0.44496); rgb(234pt)=(0.283229,0.120777,0.440584); rgb(235pt)=(0.283197,0.11568,0.436115); rgb(236pt)=(0.283091,0.110553,0.431554); rgb(237pt)=(0.28291,0.105393,0.426902); rgb(238pt)=(0.282656,0.100196,0.42216); rgb(239pt)=(0.282327,0.0949554,0.417331); rgb(240pt)=(0.281924,0.0896662,0.412415); rgb(241pt)=(0.281446,0.0843197,0.407414); rgb(242pt)=(0.280894,0.078907,0.402329); rgb(243pt)=(0.280267,0.0734172,0.397163); rgb(244pt)=(0.279566,0.0678359,0.391917); rgb(245pt)=(0.278791,0.0621454,0.386592); rgb(246pt)=(0.277941,0.0563244,0.381191); rgb(247pt)=(0.277018,0.0503444,0.375715); rgb(248pt)=(0.276022,0.0441672,0.370164); rgb(249pt)=(0.274952,0.0377518,0.364543); rgb(250pt)=(0.273809,0.0314975,0.358853); rgb(251pt)=(0.272594,0.0255631,0.353093); rgb(252pt)=(0.271305,0.0199419,0.347269); rgb(253pt)=(0.269944,0.0146249,0.341379); rgb(254pt)=(0.26851,0.00960483,0.335427); rgb(255pt)=(0,0,0)},
x dir=reverse,
xmin=-45,
xmax=45,
xlabel style={font=\color{white!15!black},yshift=0.25em},
xlabel={Azimuth $\vartheta \ (^\circ)$},
ymin=-32,
ymax=32,
axis equal,
ylabel style={font=\color{white!15!black}, yshift=-0.25em},
ylabel={Elevation $\varphi \ (^\circ)$},
axis background/.style={fill=white},
axis x line*=bottom,
axis y line*=left,
xmajorgrids,
ymajorgrids,
legend style={legend cell align=left, align=left, draw=white!15!black},
colorbar,
colorbar style={title={$\norm*{\bm{\hat{h}}_{\hat{p}}}_2^2 \, / \, T^2$ (dB)},font=\footnotesize,xshift=2.75em}
]

\addplot[scatter, only marks, mark=triangle*, mark size=2.25pt, semithick, scatter src=explicit, scatter/use mapped color={mark options={}, draw=targetcolor, fill=white}] table[row sep=crcr, meta=color]{%
	x	y	color\\
	37	-10.5	-24\\
	22	-11.0	-24\\
	6	-11		-24\\
	22 	-22		-24\\
	-7.4437   -6.5378  -24.0000\\
	-22.2474   -6.6584  -24.0000\\
	-36.3996   -6.3817  -24.0000\\
	-15.5726  -10.7206  -24.0000\\
	-28.2202  -11.0848  -24.0000\\
	-22.2474    7.9459  -24.0000\\
	-22.2474    0.1454  -24.0000\\
	-22.2474  -13.2796  -24.0000\\
	-22.2474  -20.3304  -24.0000\\
	-10.8473    3.1690  -24.0000\\
	-32.7147    3.1225  -24.0000\\
	-10.4734  -15.7311  -24.0000\\
	-33.2373  -15.4606  -24.0000\\
};

\addplot[scatter, only marks, mark=*, mark size=.95pt, scatter src=explicit, scatter/use mapped color={mark options={}, draw=mapped color, fill=mapped color}] table[row sep=crcr, meta=color]{%
x	y	color\\
-40.5916821028217	-11.0229489087965	-24.9347063968869\\
38.4709021210865	-8.71772350997695	-21.2798829128367\\
-34.0009062786496	-6.34708978880937	-21.8757793772415\\
-31.6671564904279	-14.5148107186875	-18.0925823115337\\
-24.7394724425013	-21.347797507179	-21.3601906346787\\
24.7394724425013	-21.347797507179	-23.3371076033731\\
-32.1706527855172	2.08579521353992	-17.6591198000353\\
17.3351854603275	-24.8741641950295	-19.4960504241423\\
-21.5810338411767	-9.73243196820865	-16.7161717719938\\
21.5810338411767	-9.73243196820865	-15.0230076025481\\
-22.2992933865783	7.86318410398364	-17.825308590169\\
-21.5101882668875	-0	-22.3540860961808\\
-9.46216878890615	-17.096009423715	-20.2873952519481\\
-13.8135658545711	-10.7804969483605	-17.5558964396702\\
7.3917269982062	-11.3202372131057	-19.4608107459775\\
-11.398660388876	1.79289950709647	-21.2439902800517\\
-5.36482769501651	-7.96558513627194	-17.2791873456567\\
-2.70495971641648	-2.7019503016104	-21.9341826181961\\
};

\end{axis}
\end{tikzpicture}%

%% file: figures/20240305_imaging_4pleft_peaceright_omp100_thresh10_spher_analytical_normalized_colorinv.tex
\definecolor{targetcolor}{rgb}{0.83203,0.36719,0.00000}%

\begin{tikzpicture}

\begin{axis}[%
height=.14\textheight,
at={(0.679in,0.668in)},
scale only axis,
point meta min=-28,
point meta max=-14,
colormap={mymap}{[1pt] rgb(0pt)=(0.993248,0.906157,0.143936); rgb(1pt)=(0.983868,0.904867,0.136897); rgb(2pt)=(0.974417,0.90359,0.130215); rgb(3pt)=(0.964894,0.902323,0.123941); rgb(4pt)=(0.9553,0.901065,0.118128); rgb(5pt)=(0.945636,0.899815,0.112838); rgb(6pt)=(0.935904,0.89857,0.108131); rgb(7pt)=(0.926106,0.89733,0.104071); rgb(8pt)=(0.916242,0.896091,0.100717); rgb(9pt)=(0.906311,0.894855,0.098125); rgb(10pt)=(0.89632,0.893616,0.0963354); rgb(11pt)=(0.886271,0.892374,0.0953744); rgb(12pt)=(0.876168,0.891125,0.0952505); rgb(13pt)=(0.866013,0.889868,0.0959528); rgb(14pt)=(0.85581,0.888601,0.0974519); rgb(15pt)=(0.845561,0.887322,0.0997022); rgb(16pt)=(0.83527,0.886029,0.102646); rgb(17pt)=(0.82494,0.88472,0.106217); rgb(18pt)=(0.814576,0.883393,0.110347); rgb(19pt)=(0.804182,0.882046,0.114965); rgb(20pt)=(0.79376,0.880678,0.120005); rgb(21pt)=(0.783315,0.879285,0.125405); rgb(22pt)=(0.772852,0.877868,0.131109); rgb(23pt)=(0.762373,0.876424,0.137064); rgb(24pt)=(0.751884,0.874951,0.143228); rgb(25pt)=(0.741388,0.873449,0.149561); rgb(26pt)=(0.730889,0.871916,0.156029); rgb(27pt)=(0.720391,0.87035,0.162603); rgb(28pt)=(0.709898,0.868751,0.169257); rgb(29pt)=(0.699415,0.867117,0.175971); rgb(30pt)=(0.688944,0.865448,0.182725); rgb(31pt)=(0.678489,0.863742,0.189503); rgb(32pt)=(0.668054,0.861999,0.196293); rgb(33pt)=(0.657642,0.860219,0.203082); rgb(34pt)=(0.647257,0.8584,0.209861); rgb(35pt)=(0.636902,0.856542,0.21662); rgb(36pt)=(0.626579,0.854645,0.223353); rgb(37pt)=(0.616293,0.852709,0.230052); rgb(38pt)=(0.606045,0.850733,0.236712); rgb(39pt)=(0.595839,0.848717,0.243329); rgb(40pt)=(0.585678,0.846661,0.249897); rgb(41pt)=(0.575563,0.844566,0.256415); rgb(42pt)=(0.565498,0.84243,0.262877); rgb(43pt)=(0.555484,0.840254,0.269281); rgb(44pt)=(0.545524,0.838039,0.275626); rgb(45pt)=(0.535621,0.835785,0.281908); rgb(46pt)=(0.525776,0.833491,0.288127); rgb(47pt)=(0.515992,0.831158,0.294279); rgb(48pt)=(0.506271,0.828786,0.300362); rgb(49pt)=(0.496615,0.826376,0.306377); rgb(50pt)=(0.487026,0.823929,0.312321); rgb(51pt)=(0.477504,0.821444,0.318195); rgb(52pt)=(0.468053,0.818921,0.323998); rgb(53pt)=(0.458674,0.816363,0.329727); rgb(54pt)=(0.449368,0.813768,0.335384); rgb(55pt)=(0.440137,0.811138,0.340967); rgb(56pt)=(0.430983,0.808473,0.346476); rgb(57pt)=(0.421908,0.805774,0.35191); rgb(58pt)=(0.412913,0.803041,0.357269); rgb(59pt)=(0.404001,0.800275,0.362552); rgb(60pt)=(0.395174,0.797475,0.367757); rgb(61pt)=(0.386433,0.794644,0.372886); rgb(62pt)=(0.377779,0.791781,0.377939); rgb(63pt)=(0.369214,0.788888,0.382914); rgb(64pt)=(0.360741,0.785964,0.387814); rgb(65pt)=(0.35236,0.783011,0.392636); rgb(66pt)=(0.344074,0.780029,0.397381); rgb(67pt)=(0.335885,0.777018,0.402049); rgb(68pt)=(0.327796,0.77398,0.40664); rgb(69pt)=(0.319809,0.770914,0.411152); rgb(70pt)=(0.311925,0.767822,0.415586); rgb(71pt)=(0.304148,0.764704,0.419943); rgb(72pt)=(0.296479,0.761561,0.424223); rgb(73pt)=(0.288921,0.758394,0.428426); rgb(74pt)=(0.281477,0.755203,0.432552); rgb(75pt)=(0.274149,0.751988,0.436601); rgb(76pt)=(0.266941,0.748751,0.440573); rgb(77pt)=(0.259857,0.745492,0.444467); rgb(78pt)=(0.252899,0.742211,0.448284); rgb(79pt)=(0.24607,0.73891,0.452024); rgb(80pt)=(0.239374,0.735588,0.455688); rgb(81pt)=(0.232815,0.732247,0.459277); rgb(82pt)=(0.226397,0.728888,0.462789); rgb(83pt)=(0.220124,0.725509,0.466226); rgb(84pt)=(0.214,0.722114,0.469588); rgb(85pt)=(0.20803,0.718701,0.472873); rgb(86pt)=(0.202219,0.715272,0.476084); rgb(87pt)=(0.196571,0.711827,0.479221); rgb(88pt)=(0.19109,0.708366,0.482284); rgb(89pt)=(0.185783,0.704891,0.485273); rgb(90pt)=(0.180653,0.701402,0.488189); rgb(91pt)=(0.175707,0.6979,0.491033); rgb(92pt)=(0.170948,0.694384,0.493803); rgb(93pt)=(0.166383,0.690856,0.496502); rgb(94pt)=(0.162016,0.687316,0.499129); rgb(95pt)=(0.157851,0.683765,0.501686); rgb(96pt)=(0.153894,0.680203,0.504172); rgb(97pt)=(0.150148,0.676631,0.506589); rgb(98pt)=(0.146616,0.67305,0.508936); rgb(99pt)=(0.143303,0.669459,0.511215); rgb(100pt)=(0.14021,0.665859,0.513427); rgb(101pt)=(0.137339,0.662252,0.515571); rgb(102pt)=(0.134692,0.658636,0.517649); rgb(103pt)=(0.132268,0.655014,0.519661); rgb(104pt)=(0.130067,0.651384,0.521608); rgb(105pt)=(0.128087,0.647749,0.523491); rgb(106pt)=(0.126326,0.644107,0.525311); rgb(107pt)=(0.12478,0.640461,0.527068); rgb(108pt)=(0.123444,0.636809,0.528763); rgb(109pt)=(0.122312,0.633153,0.530398); rgb(110pt)=(0.12138,0.629492,0.531973); rgb(111pt)=(0.120638,0.625828,0.533488); rgb(112pt)=(0.120081,0.622161,0.534946); rgb(113pt)=(0.119699,0.61849,0.536347); rgb(114pt)=(0.119483,0.614817,0.537692); rgb(115pt)=(0.119423,0.611141,0.538982); rgb(116pt)=(0.119512,0.607464,0.540218); rgb(117pt)=(0.119738,0.603785,0.5414); rgb(118pt)=(0.120092,0.600104,0.54253); rgb(119pt)=(0.120565,0.596422,0.543611); rgb(120pt)=(0.121148,0.592739,0.544641); rgb(121pt)=(0.121831,0.589055,0.545623); rgb(122pt)=(0.122606,0.585371,0.546557); rgb(123pt)=(0.123463,0.581687,0.547445); rgb(124pt)=(0.124395,0.578002,0.548287); rgb(125pt)=(0.125394,0.574318,0.549086); rgb(126pt)=(0.126453,0.570633,0.549841); rgb(127pt)=(0.127568,0.566949,0.550556); rgb(128pt)=(0.128729,0.563265,0.551229); rgb(129pt)=(0.129933,0.559582,0.551864); rgb(130pt)=(0.131172,0.555899,0.552459); rgb(131pt)=(0.132444,0.552216,0.553018); rgb(132pt)=(0.133743,0.548535,0.553541); rgb(133pt)=(0.135066,0.544853,0.554029); rgb(134pt)=(0.136408,0.541173,0.554483); rgb(135pt)=(0.13777,0.537492,0.554906); rgb(136pt)=(0.139147,0.533812,0.555298); rgb(137pt)=(0.140536,0.530132,0.555659); rgb(138pt)=(0.141935,0.526453,0.555991); rgb(139pt)=(0.143343,0.522773,0.556295); rgb(140pt)=(0.144759,0.519093,0.556572); rgb(141pt)=(0.14618,0.515413,0.556823); rgb(142pt)=(0.147607,0.511733,0.557049); rgb(143pt)=(0.149039,0.508051,0.55725); rgb(144pt)=(0.150476,0.504369,0.55743); rgb(145pt)=(0.151918,0.500685,0.557587); rgb(146pt)=(0.153364,0.497,0.557724); rgb(147pt)=(0.154815,0.493313,0.55784); rgb(148pt)=(0.15627,0.489624,0.557936); rgb(149pt)=(0.157729,0.485932,0.558013); rgb(150pt)=(0.159194,0.482237,0.558073); rgb(151pt)=(0.160665,0.47854,0.558115); rgb(152pt)=(0.162142,0.474838,0.55814); rgb(153pt)=(0.163625,0.471133,0.558148); rgb(154pt)=(0.165117,0.467423,0.558141); rgb(155pt)=(0.166617,0.463708,0.558119); rgb(156pt)=(0.168126,0.459988,0.558082); rgb(157pt)=(0.169646,0.456262,0.55803); rgb(158pt)=(0.171176,0.45253,0.557965); rgb(159pt)=(0.172719,0.448791,0.557885); rgb(160pt)=(0.174274,0.445044,0.557792); rgb(161pt)=(0.175841,0.44129,0.557685); rgb(162pt)=(0.177423,0.437527,0.557565); rgb(163pt)=(0.179019,0.433756,0.55743); rgb(164pt)=(0.180629,0.429975,0.557282); rgb(165pt)=(0.182256,0.426184,0.55712); rgb(166pt)=(0.183898,0.422383,0.556944); rgb(167pt)=(0.185556,0.41857,0.556753); rgb(168pt)=(0.187231,0.414746,0.556547); rgb(169pt)=(0.188923,0.41091,0.556326); rgb(170pt)=(0.190631,0.407061,0.556089); rgb(171pt)=(0.192357,0.403199,0.555836); rgb(172pt)=(0.1941,0.399323,0.555565); rgb(173pt)=(0.19586,0.395433,0.555276); rgb(174pt)=(0.197636,0.391528,0.554969); rgb(175pt)=(0.19943,0.387607,0.554642); rgb(176pt)=(0.201239,0.38367,0.554294); rgb(177pt)=(0.203063,0.379716,0.553925); rgb(178pt)=(0.204903,0.375746,0.553533); rgb(179pt)=(0.206756,0.371758,0.553117); rgb(180pt)=(0.208623,0.367752,0.552675); rgb(181pt)=(0.210503,0.363727,0.552206); rgb(182pt)=(0.212395,0.359683,0.55171); rgb(183pt)=(0.214298,0.355619,0.551184); rgb(184pt)=(0.21621,0.351535,0.550627); rgb(185pt)=(0.21813,0.347432,0.550038); rgb(186pt)=(0.220057,0.343307,0.549413); rgb(187pt)=(0.221989,0.339161,0.548752); rgb(188pt)=(0.223925,0.334994,0.548053); rgb(189pt)=(0.225863,0.330805,0.547314); rgb(190pt)=(0.227802,0.326594,0.546532); rgb(191pt)=(0.229739,0.322361,0.545706); rgb(192pt)=(0.231674,0.318106,0.544834); rgb(193pt)=(0.233603,0.313828,0.543914); rgb(194pt)=(0.235526,0.309527,0.542944); rgb(195pt)=(0.237441,0.305202,0.541921); rgb(196pt)=(0.239346,0.300855,0.540844); rgb(197pt)=(0.241237,0.296485,0.539709); rgb(198pt)=(0.243113,0.292092,0.538516); rgb(199pt)=(0.244972,0.287675,0.53726); rgb(200pt)=(0.246811,0.283237,0.535941); rgb(201pt)=(0.248629,0.278775,0.534556); rgb(202pt)=(0.250425,0.27429,0.533103); rgb(203pt)=(0.252194,0.269783,0.531579); rgb(204pt)=(0.253935,0.265254,0.529983); rgb(205pt)=(0.255645,0.260703,0.528312); rgb(206pt)=(0.257322,0.25613,0.526563); rgb(207pt)=(0.258965,0.251537,0.524736); rgb(208pt)=(0.260571,0.246922,0.522828); rgb(209pt)=(0.262138,0.242286,0.520837); rgb(210pt)=(0.263663,0.237631,0.518762); rgb(211pt)=(0.265145,0.232956,0.516599); rgb(212pt)=(0.26658,0.228262,0.514349); rgb(213pt)=(0.267968,0.223549,0.512008); rgb(214pt)=(0.269308,0.218818,0.509577); rgb(215pt)=(0.270595,0.214069,0.507052); rgb(216pt)=(0.271828,0.209303,0.504434); rgb(217pt)=(0.273006,0.20452,0.501721); rgb(218pt)=(0.274128,0.199721,0.498911); rgb(219pt)=(0.275191,0.194905,0.496005); rgb(220pt)=(0.276194,0.190074,0.493001); rgb(221pt)=(0.277134,0.185228,0.489898); rgb(222pt)=(0.278012,0.180367,0.486697); rgb(223pt)=(0.278826,0.17549,0.483397); rgb(224pt)=(0.279574,0.170599,0.479997); rgb(225pt)=(0.280255,0.165693,0.476498); rgb(226pt)=(0.280868,0.160771,0.472899); rgb(227pt)=(0.281412,0.155834,0.469201); rgb(228pt)=(0.281887,0.150881,0.465405); rgb(229pt)=(0.28229,0.145912,0.46151); rgb(230pt)=(0.282623,0.140926,0.457517); rgb(231pt)=(0.282884,0.13592,0.453427); rgb(232pt)=(0.283072,0.130895,0.449241); rgb(233pt)=(0.283187,0.125848,0.44496); rgb(234pt)=(0.283229,0.120777,0.440584); rgb(235pt)=(0.283197,0.11568,0.436115); rgb(236pt)=(0.283091,0.110553,0.431554); rgb(237pt)=(0.28291,0.105393,0.426902); rgb(238pt)=(0.282656,0.100196,0.42216); rgb(239pt)=(0.282327,0.0949554,0.417331); rgb(240pt)=(0.281924,0.0896662,0.412415); rgb(241pt)=(0.281446,0.0843197,0.407414); rgb(242pt)=(0.280894,0.078907,0.402329); rgb(243pt)=(0.280267,0.0734172,0.397163); rgb(244pt)=(0.279566,0.0678359,0.391917); rgb(245pt)=(0.278791,0.0621454,0.386592); rgb(246pt)=(0.277941,0.0563244,0.381191); rgb(247pt)=(0.277018,0.0503444,0.375715); rgb(248pt)=(0.276022,0.0441672,0.370164); rgb(249pt)=(0.274952,0.0377518,0.364543); rgb(250pt)=(0.273809,0.0314975,0.358853); rgb(251pt)=(0.272594,0.0255631,0.353093); rgb(252pt)=(0.271305,0.0199419,0.347269); rgb(253pt)=(0.269944,0.0146249,0.341379); rgb(254pt)=(0.26851,0.00960483,0.335427); rgb(255pt)=(0,0,0)},
x dir=reverse,
xmin=-45,
xmax=45,
xlabel style={font=\color{white!15!black},yshift=0.25em},
xlabel={Azimuth $\vartheta \ (^\circ)$},
ymin=-32,
ymax=32,
axis equal,
ylabel style={font=\color{white!15!black}, yshift=-0.25em},
ylabel={Elevation $\varphi \ (^\circ)$},
axis background/.style={fill=white},
axis x line*=bottom,
axis y line*=left,
xmajorgrids,
ymajorgrids,
legend style={legend cell align=left, align=left, draw=white!15!black},
colorbar,
colorbar style={title={$\norm*{\bm{\hat{h}}_{\hat{p}}}_2^2 \, / \, T^2$ (dB)},font=\footnotesize,xshift=2.75em}
]

\addplot[scatter, only marks, mark=triangle*, mark size=2.25pt, semithick, scatter src=explicit, scatter/use mapped color={mark options={}, draw=targetcolor, fill=white}] table[row sep=crcr, meta=color]{%
	x	y	color\\
	37	-10.5	-24\\
	22	-11.0	-24\\
	6	-11		-24\\
	22 	-22		-24\\
	-7.4437   -6.5378  -24.0000\\
	-22.2474   -6.6584  -24.0000\\
	-36.3996   -6.3817  -24.0000\\
	-15.5726  -10.7206  -24.0000\\
	-28.2202  -11.0848  -24.0000\\
	-22.2474    7.9459  -24.0000\\
	-22.2474    0.1454  -24.0000\\
	-22.2474  -13.2796  -24.0000\\
	-22.2474  -20.3304  -24.0000\\
	-10.8473    3.1690  -24.0000\\
	-32.7147    3.1225  -24.0000\\
	-10.4734  -15.7311  -24.0000\\
	-33.2373  -15.4606  -24.0000\\
};

\addplot[scatter, only marks, mark=*, mark size=.95pt, scatter src=explicit, scatter/use mapped color={mark options={}, draw=mapped color, fill=mapped color}] table[row sep=crcr, meta=color]{%
x	y	color\\
-38.8334622656322	-6.55526540862824	-22.4495672924847\\
38.4709021210865	-8.71772350997695	-21.5346502918602\\
-33.314118158639	-16.8012213982218	-23.8108978293187\\
-33.0740970042897	-10.4949598351109	-17.0090232347393\\
-26.0621894544297	-19.66866002054	-22.2641384301352\\
-31.9895559314697	4.16461246480653	-17.011744689536\\
15.4664754239383	-26.0298597610893	-21.173767352473\\
23.573194092791	-19.1128832291448	-22.9110978395548\\
-21.5810338411767	-9.73243196820865	-14.5667951481918\\
21.5810338411767	-9.73243196820865	-15.4589686114252\\
-21.5810338411767	9.73243196820864	-17.4125456285243\\
-21.5101882668875	-0	-18.6929507358682\\
-9.46216878890615	-17.096009423715	-18.5237504170275\\
-13.8135658545711	-10.7804969483605	-20.1721957451902\\
6.84170940572693	-9.31158746232228	-18.6337674236385\\
-10.9863785356033	3.54333208183254	-16.3155697516686\\
-6.81582191825822	-6.76810136869058	-17.1650874998933\\
};

\end{axis}
\end{tikzpicture}%

%% file: 202411XX_arxiv_rmueller_tensor_model_for_us.bbl
\begin{thebibliography}{10}
\expandafter\ifx\csname url\endcsname\relax
  \def\url#1{\texttt{#1}}\fi
\expandafter\ifx\csname urlprefix\endcsname\relax\def\urlprefix{URL }\fi
\expandafter\ifx\csname href\endcsname\relax
  \def\href#1#2{#2} \def\path#1{#1}\fi

\bibitem{huangFastReductionSpeckle2013}
J.~Huang, X.~Yang, Fast reduction of speckle noise in real ultrasound images,
  Signal Process. 93~(4) (2013) 684--694.
\newblock \href {https://doi.org/10.1016/j.sigpro.2012.09.005}
  {\path{doi:10.1016/j.sigpro.2012.09.005}}.

\bibitem{yuEnvelopeSignalBased2012}
C.~Yu, C.~Zhang, L.~Xie, An envelope signal based deconvolution algorithm for
  ultrasound imaging, Signal Process. 92~(3) (2012) 793--800.
\newblock \href {https://doi.org/10.1016/j.sigpro.2011.09.024}
  {\path{doi:10.1016/j.sigpro.2011.09.024}}.

\bibitem{kiymikUltrasoundImagingBased1997}
M.~K. Kiymik, I.~G{\"u}ler, O.~Hasekioglu, M.~Karaman, Ultrasound imaging based
  on multiple beamforming with coded excitation, Signal Process. 58~(1) (1997)
  107--113.
\newblock \href {https://doi.org/10.1016/S0165-1684(97)00016-9}
  {\path{doi:10.1016/S0165-1684(97)00016-9}}.

\bibitem{harputUltrasonicPhasedArray2008}
S.~Harput, A.~Bozkurt, Ultrasonic phased array device for acoustic imaging in
  air, IEEE Sensors J. 8~(11) (2008) 1755--1762.
\newblock \href {https://doi.org/10.1109/JSEN.2008.2004574}
  {\path{doi:10.1109/JSEN.2008.2004574}}.

\bibitem{synnevagAdaptiveBeamformingApplied2007}
J.~F. Synnevag, A.~Austeng, S.~Holm, Adaptive beamforming applied to medical
  ultrasound imaging, IEEE Open J. Ultrason., Ferroelect., Freq. Contr. 54~(8)
  (2007) 1606--1613.
\newblock \href {https://doi.org/10.1109/TUFFC.2007.431}
  {\path{doi:10.1109/TUFFC.2007.431}}.

\bibitem{jensenSyntheticApertureUltrasound2006}
J.~A. Jensen, S.~I. Nikolov, K.~L. Gammelmark, M.~H. Pedersen, Synthetic
  aperture ultrasound imaging, Ultrason. 44 (2006) e5--e15.
\newblock \href {https://doi.org/10.1016/j.ultras.2006.07.017}
  {\path{doi:10.1016/j.ultras.2006.07.017}}.

\bibitem{leightonWhatUltrasound2007}
T.~G. Leighton, What is ultrasound?, Prog. in Biophys. Mol. Biol. 93~(1) (2007)
  3--83.
\newblock \href {https://doi.org/10.1016/j.pbiomolbio.2006.07.026}
  {\path{doi:10.1016/j.pbiomolbio.2006.07.026}}.

\bibitem{mohammedPerceptionSystemIntelligent2020}
A.~S. Mohammed, A.~Amamou, F.~K. Ayevide, S.~Kelouwani, K.~Agbossou, N.~Zioui,
  The perception system of intelligent ground vehicles in all weather
  conditions: {{A}} systematic literature review, Sensors 20~(22) (2020) 6532.
\newblock \href {https://doi.org/10.3390/s20226532}
  {\path{doi:10.3390/s20226532}}.

\bibitem{patoleAutomotiveRadarsReview2017}
S.~M. Patole, M.~Torlak, D.~Wang, M.~Ali, Automotive radars: {{A}} review of
  signal processing techniques, IEEE Signal Process. Mag. 34~(2) (2017) 22--35.
\newblock \href {https://doi.org/10.1109/MSP.2016.2628914}
  {\path{doi:10.1109/MSP.2016.2628914}}.

\bibitem{fangReviewEmergingElectromagneticacoustic2022}
Z.~Fang, F.~Gao, H.~Jin, S.~Liu, W.~Wang, R.~Zhang, Z.~Zheng, X.~Xiao, K.~Tang,
  L.~Lou, K.-T. Tang, J.~Chen, Y.~Zheng, A review of emerging
  electromagnetic-acoustic sensing techniques for healthcare monitoring, IEEE
  Trans. Biomed. Circuits Syst. 16~(6) (2022) 1075--1094.
\newblock \href {https://doi.org/10.1109/TBCAS.2022.3226290}
  {\path{doi:10.1109/TBCAS.2022.3226290}}.

\bibitem{chimentiReviewAircoupledUltrasonic2014}
D.~E. Chimenti, Review of air-coupled ultrasonic materials characterization,
  Ultrason. 54~(7) (2014) 1804--1816.
\newblock \href {https://doi.org/10.1016/j.ultras.2014.02.006}
  {\path{doi:10.1016/j.ultras.2014.02.006}}.

\bibitem{haller13CompositesUltrasonic1992}
M.~Haller, B.~{Khuri-Yakub}, 1-3 composites for ultrasonic air transducers, in:
  {{IEEE}} 1992 {{Ultrason}}. {{Symp}}., Vol.~2, 1992, pp. 937--939.
\newblock \href {https://doi.org/10.1109/ULTSYM.1992.275822}
  {\path{doi:10.1109/ULTSYM.1992.275822}}.

\bibitem{lionettoAircoupledUltrasoundNovel2007}
F.~Lionetto, A.~Tarzia, A.~Maffezzoli, Air-coupled ultrasound: {{A}} novel
  technique for monitoring the curing of thermosetting matrices, IEEE Trans.
  Ultrason. Ferroelectr. Freq. Control 54~(7) (2007) 1437--1444.
\newblock \href {https://doi.org/10.1109/TUFFC.2007.404}
  {\path{doi:10.1109/TUFFC.2007.404}}.

\bibitem{bassAtmosphericAbsorptionSound1990}
H.~E. Bass, L.~C. Sutherland, A.~J. Zuckerwar, Atmospheric absorption of sound:
  {{Update}}, J. Acoust. Soc. Am. 88~(4) (1990) 2019--2021.
\newblock \href {https://doi.org/10.1121/1.400176}
  {\path{doi:10.1121/1.400176}}.

\bibitem{strakowskiUltrasonicObstacleDetector2006}
M.~Strakowski, B.~Kosmowski, R.~Kowalik, P.~Wierzba, An ultrasonic obstacle
  detector based on phase beamforming principles, IEEE Sensors J. 6~(1) (2006)
  179--186.
\newblock \href {https://doi.org/10.1109/JSEN.2005.856129}
  {\path{doi:10.1109/JSEN.2005.856129}}.

\bibitem{dahlApplicationsAirborneUltrasound2014}
T.~Dahl, J.~L. Ealo, H.~J. Bang, S.~Holm, P.~{Khuri-Yakub}, Applications of
  airborne ultrasound in human--computer interaction, Ultrason. 54~(7) (2014)
  1912--1921.
\newblock \href {https://doi.org/10.1016/j.ultras.2014.04.008}
  {\path{doi:10.1016/j.ultras.2014.04.008}}.

\bibitem{suzukiAUTD3ScalableAirborne2021}
S.~Suzuki, S.~Inoue, M.~Fujiwara, Y.~Makino, H.~Shinoda, {{AUTD3}}:
  {{Scalable}} airborne ultrasound tactile display, IEEE Trans. Haptics 14~(4)
  (2021) 740--749.
\newblock \href {https://doi.org/10.1109/TOH.2021.3069976}
  {\path{doi:10.1109/TOH.2021.3069976}}.

\bibitem{leggUltrasonicArraysRemote2020}
M.~Legg, S.~Bradley, Ultrasonic arrays for remote sensing of pasture biomass,
  Remote Sensing 12~(1) (2020) 111.
\newblock \href {https://doi.org/10.3390/rs12010111}
  {\path{doi:10.3390/rs12010111}}.

\bibitem{rekhiWirelessPowerTransfer2017}
A.~S. Rekhi, B.~T. {Khuri-Yakub}, A.~Arbabian, Wireless power transfer to
  millimeter-sized nodes using airborne ultrasound, IEEE Trans. Ultrason.
  Ferroelectr. Freq. Control 64~(10) (2017) 1526--1541.
\newblock \href {https://doi.org/10.1109/TUFFC.2017.2737620}
  {\path{doi:10.1109/TUFFC.2017.2737620}}.

\bibitem{marzoHolographicAcousticElements2015}
A.~Marzo, S.~A. Seah, B.~W. Drinkwater, D.~R. Sahoo, B.~Long, S.~Subramanian,
  Holographic acoustic elements for manipulation of levitated objects, Nat.
  Commun. 6~(1) (2015) 8661.
\newblock \href {https://doi.org/10.1038/ncomms9661}
  {\path{doi:10.1038/ncomms9661}}.

\bibitem{gellyComparisonPiezoelectricThickness2003}
J.~Gelly, F.~Lanteri, Comparison of piezoelectric (thickness mode) and {{MEMS}}
  transducers, in: {{IEEE Symp}}. {{Ultrason}}., Vol.~2, 2003, pp. 1965--1974
  Vol.2.
\newblock \href {https://doi.org/10.1109/ULTSYM.2003.1293302}
  {\path{doi:10.1109/ULTSYM.2003.1293302}}.

\bibitem{jagerAircoupled40KHZUltrasonic2017a}
A.~J{\"a}ger, D.~Gro{\ss}kurth, M.~Rutsch, A.~Unger, R.~Golinske, H.~Wang,
  S.~Dixon, K.~Hofmann, M.~Kupnik, Air-coupled 40-{{KHZ}} ultrasonic
  {{2D-phased}} array based on a {{3D-printed}} waveguide structure, in: {{IEEE
  Int}}. {{Ultrason}}. {{Symp}}. ({{IUS}}), 2017, pp. 1--4.
\newblock \href {https://doi.org/10.1109/ULTSYM.2017.8091892}
  {\path{doi:10.1109/ULTSYM.2017.8091892}}.

\bibitem{vibergCalibrationArrayProcessing2009}
M.~Viberg, M.~Lanne, A.~Lundgren, Calibration in array processing, in: T.~E.
  Tuncer, B.~Friedlander (Eds.), Classical and {{Modern Direction-of-Arrival
  Estimation}}, Elsevier Inc., Burlington, 2009, pp. 93--124.
\newblock \href {https://doi.org/10.1016/B978-0-12-374524-8.00003-9}
  {\path{doi:10.1016/B978-0-12-374524-8.00003-9}}.

\bibitem{wuSelfCalibrationDirectPosition2020}
G.~Wu, M.~Zhang, F.~Guo, Self-{{Calibration}} direct position determination
  using a single moving array with sensor gain and phase errors, Signal
  Process. 173 (2020) 107587.
\newblock \href {https://doi.org/10.1016/j.sigpro.2020.107587}
  {\path{doi:10.1016/j.sigpro.2020.107587}}.

\bibitem{ngSensorarrayCalibrationUsing1996}
B.~C. Ng, C.~M.~S. See, Sensor-array calibration using a maximum-likelihood
  approach, IEEE Trans. Antennas Propag. 44~(6) (1996) 827--835.
\newblock \href {https://doi.org/10.1109/8.509886}
  {\path{doi:10.1109/8.509886}}.

\bibitem{vibergBayesianApproachAutocalibration1994}
M.~Viberg, A.~L. Swindlehurst, A {{Bayesian}} approach to auto-calibration for
  parametric array signal processing, IEEE Trans. Signal Process. 42~(12)
  (1994) 3495--3507.
\newblock \href {https://doi.org/10.1109/78.340783}
  {\path{doi:10.1109/78.340783}}.

\bibitem{weissArrayShapeCalibration1989}
A.~J. Weiss, B.~Friedlander, Array shape calibration using sources in unknown
  locations---a maximum likelihood approach, IEEE Trans. Acoust. Speech Signal
  Process. 37~(12) (1989) 1958--1966.
\newblock \href {https://doi.org/10.1109/29.45542}
  {\path{doi:10.1109/29.45542}}.

\bibitem{hanCalibratingNestedSensor2015}
K.~Han, P.~Yang, A.~Nehorai, Calibrating nested sensor arrays with model
  errors, IEEE Trans. Antennas Propag. 63~(11) (2015) 4739--4748.
\newblock \href {https://doi.org/10.1109/TAP.2015.2477411}
  {\path{doi:10.1109/TAP.2015.2477411}}.

\bibitem{liuEigenstructureMethodEstimating2011}
A.~Liu, G.~Liao, C.~Zeng, Z.~Yang, Q.~Xu, An eigenstructure method for
  estimating {{DOA}} and sensor gain-phase errors, IEEE Trans. Signal Process.
  59~(12) (2011) 5944--5956.
\newblock \href {https://doi.org/10.1109/TSP.2011.2165064}
  {\path{doi:10.1109/TSP.2011.2165064}}.

\bibitem{liuDOAEstimationUniform2009}
Z.~Liu, Z.~Huang, F.~Wang, Y.~Zhou, {{DOA}} estimation with uniform linear
  arrays in the presence of mutual coupling via blind calibration, Signal
  Process. 89~(7) (2009) 1446--1456.
\newblock \href {https://doi.org/10.1016/j.sigpro.2009.01.017}
  {\path{doi:10.1016/j.sigpro.2009.01.017}}.

\bibitem{liuClutterbasedGainPhase2019}
Y.~Liu, B.~Jiu, H.~Liu, Clutter-based gain and phase calibration for monostatic
  {{MIMO}} radar with partly calibrated array, Signal Process. 158 (2019)
  219--226.
\newblock \href {https://doi.org/10.1016/j.sigpro.2019.01.011}
  {\path{doi:10.1016/j.sigpro.2019.01.011}}.

\bibitem{liaoDirectionFindingPartly2012}
B.~Liao, S.~C. Chan, Direction finding with partly calibrated uniform linear
  arrays, IEEE Trans. Antennas Propag. 60~(2) (2012) 922--929.
\newblock \href {https://doi.org/10.1109/TAP.2011.2173144}
  {\path{doi:10.1109/TAP.2011.2173144}}.

\bibitem{parvaziDirectionofarrivalEstimationArray2011}
P.~Parvazi, M.~Pesavento, A.~B. Gershman, Direction-of-arrival estimation and
  array calibration for partly-calibrated arrays, in: 2011 {{IEEE Int}}.
  {{Conf}}. {{Acoust}}. {{Speech Signal Process}}. ({{ICASSP}}), 2011, pp.
  2552--2555.
\newblock \href {https://doi.org/10.1109/ICASSP.2011.5947005}
  {\path{doi:10.1109/ICASSP.2011.5947005}}.

\bibitem{wanFourthorderDirectionFinding2020}
H.~Wan, B.~Liao, Fourth-order direction finding in antenna arrays with partial
  channel gain/phase calibration, Signal Process. 169 (2020) 107380.
\newblock \href {https://doi.org/10.1016/j.sigpro.2019.107380}
  {\path{doi:10.1016/j.sigpro.2019.107380}}.

\bibitem{huangLowrankRowsparseDecomposition2023}
H.~Huang, Q.~Liu, H.~C. So, A.~M. Zoubir, Low-rank and row-sparse decomposition
  for joint {{DOA}} estimation and distorted sensor detection, IEEE Trans.
  Aerosp. Electron. Syst. 59~(4) (2023) 4763--4773.
\newblock \href {https://doi.org/10.1109/TAES.2023.3241886}
  {\path{doi:10.1109/TAES.2023.3241886}}.

\bibitem{geissAntennaArrayCalibration2021}
J.~Geiss, E.~Sippel, M.~Hehn, M.~Vossiek, Antenna array calibration using a
  sparse scene, IEEE Open J. Antennas Propag. 2 (2021) 349--361.
\newblock \href {https://doi.org/10.1109/OJAP.2021.3061935}
  {\path{doi:10.1109/OJAP.2021.3061935}}.

\bibitem{taghizadehAdHocMicrophone2015}
M.~J. Taghizadeh, R.~Parhizkar, P.~N. Garner, H.~Bourlard, A.~Asaei, Ad hoc
  microphone array calibration: {{Euclidean}} distance matrix completion
  algorithm and theoretical guarantees, Signal Process. 107 (2015) 123--140.
\newblock \href {https://doi.org/10.1016/j.sigpro.2014.07.016}
  {\path{doi:10.1016/j.sigpro.2014.07.016}}.

\bibitem{ramamohanSelfcalibrationAcousticScalar2023}
K.~N. Ramamohan, S.~P. Chepuri, D.~F. Comesa{\~n}a, G.~Leus, Self-calibration
  of acoustic scalar and vector sensor arrays, IEEE Trans. Signal Process. 71
  (2023) 61--75.
\newblock \href {https://doi.org/10.1109/TSP.2022.3214383}
  {\path{doi:10.1109/TSP.2022.3214383}}.

\bibitem{ollierRobustCalibrationRadio2017}
V.~Ollier, M.~N.~E. Korso, R.~Boyer, P.~Larzabal, M.~Pesavento, Robust
  calibration of radio interferometers in non-{{Gaussian}} environment, IEEE
  Trans. Signal Process. 65~(21) (2017) 5649--5660.
\newblock \href {https://doi.org/10.1109/TSP.2017.2733496}
  {\path{doi:10.1109/TSP.2017.2733496}}.

\bibitem{pierreExperimentalPerformanceCalibration1991}
J.~Pierre, M.~Kaveh, Experimental performance of calibration and
  direction-finding algorithms, in: 1991 {{Int}}. {{Conf}}. {{Acoust}}.
  {{Speech Signal Process}}. ({{ICASSP}}), 1991, pp. 1365--1368 vol.2.
\newblock \href {https://doi.org/10.1109/ICASSP.1991.150676}
  {\path{doi:10.1109/ICASSP.1991.150676}}.

\bibitem{paulrajDirectionArrivalEstimation1985}
A.~Paulraj, T.~Kailath, Direction of arrival estimation by eigenstructure
  methods with unknown sensor gain and phase, in: 1985 {{IEEE Int}}. {{Conf}}.
  {{Acoust}}. {{Speech Signal Process}}. ({{ICASSP}}), Vol.~10, 1985, pp.
  640--643.
\newblock \href {https://doi.org/10.1109/ICASSP.1985.1168341}
  {\path{doi:10.1109/ICASSP.1985.1168341}}.

\bibitem{nanzerDistributedPhasedArrays2021}
J.~A. Nanzer, S.~R. Mghabghab, S.~M. Ellison, A.~Schlegel, Distributed phased
  arrays: {{Challenges}} and recent advances, IEEE Trans. Microw. Theory Tech.
  69~(11) (2021) 4893--4907.
\newblock \href {https://doi.org/10.1109/TMTT.2021.3092401}
  {\path{doi:10.1109/TMTT.2021.3092401}}.

\bibitem{heidenreichJoint2DDOA2012}
P.~Heidenreich, A.~M. Zoubir, M.~Rubsamen, Joint 2-{{D DOA}} estimation and
  phase calibration for uniform rectangular arrays, IEEE Trans. Signal Process.
  60~(9) (2012) 4683--4693.
\newblock \href {https://doi.org/10.1109/TSP.2012.2203125}
  {\path{doi:10.1109/TSP.2012.2203125}}.

\bibitem{wijnholdsMultisourceSelfcalibrationSensor2009}
S.~J. Wijnholds, A.-J. {van der Veen}, Multisource self-calibration for sensor
  arrays, IEEE Trans. Signal Process. 57~(9) (2009) 3512--3522.
\newblock \href {https://doi.org/10.1109/TSP.2009.2022894}
  {\path{doi:10.1109/TSP.2009.2022894}}.

\bibitem{tosicDictionaryLearning2011}
I.~To{\v s}i{\'c}, P.~Frossard, Dictionary learning, IEEE Signal Process. Mag.
  28~(2) (2011) 27--38.
\newblock \href {https://doi.org/10.1109/MSP.2010.939537}
  {\path{doi:10.1109/MSP.2010.939537}}.

\bibitem{guoTensorbasedAngleArray2019}
Y.~Guo, X.~Wang, L.~Wan, M.~Huang, C.~Shen, K.~Zhang, Y.~Yang, Tensor-based
  {{Angle}} and array gain-phase error estimation scheme in bistatic {{MIMO}}
  radar, IEEE Access 7 (2019) 47972--47981.
\newblock \href {https://doi.org/10.1109/ACCESS.2019.2909760}
  {\path{doi:10.1109/ACCESS.2019.2909760}}.

\bibitem{sunSpaceTimerangeClutter2024}
Y.~Sun, W.-Q. Wang, C.~Jiang, Space--time-range clutter suppression via
  tensor-based {{STAP}} for airborne {{FDA-MIMO}} radar, Signal Process. 214
  (2024) 109235.
\newblock \href {https://doi.org/10.1016/j.sigpro.2023.109235}
  {\path{doi:10.1016/j.sigpro.2023.109235}}.

\bibitem{chenAngleEstimationBased2024}
J.~Chen, Y.~Tang, X.~Zhu, J.~Li, Angle estimation based on {{Vandermonde}}
  constrained {{CP}} tensor decomposition for bistatic {{MIMO}} radar under
  spatially colored noise, Signal Process. 220 (2024) 109429.
\newblock \href {https://doi.org/10.1016/j.sigpro.2024.109429}
  {\path{doi:10.1016/j.sigpro.2024.109429}}.

\bibitem{mullerDictionarybasedLearning3Dimaging2020}
R.~M{\"u}ller, D.~Schenck, G.~Allevato, M.~Rutsch, J.~Hinrichs, M.~Kupnik,
  M.~Pesavento, Dictionary-based learning for {{3D-imaging}} with air-coupled
  ultrasonic phased arrays, in: {{IEEE Int}}. {{Ultrason}}. {{Symp}}.
  ({{IUS}}), 2020, pp. 1--4.
\newblock \href {https://doi.org/10.1109/IUS46767.2020.9251726}
  {\path{doi:10.1109/IUS46767.2020.9251726}}.

\bibitem{kusheParallelSparseRegularization2019}
G.~Kushe, Y.~Yang, C.~Steffens, M.~Pesavento, A parallel sparse regularization
  method for structured multilinear low-rank tensor decomposition, in: 27th
  {{Eur}}. {{Signal Process}}. {{Conf}}. ({{EUSIPCO}}), 2019, pp. 1--5.
\newblock \href {https://doi.org/10.23919/EUSIPCO.2019.8902569}
  {\path{doi:10.23919/EUSIPCO.2019.8902569}}.

\bibitem{liuRobustDetectionMIMO2019}
J.~Liu, J.~Li, Robust detection in {{MIMO}} radar with steering vector
  mismatches, IEEE Trans. Signal Process. 67~(20) (2019) 5270--5280.
\newblock \href {https://doi.org/10.1109/TSP.2019.2939078}
  {\path{doi:10.1109/TSP.2019.2939078}}.

\bibitem{rutschWaveguideAircoupledUltrasonic2021}
M.~Rutsch, A.~Unger, G.~Allevato, J.~Hinrichs, A.~J{\"a}ger, T.~Kaindl,
  M.~Kupnik, Waveguide for air-coupled ultrasonic phased-arrays with
  propagation time compensation and plug-in assembly, J. Acoust. Soc. Am.
  150~(5) (2021) 3228--3237.
\newblock \href {https://doi.org/10.1121/10.0006969}
  {\path{doi:10.1121/10.0006969}}.

\bibitem{vibergIntroductionArrayProcessing2014}
M.~Viberg, Introduction to array processing, in: A.~M. Zoubir, M.~Viberg,
  R.~Chellappa, S.~Theodoridis (Eds.), Array and {{Statistical Signal
  Processing}}, Vol.~3 of Academic {{Press Library}} in {{Signal Processing}},
  Elsevier Ltd., Kidlington, UK, 2014, pp. 463--502.
\newblock \href {https://doi.org/10.1016/B978-0-12-411597-2.00011-4}
  {\path{doi:10.1016/B978-0-12-411597-2.00011-4}}.

\bibitem{allevatoRealtime3DImaging2021}
G.~Allevato, J.~Hinrichs, M.~Rutsch, J.~P. Adler, A.~J{\"a}ger, M.~Pesavento,
  M.~Kupnik, Real-time 3-{{D}} imaging using an air-coupled ultrasonic
  phased-array, IEEE Trans. Ultrason. Ferroelectr. Freq. Control 68~(3) (2021)
  796--806.
\newblock \href {https://doi.org/10.1109/TUFFC.2020.3005292}
  {\path{doi:10.1109/TUFFC.2020.3005292}}.

\bibitem{melvinSpacetimeAdaptiveProcessing2014}
W.~L. Melvin, Space-time adaptive processing for radar, in: N.~D. Sidiropoulos,
  F.~Gini, R.~Chellappa, S.~Theodoridis (Eds.), Communications and {{Radar
  Signal Processing}}, Vol.~2 of Academic {{Press Library}} in {{Signal
  Processing}}, Elsevier Ltd., Kidlington, UK, 2014, pp. 595--665.
\newblock \href {https://doi.org/10.1016/B978-0-12-396500-4.00012-0}
  {\path{doi:10.1016/B978-0-12-396500-4.00012-0}}.

\bibitem{allevatoUltrasonicPhasedArrays2023}
G.~Allevato, Ultrasonic phased arrays for {{3D}} sonar imaging in air, Ph.D.
  thesis, Technische Universit{\"a}t Darmstadt, Darmstadt, Germany (Sep. 2023).
\newblock \href {https://doi.org/10.26083/tuprints-00024425}
  {\path{doi:10.26083/tuprints-00024425}}.

\bibitem{carrollAnalysisIndividualDifferences1970}
J.~D. Carroll, J.-J. Chang, Analysis of individual differences in
  multidimensional scaling via an $n$-way generalization of
  ``{E}ckart-{Y}oung'' decomposition, Psychometrika 35~(3) (1970) 283--319.
\newblock \href {https://doi.org/10.1007/BF02310791}
  {\path{doi:10.1007/BF02310791}}.

\bibitem{harshmanFoundationsPARAFACProcedure1970}
R.~A. Harshman, Foundations of the {PARAFAC} procedure: Models and conditions
  for an ``explanatory'' multi-modal factor analysis, {{UCLA Working Papers}}
  in {{Phonetics}}~16, Ann Arbor, Michigan (1970).

\bibitem{delathauwerDecompositionsHigherorderTensor2008}
L.~De~Lathauwer, Decompositions of a higher-order tensor in block
  terms---{{Part II}}: {{Definitions}} and uniqueness, SIAM J. Matrix Anal.
  Appl. 30~(3) (2008) 1033--1066.
\newblock \href {https://doi.org/10.1137/070690729}
  {\path{doi:10.1137/070690729}}.

\bibitem{cichockiTensorDecompositionsSignal2015}
A.~Cichocki, D.~Mandic, L.~De~Lathauwer, G.~Zhou, Q.~Zhao, C.~Caiafa, H.~A.
  PHAN, Tensor {{Decompositions}} for {{Signal Processing Applications}}:
  {{From}} two-way to multiway component analysis, IEEE Signal Processing
  Magazine 32~(2) (2015) 145--163.
\newblock \href {https://doi.org/10.1109/MSP.2013.2297439}
  {\path{doi:10.1109/MSP.2013.2297439}}.

\bibitem{koldaTensorDecompositionsApplications2009}
T.~G. Kolda, B.~W. Bader, Tensor decompositions and applications, SIAM Rev.
  51~(3) (2009) 455--500.
\newblock \href {https://doi.org/10.1137/07070111X}
  {\path{doi:10.1137/07070111X}}.

\bibitem{kiersStandardizedNotationTerminology2000}
H.~A.~L. Kiers, Towards a standardized notation and terminology in multiway
  analysis, J. Chemom. 14~(3) (2000) 105--122.
\newblock \href
  {https://doi.org/10.1002/1099-128X(200005/06)14:3<105::AID-CEM582>3.0.CO;2-I}
  {\path{doi:10.1002/1099-128X(200005/06)14:3<105::AID-CEM582>3.0.CO;2-I}}.

\bibitem{sidiropoulosTensorDecompositionSignal2017}
N.~D. Sidiropoulos, L.~De~Lathauwer, X.~Fu, K.~Huang, E.~E. Papalexakis,
  C.~Faloutsos, Tensor decomposition for signal processing and machine
  learning, IEEE Trans. Signal Process. 65~(13) (2017) 3551--3582.
\newblock \href {https://doi.org/10.1109/TSP.2017.2690524}
  {\path{doi:10.1109/TSP.2017.2690524}}.

\bibitem{paulrajSubspaceMethodsDirectionsofarrival1993}
A.~Paulraj, B.~Ottersten, R.~Roy, A.~Swindlehurst, G.~Xu, T.~Kailath, Subspace
  methods for directions-of-arrival estimation, in: N.~K. Bose, C.~R. Rao
  (Eds.), Handbook of {{Statistics}}, Vol.~10, Elsevier Science Publishers
  B.V., 1993, pp. 693--739.

\bibitem{yangInexactBlockCoordinate2020}
Y.~Yang, M.~Pesavento, Z.-Q. Luo, B.~Ottersten, Inexact block coordinate
  descent algorithms for nonsmooth nonconvex optimization, IEEE Trans. Signal
  Process. 68 (2020) 947--961.
\newblock \href {https://doi.org/10.1109/TSP.2019.2959240}
  {\path{doi:10.1109/TSP.2019.2959240}}.

\bibitem{bertsekasNonlinearProgramming2016}
D.~P. Bertsekas, Nonlinear Programming, 3rd Edition, Athena Scientific,
  Belmont, 2016.

\bibitem{richardsPrinciplesModernRadar2010}
M.~A. Richards, J.~A. Scheer, W.~A. Holm (Eds.), Principles of {{Modern
  Radar}}: {{Basic Principles}}, Vol.~1, SciTech Publishing, Raleigh (NC),
  2010.

\bibitem{royESPRITestimationSignalParameters1989}
R.~Roy, T.~Kailath, {{ESPRIT-estimation}} of signal parameters via rotational
  invariance techniques, IEEE Trans. Acoust. Speech Signal Process. 37~(7)
  (1989) 984--995.
\newblock \href {https://doi.org/10.1109/29.32276}
  {\path{doi:10.1109/29.32276}}.

\bibitem{kerstensERTISFullyEmbedded2019}
R.~Kerstens, D.~Laurijssen, J.~Steckel, {{eRTIS}}: {{A}} fully embedded real
  time {{3D}} imaging sonar sensor for robotic applications, in: Int. {{Conf}}.
  {{Robotics Autom}}. ({{ICRA}}), 2019, pp. 1438--1443.
\newblock \href {https://doi.org/10.1109/ICRA.2019.8794419}
  {\path{doi:10.1109/ICRA.2019.8794419}}.

\bibitem{schmidtMultipleEmitterLocation1986}
R.~O. Schmidt, Multiple emitter location and signal parameter estimation, IEEE
  Trans. Antennas Propag. 34~(3) (1986) 276--280.
\newblock \href {https://doi.org/10.1109/TAP.1986.1143830}
  {\path{doi:10.1109/TAP.1986.1143830}}.

\bibitem{zhangSurveySparseRepresentation2015c}
Z.~Zhang, Y.~Xu, J.~Yang, X.~Li, D.~Zhang, A survey of sparse representation:
  {{Algorithms}} and applications, IEEE Access 3 (2015) 490--530.
\newblock \href {https://doi.org/10.1109/ACCESS.2015.2430359}
  {\path{doi:10.1109/ACCESS.2015.2430359}}.

\bibitem{pesaventoThreeMoreDecades2023}
M.~Pesavento, M.~{Trinh-Hoang}, M.~Viberg, Three more decades in array signal
  processing research: {{An}} optimization and structure exploitation
  perspective, IEEE Signal Process. Mag. 40~(4) (2023) 92--106.
\newblock \href {https://doi.org/10.1109/MSP.2023.3255558}
  {\path{doi:10.1109/MSP.2023.3255558}}.

\bibitem{patiOrthogonalMatchingPursuit1993}
Y.~Pati, R.~Rezaiifar, P.~Krishnaprasad, Orthogonal matching pursuit:
  {{Recursive}} function approximation with applications to wavelet
  decomposition, in: Proc. 27th {{Asilomar Conf}}. {{Signals Syst}}.
  {{Comput}}., 1993, pp. 40--44 vol.1.
\newblock \href {https://doi.org/10.1109/ACSSC.1993.342465}
  {\path{doi:10.1109/ACSSC.1993.342465}}.

\bibitem{donohoUncertaintyPrinciplesIdeal2001}
D.~L. Donoho, X.~Huo, Uncertainty principles and ideal atomic decomposition,
  IEEE Trans. Inform. Theory 47~(7) (Nov./2001) 2845--2862.
\newblock \href {https://doi.org/10.1109/18.959265}
  {\path{doi:10.1109/18.959265}}.

\bibitem{rutschDuctAcousticsAircoupled2023}
M.~Rutsch, Duct acoustics for air-coupled ultrasonic phased arrays, Ph.D.
  thesis, Technische Universit{\"a}t Darmstadt, Darmstadt, Germany (2023).
\newblock \href {https://doi.org/10.26083/tuprints-00023129}
  {\path{doi:10.26083/tuprints-00023129}}.

\bibitem{richardsFundamentalsradarSignal2022}
M.~A. Richards, Fundamentals of Radar Signal Processing, 3rd Edition, McGraw
  Hill, New York, 2022.

\bibitem{aumannPhasedarrayCalibrationAdaptive1991}
H.~M. Aumann, F.~G. Willwerth, Phased-array calibration by adaptive nulling,
  Tech. Rep. 915, Lincoln Laboratory/Massachusetts Institute of Technology,
  Lexington/Massachusetts (May 1991).

\end{thebibliography}
